\documentclass[12pt]{article}

\usepackage{times}
\usepackage[T1]{fontenc}
\usepackage[utf8]{inputenc}
\usepackage{scicite}
\makeatletter
\def\input@path{{common/}}
\makeatother

\usepackage{amsmath,amssymb,amsthm}
\usepackage{graphicx}
\usepackage{longtable}
\usepackage{booktabs}
\usepackage{xcolor}
\usepackage{hyperref}
\hypersetup{colorlinks=true, linkcolor=blue!50!black, urlcolor=blue!50!black, unicode=true}

\graphicspath{{figures/}{common/figures/}}

\theoremstyle{definition}
\newtheorem{defn}{定義}[section]

\newcommand{\cX}{\Xi}                  

\newcommand{\Pideal}{P^{\mathrm{ideal}}}  

\newcommand{\E}{\mathbb{E}}

\usepackage{placeins}
\usepackage{array}
\usepackage{algorithm}
\usepackage{algpseudocode}
\usepackage[normalem]{ulem}
\newtheorem{assumption}[defn]{Assumption}
\newtheorem{proposition}[defn]{Proposition}

\newenvironment{sciabstract}{\begin{quote}\setlength{\emergencystretch}{3em}}{\end{quote}}

\title{From joint attention to shared symbols:\\ A free-energy principle account}
\author{Tadahiro Taniguchi$^{1,3,*}$, Koji Hashimoto$^{2}$, and Ryuichiro Ishikawa$^{4}$}
\date{}

\begin{document}
\maketitle
\begin{center}
\small
$^{1}$Graduate School of Informatics, Kyoto University, Kyoto, Japan\\
$^{2}$Department of Physics, Graduate School of Science, Kyoto University, Kyoto, Japan\\
$^{3}$Ritsumeikan University, Shiga, Japan\\
$^{4}$Faculty of International Research and Education, Waseda University\\
$^{*}$Corresponding author: taniguchi@i.kyoto-u.ac.jp
\end{center}

\noindent\textbf{Short title:} Joint attention and shared symbols

\noindent\textbf{Abstract:}
\begin{sciabstract}
Joint attention precedes vocabulary acquisition. How can attending to the same object help two people with different sensory experiences come to use a shared word? In a free-energy principle account, marginalizing the common object induces a statistical dependence between private perceptual states. In a categorical world, the joint prior inherits a mixture structure whose component index provides a candidate for shared reference. The collective predictive coding model approximates this prior by representing the dependence through that shared index. An established Metropolis--Hastings naming game lets agents infer and exchange words using private evidence. Toy and multiview image experiments illustrate how word exchange can coordinate private categories and improve the fit of collective models. The account connects joint attention to shared symbols through the coupling of private inference.
\end{sciabstract}
 
\noindent\textbf{Teaser:} A common object couples private inference, supporting a free-energy account of how joint attention can enable shared symbols.
\section{Introduction}
\label{sec:introduction}
How can individuals with different private experiences come to use shared symbols for the same world? Accounts of human communication place joint attention and shared intentionality near the center of this question \cite{tomasello2005understanding,tomasello2008origins}. Attending together to an object gives individuals a shared referential context,  even though their distinct sensory perspectives provide them with different observations of that same object. What remains less explicit is the computational mechanism by which this relation can support a reusable system of shared symbols.  Here, we develop a free-energy principle account \cite{friston2010free} that connects joint attention to the collective predictive coding (CPC) hypothesis \cite{taniguchi2024collective,taniguchi2026generative,taniguchi2025systems}. We examine how a shared object couples private inference and, in a categorical world, provides structure that can support a shared referential symbol.

Developmental psychology and comparative cognition motivate this starting point. Joint attentional frames, pointing, and the motivation to share attention have been proposed to support cooperative communication before children acquire conventional vocabularies \cite{tomasello2005understanding,carpenter2007shared,tomasello2008origins}. In this view, communication builds on the ability to coordinate perspectives and interpret intentions against a shared background, not merely to detect correlations between sensory inputs. Accounts integrating biological and cultural processes likewise situate language within interacting cognitive capacities, social relations, and cultural transmission \cite{arnon2025biocultural}. Observational, longitudinal, and word-learning studies link joint engagement and caregivers' references within the child's attentional focus to early communication and vocabulary development \cite{tomasello1986joint,carpenter1998social}. Infant gaze following and pointing also predict subsequent vocabulary growth \cite{brooks2008gaze}. Infants also point to share attention and interest, supporting accounts of prelinguistic communication that go beyond object co-occurrence \cite{liszkowski2004point,tomasello2007pointing}.  These findings support joint attention as a developmental scaffold for vocabulary acquisition, while leaving open the computational dynamics through which a shared attentional context supports the formation of shared symbols.

Experiments on cross-situational learning show that infants can learn word--referent mappings from statistical regularities across individually ambiguous episodes \cite{smith2008infants}. Computational models have examined how learners combine these regularities with social cues such as gaze and prosody \cite{yu2007unified}, or jointly infer a speaker's referential intentions and the lexicon from utterances and their contexts \cite{frank2009referential}. Robotic models also demonstrate cross-situational learning of multimodal categories and word meanings from perceptual experience and spoken descriptions \cite{taniguchi2017crosssituational}.

Evolutionary proposals address a different timescale. Comparative gaze-following experiments motivate the cooperative-eye hypothesis, while the interdependence hypothesis places shared intentionality within the evolution of human cooperation \cite{tomasello2007eye,tomasello2012cooperation}. Such accounts concern the conditions under which shared attention and cooperative communication could evolve. The present model instead asks what computational structure becomes available once a common referent is given.

 The free-energy principle offers a potential computational framework for understanding how joint attention can support the emergence of shared symbols. In its variational formulation, perception and learning can be described through a generative model that explains an individual's sensory observations \cite{friston2010free,buckley2017mathematical,taniguchi2023world}. In a canonical cortical predictive-coding model, higher levels predict lower-level activity, while feedforward signals convey residual prediction errors \cite{rao1999predictive}. Under specified hierarchical generative-model assumptions, predictive coding can be formulated as variational inference under the free-energy principle \cite{friston2009predictive}. The challenge for communication is to connect these individual inferences without assuming a shared symbolic system in advance. Each agent has direct access only to its own observations and internal computations, a restriction we call cognitive closure. If neither agent can inspect the other's internal states, what couples their inferences, and why should a shared symbol become useful? An account of collective inference must specify that coupling rather than append an agreement reward or consistency penalty to otherwise independent agents.

Active-inference accounts already explain aspects of communication through coupled generative models, reciprocal inference, and social expectations \cite{friston2015communication,vasil2020world,friston2020linguistic}.  Federated inference starts from agents with a common world model and examines how belief sharing supports learning, including the emergence of a shared lexicon \cite{friston2024federated}. We start from a common referent and agent-specific recognition channels, derive the joint prior that couples their private states, and show how a categorical world supplies a candidate shared variable. Variational accounts of culture and niche construction connect socially patterned expectations to the environments that agents inherit and modify \cite{veissiere2020thinking,constant2018niche}. Our contribution is therefore not the first application of free energy to communication or meaning. It is the derivation of a particular interaction from a common-cause constraint, followed by the identification of a candidate shared variable in the induced prior.

\begingroup\setlength{\emergencystretch}{3em}
\par\endgroup

Constructive approaches address related parts of this problem. Symbol Emergence Systems Theory connects embodied category formation with the development of shared sign systems through social interaction \cite{taniguchi2016survey,taniguchi2019developmental,taniguchi2026systems}. Situated language games, exemplified by the Talking Heads experiment, provide a complementary constructive tradition \cite{steels2015talking}. Signaling and referential games investigate the formation of communication conventions within specified message channels and interaction tasks \cite{lewis1969convention,lazaridou2017multi,galke2022missing}. Iterated-learning experiments show how transmission across generations can shape linguistic structure \cite{kirby2008cumulative}. Interaction within a closed community can also yield compositional structure without generational turnover \cite{raviv2019compositional}. Naming games describe how local interactions can produce population-level vocabulary agreement \cite{baronchelli2006sharp}.

Within this constructive tradition, the collective predictive coding (CPC) hypothesis connects symbol emergence to the free-energy principle through a collective generative model \cite{taniguchi2024collective}. In this model, a shared symbolic variable conditions the priors over agents' private perceptual states, embedding their coupling within the generative model itself. The Metropolis--Hastings naming game (MHNG) provides a decentralized Bayesian inference procedure based on local proposals and listener acceptance \cite{hagiwara2019interpersonal,taniguchi2023emergent}. Human joint-attention naming experiments test whether acceptance behavior is consistent with MH-based predictions \cite{okumura2023joint}. Recursive naming games extend MHNG to multiple agents \cite{inukai2023recursive}, and deep generative versions demonstrate communication between private multimodal representations \cite{hoang2024multimodal}. The MH Captioning Game extends this inference-based approach to token sequences and private visual representations \cite{matsui2025mhcg,ochiai2026emergent}. The broader generative emergent communication framework connects CPC to collective world models and language models \cite{taniguchi2026generative}. Joint attention supplies a shared referential context in such naming-game settings \cite{hagiwara2019interpersonal,okumura2023joint}, but this experimental role does not by itself explain the shared-variable structure of CPC. The mathematical relationship between attending to a common object and that model structure has remained implicit. Our central question is how the common-cause constraint supplied by joint attention can give rise to the structure underlying the CPC generative model, rather than taking a shared symbolic variable as a starting assumption.

We begin with the world's generative process: in each episode $d$, an object $\xi_d$ is drawn from the world distribution $\Pideal$, and generates a private observation $o_{i,d}$ for each agent through its sensory channel $P_i(o_{i,d}\mid\xi_d)$. Each agent then infers its internal state $z_{i,d}$ from its own observation. We model the shared-object constraint supplied by joint attention: different private observations originate from the same external object. This constraint captures one aspect of joint attention, not its full psychological organization; the model does not explain how mutual awareness, shared goals, or communicative intentions are established. For specified world and recognition models satisfying a marginal-calibration condition, marginalizing the common object induces a pairwise interaction between internal states. The interaction is the log Bayes factor comparing a shared cause with independent causes. For a population, we specify a collective prior by composing these pairwise relations with a population-normalized coupling. Its mean-field variational free energy combines the local free energies and the expected interaction.  The coupling between private inferences is thus expressed within the collective generative model, with its interaction term derived from the shared-cause constraint.

The next step concerns what can be shared. For two agents, a categorical world induces an exact mixture structure in the joint-attention prior. Its component index provides a candidate shared variable $w$, while the component marginals provide its agent-specific meanings. The index need not convey every aspect of the common object: within-category correlations can remain. Neglecting this residual dependence gives an effective CPC generative model, and Bayes' rule then specifies how the agents' evidence combines. Thus CPC enters as a model obtained from the joint-attention construction, not as a symbolic architecture postulated at the outset. A Metropolis--Hastings naming game provides a decentralized procedure for exchanging and inferring the resulting symbols.

Our contributions are threefold.
\begin{enumerate}
\item We derive how a common referent couples private perceptual inference. Marginalizing the shared cause induces a log Bayes factor between private states, which enters the variational free energy of a collective generative model.
\item Under a categorical-world assumption, the two-agent joint-attention prior inherits an exact mixture structure. Its component index provides a candidate for shared reference. Factoring the private states within each component yields an effective collective predictive coding model.
\item We show how sign coordination relates to collective free energy in two experimental settings. Using the established Metropolis--Hastings naming game, we find that, relative to the communication controls, better category recovery and word agreement are accompanied by lower world-referenced free energy in the toy experiment and  a lower sum of individual and collective regularization in the image experiment with fixed visual representations.
\end{enumerate}

\begingroup\setlength{\emergencystretch}{3em}
Together, these contributions provide a conditional mechanistic account of the transition from joint attention to shared referential symbols.  The free-energy principle supplies the variational framework, and the CPC hypothesis provides a model linking private inference through a shared symbolic variable. Our derivation shows how the shared-cause and categorical-world assumptions give rise to structure underlying this model.\par\endgroup

We focus on how joint attention can support the emergence of shared referential symbols by coupling private inference. Figure~\ref{fig:mhng-triad} illustrates the route from independent prediction through joint attention to the effective CPC model.


\section{Results}
\label{sec:results}

\subsection{A common referent couples private inference}
\label{sec:derivation}
\label{results:common-cause}

Two people looking at the same object receive different sensory evidence. We first show how their common referent induces a dependence between private perceptual states. We then show how a categorical world supplies a shared index that can carry part of that dependence. The toy and image experiments illustrate how agents can use this structure to coordinate words and private categories.

Let $\Pideal$ denote the world distribution of objects. An object is drawn as $\xi_d\sim\Pideal$, and each agent $i$ receives a private observation $o_{i,d}\sim P_i(\cdot\mid\xi_d)$, such as an image or a sensory feature vector. The same object can therefore generate different observations across agents because their sensory perspectives differ. Uppercase $P$ denotes distributions in the external world, including its observation process; lowercase $p$ denotes distributions over private states and the agents' generative models. In particular, $P_i(o\mid\xi)$ describes world-side observations, whereas $p_i(o\mid z_i)$ is the agent's internal observation model.

\begin{figure}[tb!p]
\centering
{\small\textbf{(a)  One object, two perspectives}\par}
\includegraphics[width=0.70\linewidth,trim=0 10bp 0 59bp,clip]{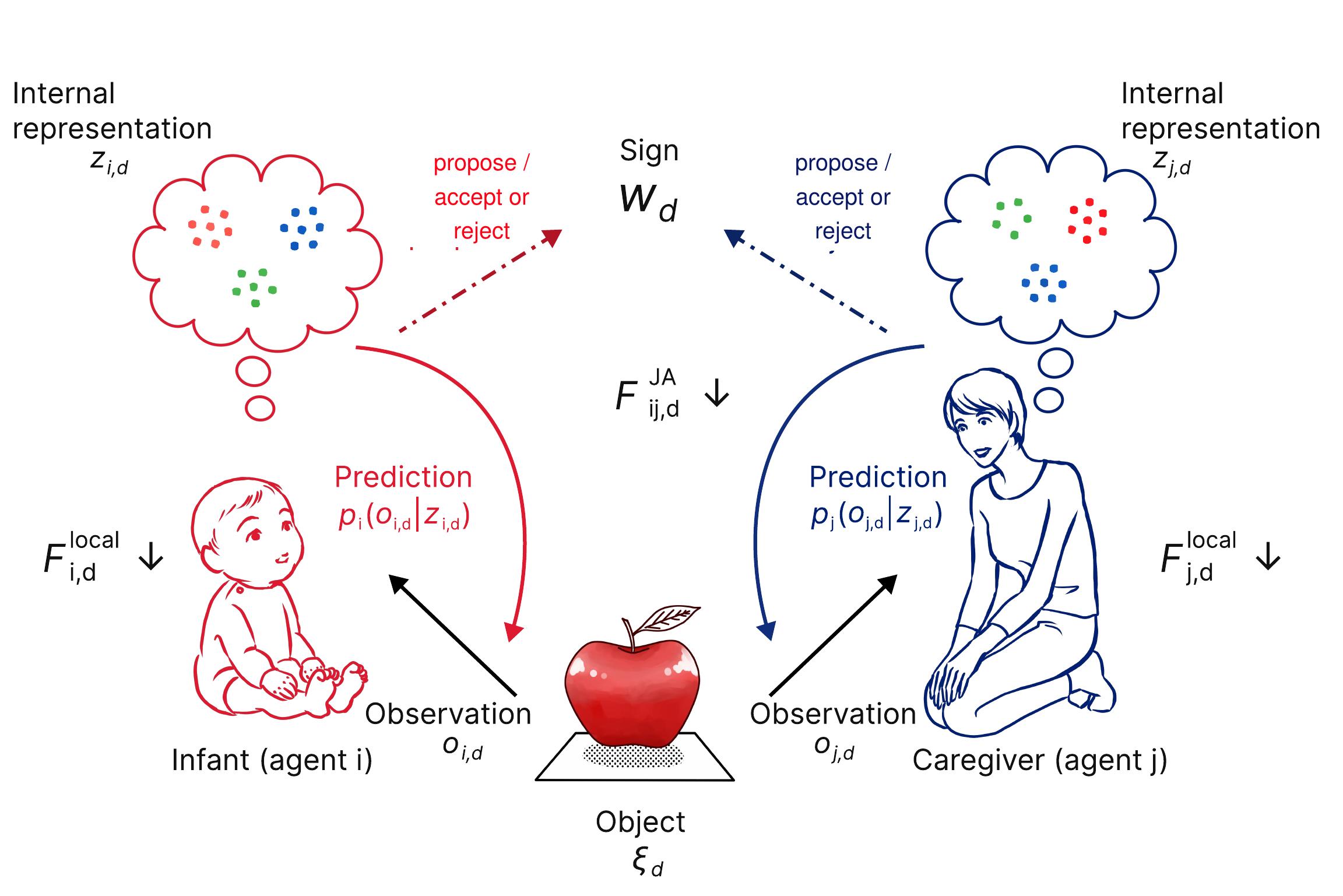}
\par\vspace{5pt}
\begin{minipage}[t]{0.30\linewidth}
\centering\small
\parbox[t][2\baselineskip][t]{\linewidth}{\centering\strut\textbf{(b)  Independent private priors}\par}\par\strut$H_{\mathrm{IND}}$\par\vspace{3pt}
\includegraphics[width=0.90\linewidth]{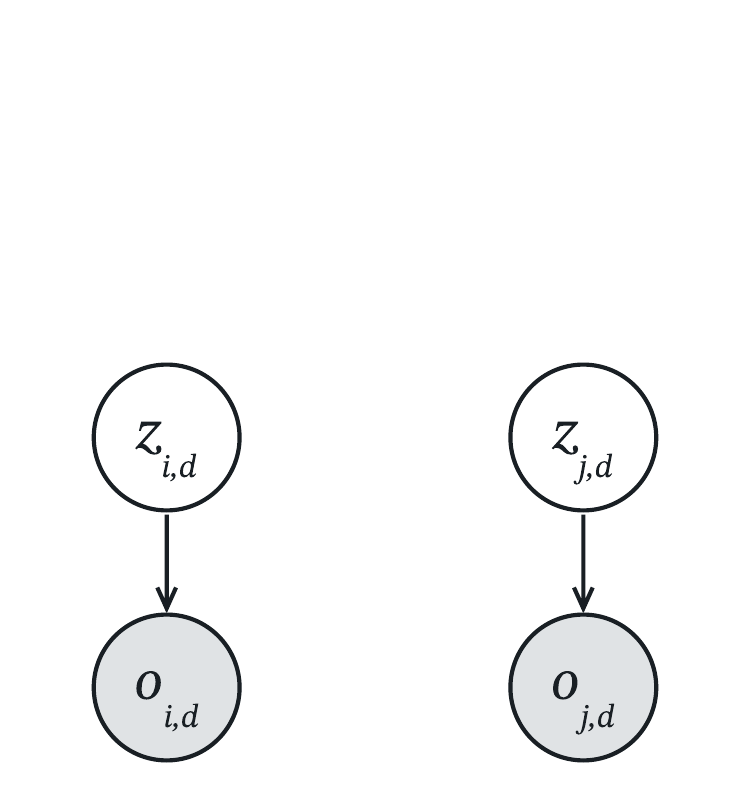}
\end{minipage}\hfill
\begin{minipage}[t]{0.30\linewidth}
\centering\small
\parbox[t][2\baselineskip][t]{\linewidth}{\centering\strut\textbf{(c)  Coupled private priors}\par}\par\strut$H_{\mathrm{JA}}$\par\vspace{3pt}
\includegraphics[width=0.90\linewidth]{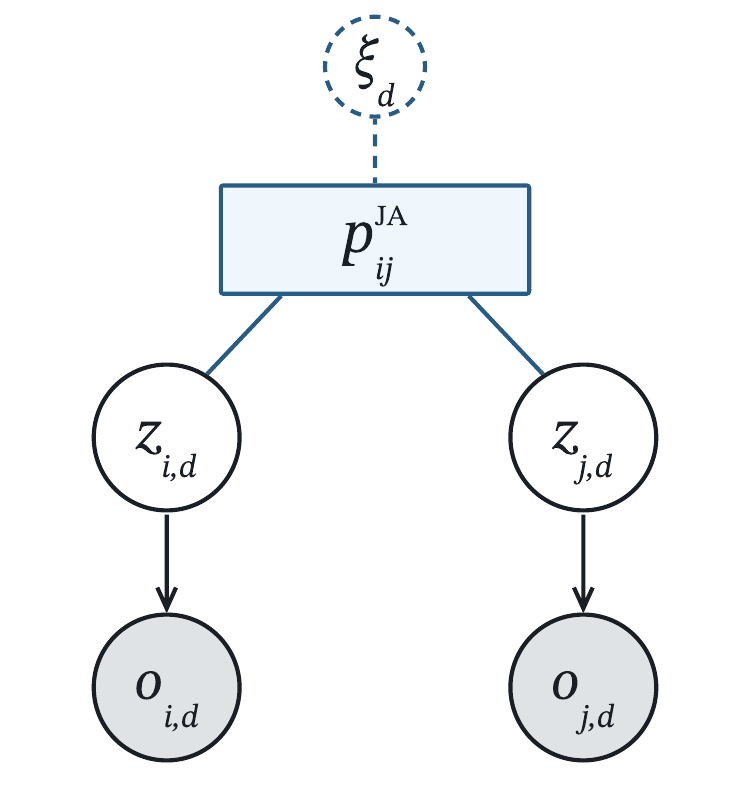}
\end{minipage}\hfill
\begin{minipage}[t]{0.30\linewidth}
\centering\small
\parbox[t][2\baselineskip][t]{\linewidth}{\centering\strut\textbf{(d)  CPC via a shared index}\par}\par\strut Shared category $w_d$\par\vspace{3pt}
\includegraphics[width=0.90\linewidth]{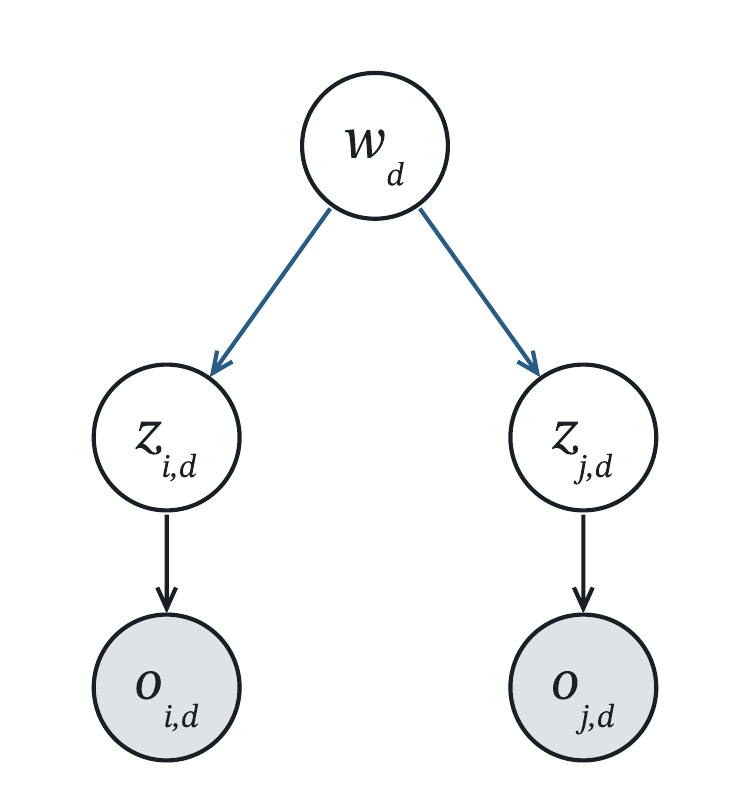}
\end{minipage}

\caption[From a common referent to a candidate shared word]{From a common referent to a candidate shared word. (a) Two agents attend to the same object $\xi_d$ through different private observations. The illustrated word exchange is the process to be explained. Both agents alternate between proposing a sign and accepting or rejecting a proposal. (b) Independent predictive coding ($H_{\mathrm{IND}}$) gives each agent a separate prior over its private state. (c) Under joint attention ($H_{\mathrm{JA}}$), marginalizing the common object (dashed) induces the joint prior $p_{ij}^{\mathrm{JA}}$ (rectangle), which couples the private states. (d) In a categorical world, the mixture index $w_d$ provides a candidate shared word. The effective CPC model represents the coupling through $w_d$ by neglecting within-category dependence (Eq.~\eqref{eq:results-cpc}). Gray circles denote observations, white circles denote latent variables, and arrows denote generative conditionals.}
\label{fig:mhng-triad}
\end{figure}

 Under marginal calibration, a common referent induces an interaction between private states that can be written as a shared-cause log Bayes factor. Figure~\ref{fig:mhng-triad}(a) illustrates the setting: two agents attend to the same external object $\xi_d$ but receive different private observations $o_{i,d}$ and infer their own perceptual states $z_{i,d}$. The exchange of signs shown in this panel is the process to be explained; a shared symbolic variable is not assumed at the start of the derivation. The shared-object constraint captures a computational aspect of joint attention (Materials and Methods, Sec.~\ref{subsec:world-generation}).

Figure~\ref{fig:mhng-triad}(b) gives the independent predictive-coding baseline, $H_{\mathrm{IND}}$. Each agent has a local prior $p_i(z_i)$ and an observation model $p_i(o_{i,d}\mid z_i)$; the prior over the two private states factorizes as $p_i(z_i)p_j(z_j)$. Figure~\ref{fig:mhng-triad}(c) introduces the shared-cause constraint, $H_{\mathrm{JA}}$: the two private states are related through observations of a common object.  Marginalizing that object through the private recognition channels yields a joint prior $p_{ij}^{\mathrm{JA}}$. We combine this prior with the local observation models to define an effective cognitive model.

 For specified world and recognition models, let $r_i(z_i\mid\xi)$ be the recognition channel obtained by averaging private recognition over the observations generated by $\xi$. We use $r_i$ to distinguish this induced object-to-state channel from the external observation process $P_i(o\mid\xi)$, the internal model posterior $p_i(z_i\mid o)$, and its variational approximation $q_i(z_i\mid o)$. It combines the external observation process with private recognition, rather than denoting an additional conditional in the internal generative model or a variational factor optimized by the agent. Marginal calibration requires $\int\Pideal(\xi)r_i(z_i\mid\xi)d\xi=p_i(z_i)$. Marginalizing one common object gives
\begin{equation}
p_{ij}^{\mathrm{JA}}(z_i,z_j)
=\int\Pideal(\xi)r_i(z_i\mid\xi)r_j(z_j\mid\xi)d\xi,
\qquad
V_{ij}(z_i,z_j)=\log\frac{p_{ij}^{\mathrm{JA}}(z_i,z_j)}{p_i(z_i)p_j(z_j)}.
\label{eq:results-common-cause}
\end{equation}
The product of recognition channels in Eq.~\eqref{eq:results-common-cause} assumes conditionally independent observations given $\xi$ and separate local recognition draws given those observations.
 The density ratio compares the joint prior in Fig.~\ref{fig:mhng-triad}(c) with the independent priors in Fig.~\ref{fig:mhng-triad}(b). Its logarithm $V_{ij}$ therefore expresses the interaction induced by the common referent. The rectangular factor in panel (c) represents this joint prior; the dashed object and link indicate the common cause marginalized in its derivation, not an additional latent variable retained in the effective model.

For $N\geq2$ agents, let $q_i(z_i\mid o_{i,d})$ be the local variational recognition distribution and $q_{1:N,d}=\prod_iq_i(z_i\mid o_{i,d})$ its product across agents. Define the collective functional for object $d$ as
\begin{equation}
F_d[q_{1:N,d}]
:=\sum_{i=1}^{N}F_{i,d}^{\mathrm{local}}[q_i]
-\beta_N\sum_{i<j}\mathbb E_{q_iq_j}[V_{ij}],
\qquad \beta_N=\frac{1}{N-1}.
\label{eq:results-Fd}
\end{equation}
Here $F_{i,d}^{\mathrm{local}}=\mathbb E_{q_i}[-\log p_i(o_{i,d}\mid z_i)]+\mathrm{KL}(q_i\|p_i)$ is the individual variational free energy, $\mathrm{KL}$ denotes Kullback--Leibler divergence, and $d$ indexes a jointly observed object.  For two agents, $\beta_2=1$ and Eq.~\eqref{eq:results-Fd} is the negative evidence lower bound (ELBO) of the effective cognitive model with prior $p_{ij}^{\mathrm{JA}}$ and local likelihoods $p_i(o_i\mid z_i)$ and $p_j(o_j\mid z_j)$. Marginal calibration alone does not imply that this model reproduces the world's joint observation distribution. For $N>2$, Supplementary Text~\ref{subsec:multi-agent-ja-free-energy} and~\ref{sec:global-joint-attention} describe the pairwise extension and distinguish it from the model in which all agents share a single object.

 The transition from Fig.~\ref{fig:mhng-triad}(b) to (c) thus couples private inference without first requiring a common symbolic system. Figure~\ref{fig:mhng-triad}(d) previews the next step: in a categorical world, the induced joint prior has a mixture structure whose index provides a candidate shared variable $w_d$. Neglecting dependence remaining within each category yields the effective CPC model shown in panel (d). The following section derives this step and distinguishes the exact mixture representation from the CPC approximation.

\subsection{Categorical structure provides a candidate shared variable}
\label{sec:shared-variable}
\label{results:shared-variable}

In a categorical world, the two-agent joint-attention prior admits an exact mixture-index representation, while the effective CPC model neglects within-category dependence. Suppose that the world is an episode-independent mixture $\Pideal(\xi)=\sum_k\pi_kP_k(\xi)$. Linearity of marginalization gives
\begin{equation}
p_{ij}^{\mathrm{JA}}(z_i,z_j)=\sum_w\pi_w p_{ij}(z_i,z_j\mid w).
\label{eq:results-mixture}
\end{equation}
Each component is $p_{ij}(z_i,z_j\mid w)=\int P_w(\xi)r_i(z_i\mid\xi)r_j(z_j\mid\xi)\,d\xi$, obtained by marginalizing the common object within category $w$. The induced prior inherits the index $w$ of the specified world mixture. The construction does not identify a unique mixture decomposition from private observations. Its component marginals define the agent-specific distributions $p_i(z_i\mid w)$ and $p_j(z_j\mid w)$, which are reusable across objects under the assumed stationary world statistics. The construction and its assumptions are specified in Supplementary Text~\ref{supp:shared-variable}.

The exact mixture representation can retain dependence within categories. Neglecting that dependence yields the effective CPC model
\begin{equation}
p(w_d,z_{1:N,d},o_{1:N,d})
=p(w_d)\prod_i p_i(z_{i,d}\mid w_d)p_i(o_{i,d}\mid z_{i,d}).
\label{eq:results-cpc}
\end{equation}
The right-hand probabilistic graphical model (PGM) in Fig.~\ref{fig:mhng-triad} represents this approximation, not a general graphical equivalence. Summing out $w$ recovers the original joint-attention prior before factorization, but need not recover it after the CPC approximation. Supplementary Text~\ref{subsec:category-residual} defines and quantifies the omitted dependence.

With a mean-field recognition distribution $q_d(w_d\mid o_{1:N,d})\prod_i q_i(z_{i,d}\mid o_{i,d})$, the collective free energy of Eq.~\eqref{eq:results-cpc} decomposes as
\begin{equation}
\ensuremath{F_d^{\mathrm{CPC}}=R_d+I_d+C_d.}
\label{eq:results-cfe-decomposition}
\end{equation}
The prediction error $R_d$ sums the expected negative log likelihoods of the private observations. Individual regularization $I_d$ sums the KL divergences from each private recognition distribution to its sign-conditioned category distribution, averaged over $q_d$. Collective regularization $C_d=\mathrm{KL}(q_d\|p(w_d))$ measures the departure of the shared-sign distribution from its prior. Supplementary Text~\ref{subsec:cfe} gives the derivation.

\begingroup\setlength{\emergencystretch}{3em}
The effective model connects this construction to existing decentralized Bayesian inference.  For fixed private states and category parameters, its sign posterior satisfies $p(w\mid z_{1:N})\propto p(w)\prod_i p_i(z_i\mid w)$. For two agents, the established MHNG procedure proposes a word from the speaker's local posterior and accepts or rejects it using the listener's likelihood ratio \cite{taniguchi2023emergent,taniguchi2024collective}. The cancellation in the Metropolis--Hastings ratio allows the exchange of a discrete sign without transmitting private observations, features, or parameters. This is a known implementation mechanism; the present contribution is the route from the common-cause constraint to the shared-variable model and the boundary of its approximation.\par\endgroup

The shared index is a candidate for a referential symbol, not by itself a social symbol. Correspondence between private signs must still be established through interaction and learning. The following experiments examine that process while distinguishing category recovery, word-name agreement, and generative-model fit.

\subsection{Private sign coordination and world-referenced inference}
\label{sec:numerical-experiment}
\label{results:toy}

\begin{figure}[tb!p]
\centering
\includegraphics[width=\linewidth]{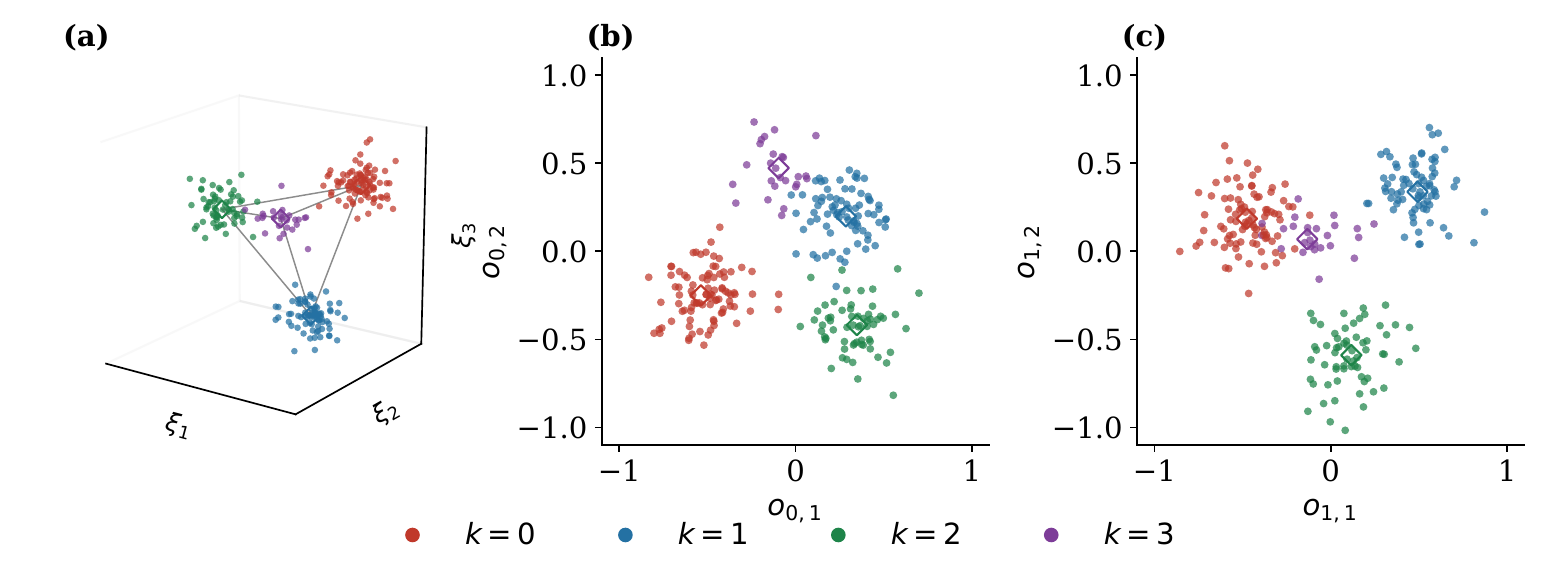}
\caption{World and observations in the primary cell of  the two-agent toy experiment ($\sigma_\xi=0.1$, $\sigma_{\mathrm{obs}}=0.10$), seed 0. Here $\sigma_\xi$ and $\sigma_{\mathrm{obs}}$ are the within-component and observation-noise standard deviations, respectively. Colors denote the true component $k$. (a) $\xi\in\mathbb R^3$; diamonds are tetrahedron vertices. (b) Agent 0 observation coordinates $(o_{0,1},o_{0,2})$; (c) agent 1 observation coordinates $(o_{1,1},o_{1,2})$. Each observes $o_i=A_i\xi+$ noise. The three world axes are $\xi_1,\xi_2,\xi_3$. The two observation panels share axis ranges; the same color denotes the same object.}
\label{fig:exp1-world}
\end{figure}

In the tested two-agent setting, MHNG  starting from independently initialized, agent-specific category distributions improves mean world-category recovery and world-referenced free energy relative to the communication controls.  The two-agent toy experiment uses two noisy, agent-specific projections of 256 objects from a four-component world, with 16 seeds (Fig.~\ref{fig:exp1-world}). Each agent initializes its own private clusters without cross-agent label matching. All learned conditions use the same latent Gaussian model and prior; the samplers receive 1,000 parameter/perception updates per agent, whereas centralized variational inference (Central VI) uses its ELBO stopping criterion. Full conditions, readouts, and uncertainty estimates are given in Materials and Methods. Here a private sign is the locally stored categorical variable $w_{i,d}$; the continuous state $z_{i,d}$ is its private perceptual representation.

The primary score is $F_d$ as defined in Eq.~\eqref{eq:results-Fd}, first averaged over the 256 objects within each seed and then summarized across seeds. The population extension reports $F_d/N$ to express the same pairwise-composition score per agent. Lower values indicate better fit to this fixed world-referenced evaluation model; they do not define the learned CPC model's training objective.

\label{subsec:exp1-main}

\begin{figure}[!tb]
\centering
\includegraphics[width=\linewidth]{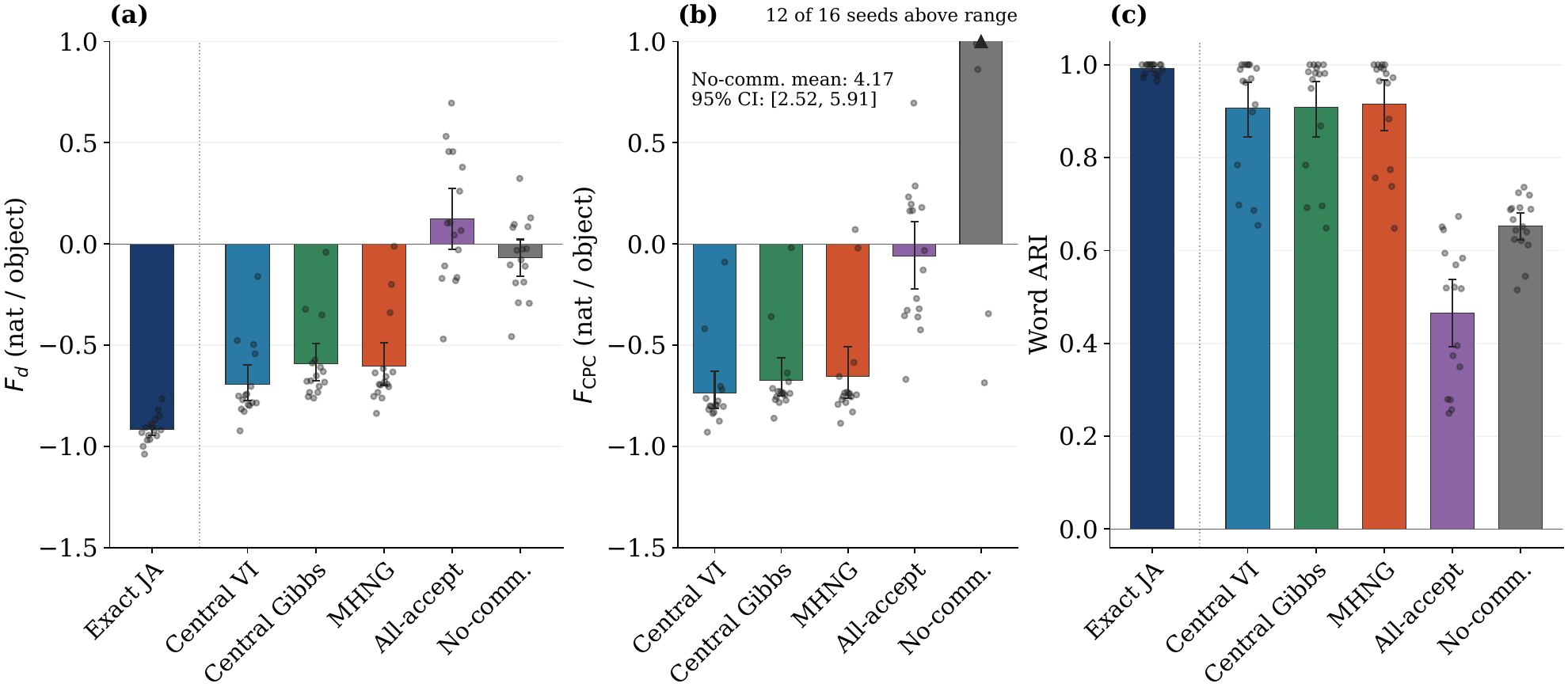}

\caption[The two-agent toy experiment]{The two-agent toy experiment starts from independently initialized private category distributions. (a) World-referenced $F_d$ (lower is better). (b) $F_{\mathrm{CPC}}$, the negative log marginal likelihood on the observed training pairs with fitted parameters held fixed (lower is better). Each pair corresponds to one object. (c) Word ARI against world categories (higher is better). Exact JA, to the left of the divider in (a) and (c), is the supplied-model reference. Sampling methods receive 1,000 parameter/perception updates per agent. Central VI runs to its ELBO stopping criterion. Bars show 16-seed means, dots individual seeds, and whiskers 95\% percentile seed-bootstrap intervals (20,000 resamples), computed using all 16 seeds for each method. Panels (a) and (b) share the range $[-1.5,1.0]$ nats per object. In (b), the triangle denotes the 12 No-comm.\ seed values above 1.0 (maximum 12.49). The No-comm.\ bar is clipped at the upper axis limit. Its mean and 95\% interval are given as text. Private-word ARIs are agent-averaged within seed.}
\label{fig:exp1-main}
\end{figure}

Figure~\ref{fig:exp1-main} compares $F_d$ and the adjusted Rand index (ARI) against world categories in the primary cell, $(\sigma_\xi,\sigma_{\mathrm{obs}})=(0.1,0.10)$ and vocabulary size $M=4$. The five learned methods are also evaluated by the negative log marginal likelihood of the paired observations under the fitted CPC model. MHNG gives a lower mean negative log marginal likelihood than either communication control across the 16 seeds. The $F_d$ estimates use common quasi-Monte Carlo points.

\begin{figure}[tb!p]
\centering
\begingroup
\setlength{\tabcolsep}{2pt}
\definecolor{truthzero}{HTML}{333333}
\definecolor{truthone}{HTML}{666666}
\definecolor{truthtwo}{HTML}{999999}
\definecolor{truththree}{HTML}{BBBBBB}
\definecolor{wordzero}{HTML}{4477AA}
\definecolor{wordone}{HTML}{EE6677}
\definecolor{wordtwo}{HTML}{228833}
\definecolor{wordthree}{HTML}{CCBB44}
\newcommand{\formationpanel}[1]{\parbox[c]{0.25\linewidth}{\includegraphics[width=\linewidth]{#1}}}
\newcommand{\formationagent}[2]{\parbox[c]{0.105\linewidth}{\centering\footnotesize\textbf{Agent #1}\\#2}}
\begin{tabular}{@{}cccc@{}}
& \small\shortstack{\textbf{True category}\\$k_d$}
& \small\shortstack{\textbf{Initial private word}\\$w_{i,d}$: update 0}
& \small\shortstack{\textbf{Final private word}\\$w_{i,d}$: update 1,000} \\[2pt]
\formationagent{0}{World 3D}
& \formationpanel{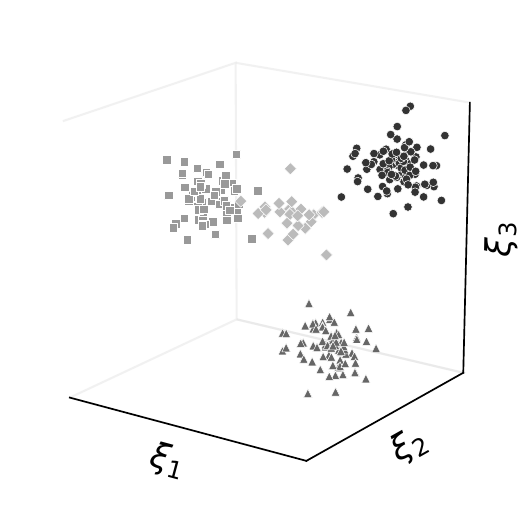}
& \formationpanel{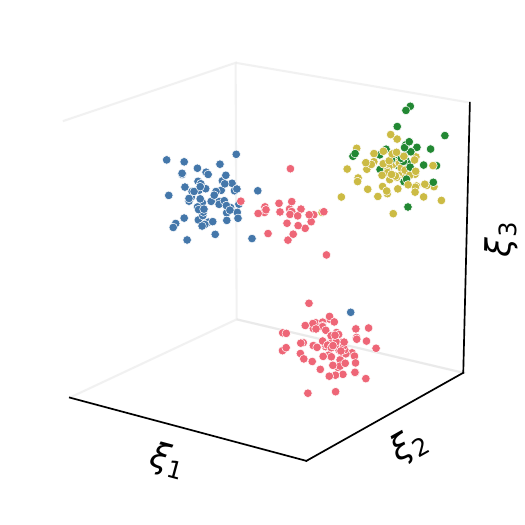}
& \formationpanel{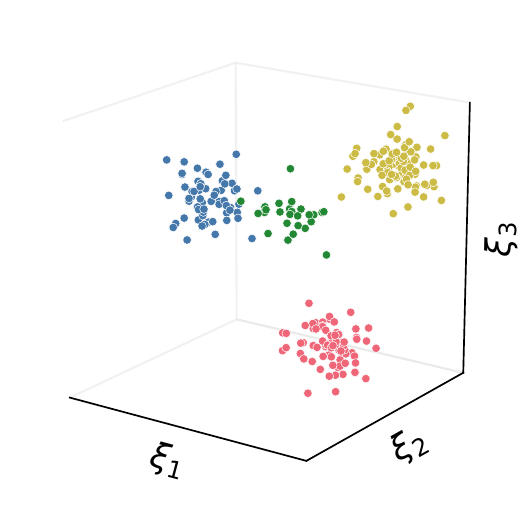} \\[2pt]
\formationagent{1}{World 3D}
& \formationpanel{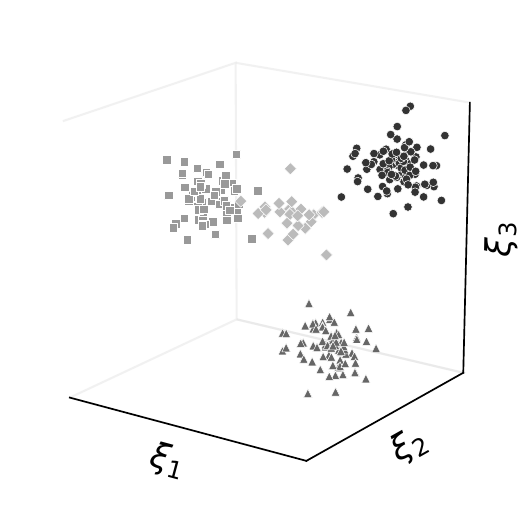}
& \formationpanel{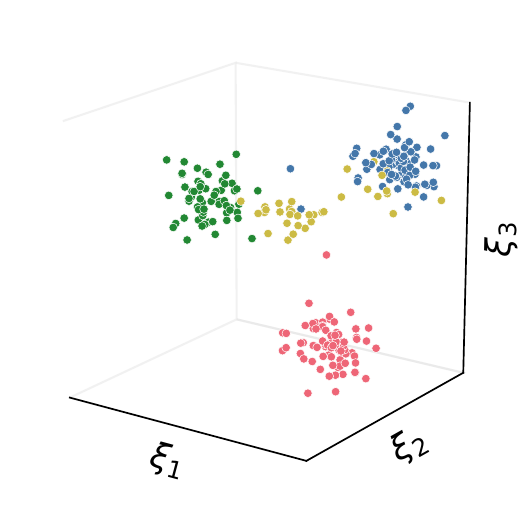}
& \formationpanel{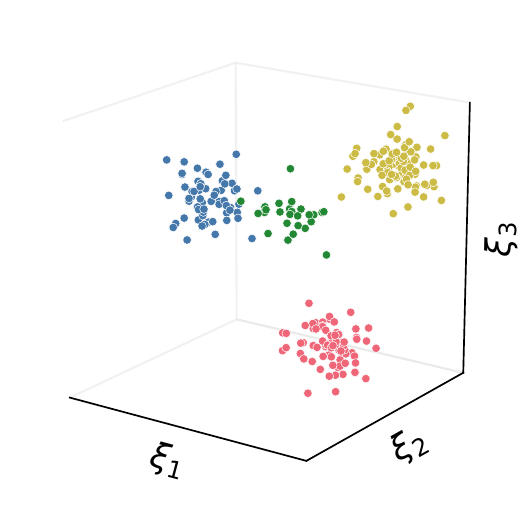} \\[2pt]
\cmidrule(lr){2-4}
\formationagent{0}{Private 2D}
& \formationpanel{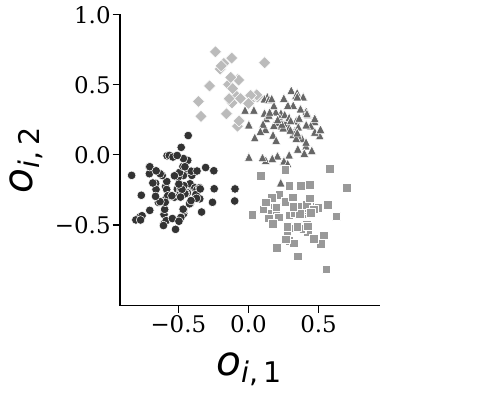}
& \formationpanel{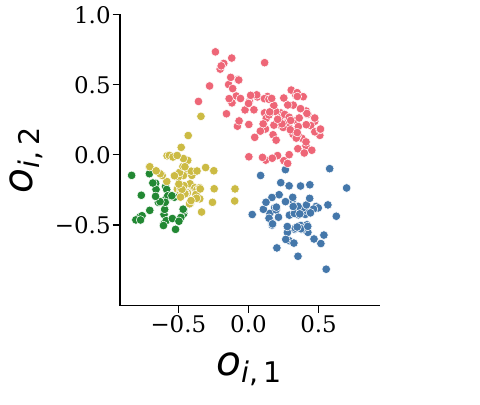}
& \formationpanel{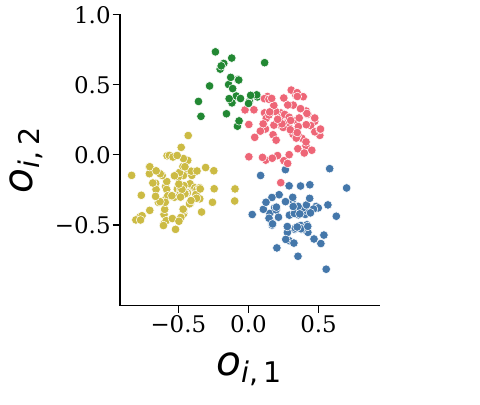} \\[2pt]
\formationagent{1}{Private 2D}
& \formationpanel{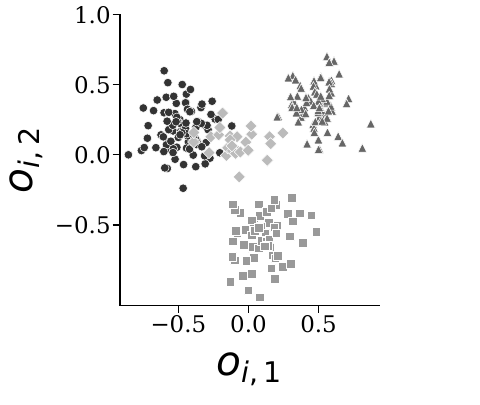}
& \formationpanel{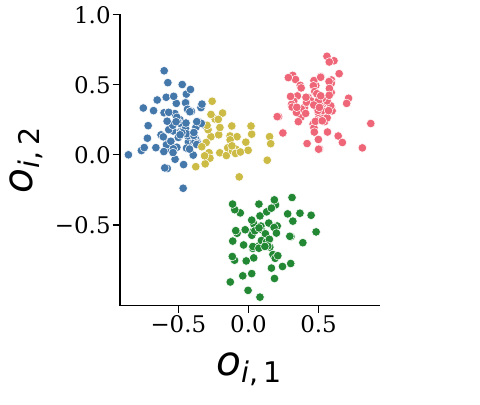}
& \formationpanel{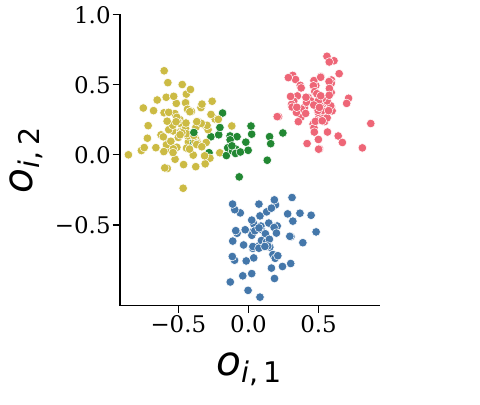} \\[3pt]
& \footnotesize\shortstack{True categories (gray symbols)\\
\textcolor{truthzero}{$\bullet$} 0\quad
\textcolor{truthone}{$\blacktriangle$} 1\quad
\textcolor{truthtwo}{$\blacksquare$} 2\quad
\textcolor{truththree}{$\blacklozenge$} 3}
& \multicolumn{2}{c}{\footnotesize\shortstack{Word labels (colored circles; same key for both agents)\\
\textcolor{wordzero}{$\bullet$} 0\qquad
\textcolor{wordone}{$\bullet$} 1\qquad
\textcolor{wordtwo}{$\bullet$} 2\qquad
\textcolor{wordthree}{$\bullet$} 3}} \\
\end{tabular}
\endgroup
\caption{Private clustering and shared word use in  the two-agent toy experiment (seed 0). The upper two rows show agents 0 and 1 at common world coordinates $\xi_d\in\mathbb R^3$, with identical positions and viewing angle. The lower two rows show the same agents in their own observation spaces $o_i\in\mathbb R^2$. Columns show true categories, independent initialization, and 1,000 updates per agent. Gray shapes encode true categories; colored circles encode raw word integers with a common key and no relabeling. World coordinates and true categories are for visualization and evaluation only, not supplied to learned inference.}
\label{fig:exp1-w-formation}
\end{figure}

\begin{figure}[!tb]
\centering
\includegraphics[width=\linewidth]{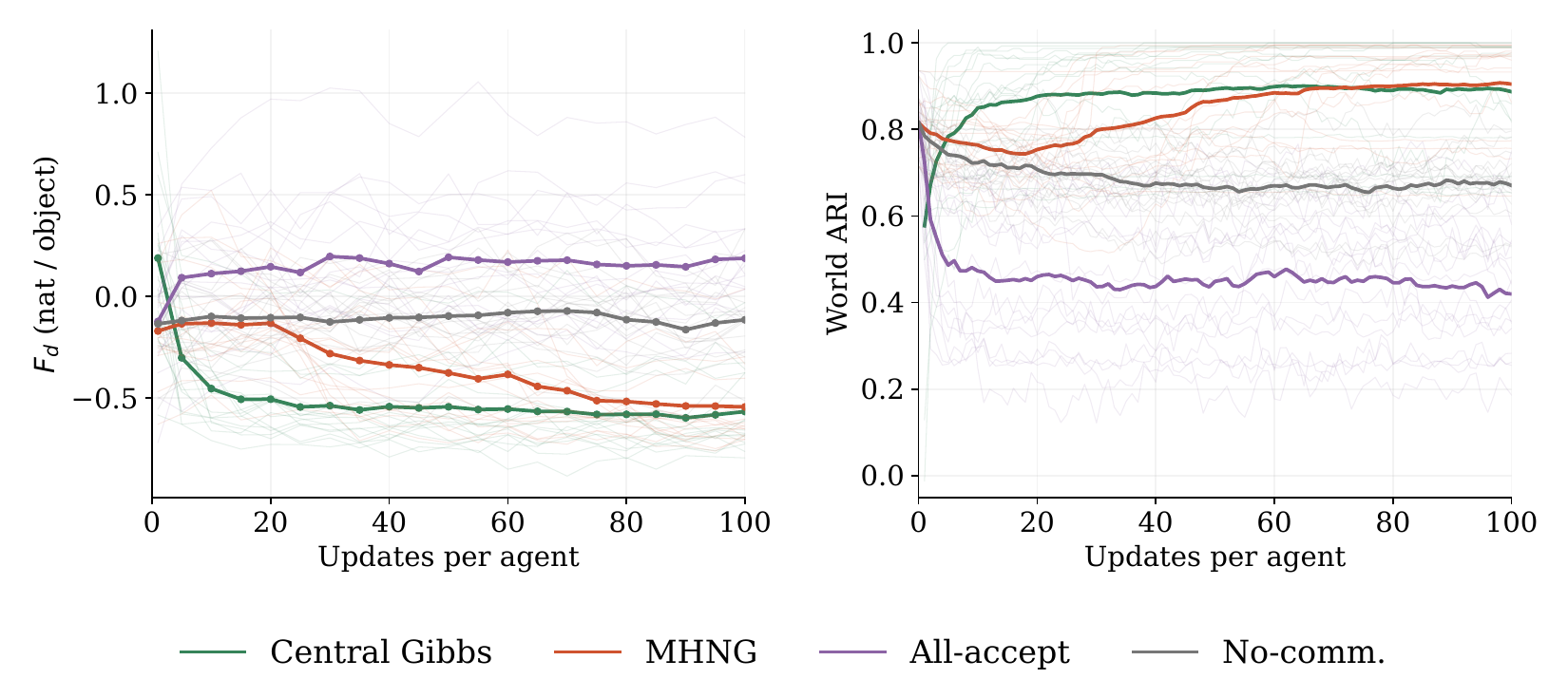}
\caption{Learning trajectories under the matched latent model during the first 100 updates per agent. Left: $F_d$ (lower is better), evaluated at update 1 and every five updates thereafter; markers show evaluated means and connecting lines guide the eye. Right: world ARI (higher is better), recorded at every update and averaged over agents within each seed. Thick lines show 16-seed means and faint lines the individual seeds. The $F_d$ readout is the word/parameter-conditioned Gaussian at each evaluation time, not an empirical posterior mixture. Curves use private words for MHNG, All-accept, and No-comm., and the common sampled word for Central Gibbs. Both horizontal axes span 0--100 local parameter/perception updates, not equal communication or computational cost. All seeds are retained. Training continues to 1,000 updates per agent for Fig.~\ref{fig:exp1-main}.}
\label{fig:exp1-learning}
\end{figure}

 Across the 16 seeds, MHNG achieved lower mean $F_d$ and $F_{\mathrm{CPC}}$ and higher mean word ARI than either communication control (Fig.~\ref{fig:exp1-main}). It achieved lower $F_d$ and higher word ARI than All-accept in every seed. The paired 95\% seed-bootstrap intervals for MHNG minus Central Gibbs included zero for all three scores. Exploratory paired sign-flip tests detected differences from both communication controls on all three scores (Holm-adjusted $p<0.001$). Supplementary Table~\ref{tab:toy-learning-summary} reports the mean scores and paired differences.

 MHNG achieved higher mean agreement between private signs than All-accept and No-comm., without centralized control. The two readouts capture distinct aspects of coordination (Supplementary Table~\ref{tab:toy-learning-summary}).

\begingroup
We compute $F_d$ from Eq.~\eqref{eq:Fd} with the same world-derived joint prior for all methods. For each sampler, we integrate over the Gaussian perceptual distribution conditioned on the observation, final sign, and final category parameters. For Central VI, we use the fitted variational Gaussian mixture (Sec.~\ref{methods:toy-readout}).\par\endgroup

\label{subsec:exp1-mechanism}

Figure~\ref{fig:exp1-w-formation} compares each agent's initial and final private words both in its own observation space and at common three-dimensional world coordinates $\xi_d$. The private-space views show local clustering, while the common-coordinate views permit direct comparison of word use across agents. The world coordinates are used only for visualization, not supplied to learned inference. The initial state already contains private clusters; this is correspondence learning, not clustering from an unstructured state. Figure~\ref{fig:exp1-learning} shows $F_d$ and world ARI during the first 100 updates. These finite trajectories do not certify equilibrium.

\FloatBarrier
\paragraph{Population extension.}
 For the population extension to $N=2,4,8,16$, we use a common initial label assignment to reduce initialization variability across agents. All compared methods receive labels from agent 0's clustering, while each agent initializes its category parameters using its own observations. Under that protocol, random-pair MHNG (labeled RMHNG in the population comparisons) improves mean pairwise $F_d/N$ and world ARI relative to No-comm. at each tested population size. Its detailed results are reported in Supplementary Text~\ref{supp:population-results}. Initialization and the balanced local-update schedule are described in Sec.~\ref{subsec:exp2}. Recursive MHNG extends naming games to multi-agent inference \cite{inukai2023recursive}; the population comparison here uses a random-pair schedule.

Together, the toy results provide a constructive example of private-sign coordination under the shared-object constraint. They do not establish convergence in every seed, equivalence to centralized inference, or sampling of the full collective posterior for $N>2$.

\subsection{Shared-symbol learning with fixed visual representations}
\label{subsec:exp3-images}
\label{results:images}
\begin{figure}[!tb]
\centering
{\small\textbf{(a) One object, two private views}\par\smallskip}
\makebox[0pt][l]{\raisebox{0.475\linewidth}[0pt][0pt]{%
  \makebox[\linewidth]{\small\textbf{(b) Visual naming with fixed VAEs}}}}%
\includegraphics[width=\linewidth]{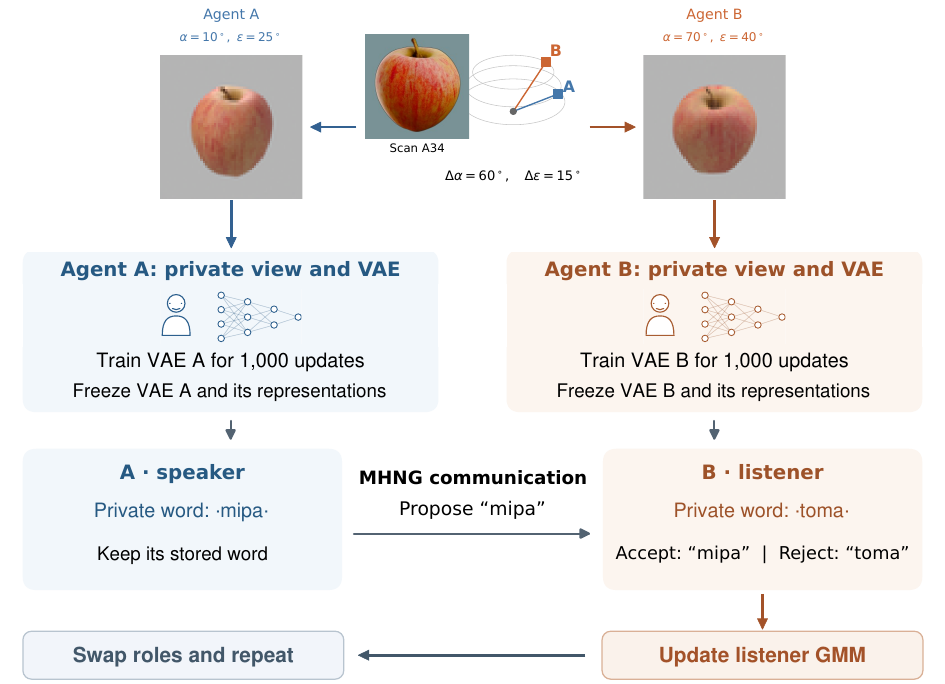}

\caption[Two-agent visual naming with fixed private representations]{Two-agent visual naming with fixed private representations. (a) Two orthographic cameras view the same textured apple scan (A34) from different angles, producing separate RGB inputs. This is an example camera pair. During naming, views are independently reshuffled within each scan at each round. (b) Each agent first trains its VAE, then freezes it along with its recognition distributions and sampled latent states. In MHNG, the speaker proposes a sign and the listener accepts or rejects it. The listener then updates its GMM from all 720 remembered latent states and their current signs before the agents swap roles.}
\label{fig:exp3-design}
\end{figure}

\begingroup\setlength{\emergencystretch}{4em}

The visual naming experiment instantiates the two-agent setting in Fig.~\ref{fig:mhng-triad}(a). The agents learn to name the same objects from images rendered from different viewpoints. Image-based MHNG and mutual learning between VAEs and category models are established approaches \cite{taniguchi2023emergent,hoang2024multimodal}. Alongside category recovery and word agreement, we use the regularization terms of collective free energy (CFE) to assess how well the learned category models fit the private visual representations of both agents.

\begingroup
To separate sign and category learning from visual representation learning, we kept the private visual representations fixed during naming. Each agent first trained its own VAE for 1,000 updates. During GMM and sign learning, the encoder and decoder remained fixed, as did the recognition distributions and one latent sample per training image. Within each of ten independent seeds, all four methods shared these representations, their initialization, and their interaction schedule.\par\endgroup

Here, a word is the integer-valued sign used in the visual naming task. The dataset contains ten textured 3D scans, one per object type, observed from different camera viewpoints by the two agents. The comparison includes MHNG, All-accept, Central Gibbs, and No-comm.\ (Fig.~\ref{fig:exp3-design}). A directional turn concerns a random subset of 128 training targets, after which the listener updates its GMM from all 720 remembered latent states and current private signs before taking the speaker role. The final subset has 64 targets, giving 72,000 target visits and 563 GMM updates per agent. Images, latent states, model parameters, and gradients are not communicated by MHNG or All-accept. The full protocol is specified in Sec.~\ref{methods:images}.\par\endgroup

Supplementary Fig.~\ref{fig:exp3-images} shows example inputs and reconstructions from the fixed VAEs used in all four conditions.

\begin{figure}[!tb]
\centering
\includegraphics[width=\linewidth]{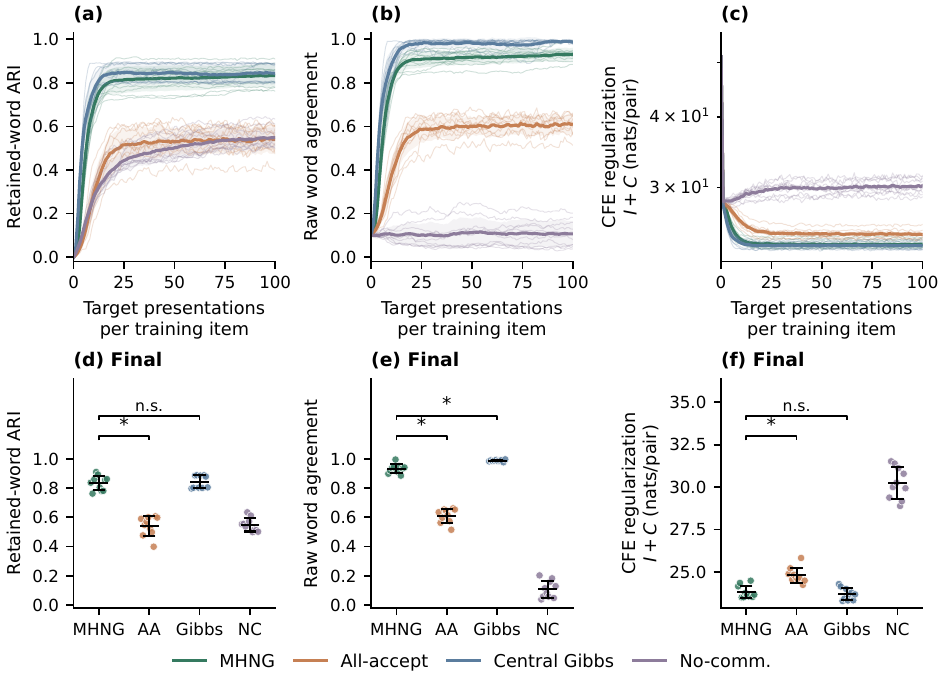}

\caption[Word coordination and CFE regularization]{Word coordination and CFE regularization with fixed visual representations (ten seeds). (a--c) Seed trajectories and means, with sample SD shaded in (a,b). The horizontal axis shows target presentations per training item, calculated as cumulative target presentations per agent divided by the 720 training items. Panel (c) uses a log scale. (d--f) Final seed values and mean $\pm$ sample SD. Brackets show the MHNG comparisons with All-accept and Central Gibbs. $^{*}$ marks adjusted $p<0.05$ and n.s. marks $p\geq0.05$ (two-sided paired sign-flip tests with Holm correction across 18 contrasts). AA, Gibbs, and NC denote All-accept, Central Gibbs, and No-comm. The sum $I+C$ captures mean-field CFE differences within a seed under fixed recognition distributions and a common normalized observation model.}
\label{fig:exp3-results}
\end{figure}

\begingroup
We assess category recovery and agreement on word identities separately (Fig.~\ref{fig:exp3-results}). Retained-word ARI measures how closely each agent groups images by object category. Raw word agreement measures how often the two agents assign the same sign to different views of the same object. Both measures use training views. Supplementary Table~\ref{tab:exp3-results} reports the means and standard deviations across seeds.\par\endgroup

\begingroup\setlength{\emergencystretch}{4em}
To assess the learned category models separately from sign accuracy, we evaluate the sum of individual and collective regularization, $I+C$, from the effective CPC decomposition (Eq.~\eqref{eq:results-cfe-decomposition}). The individual term averages the Gaussian recognition-to-category KL divergences under an optimized evaluation-only shared-sign distribution; the collective term is that distribution's KL divergence from the uniform sign prior. Neither term substitutes an agent's retained sign for the shared-sign distribution (Sec.~\ref{methods:image-regularization}).

With ten paired seeds and Holm correction across 18 comparisons, the smallest attainable adjusted $p$ value is $18(2/2^{10})\approx 0.0352$. After correction, MHNG yielded higher retained-word ARI and raw word agreement and lower $I+C$ than All-accept (Fig.~\ref{fig:exp3-results}(d--f)). Central Gibbs achieved higher raw word agreement than MHNG. The tests detected no difference between MHNG and Central Gibbs in retained-word ARI or $I+C$. MHNG also had significantly higher retained-word ARI and raw word agreement and lower $I+C$ than No-comm.\ (Supplementary Table~\ref{tab:exp3-all-contrasts}).

The MHNG--All-accept regularization difference combines a lower individual term ($21.536$ versus $22.844$) with a higher collective term ($2.273$ versus $1.948$). Thus, a smaller collective term alone does not imply better category-model compatibility. The shared integer indices are retained without post-hoc alignment, so No-comm.'s score can reflect uncoordinated sign permutations as well as different category partitions.

We report $I+C$ as a regularization score rather than an absolute image CFE. For fixed recognition distributions and a common normalized observation model, the reconstruction contribution is identical across methods within a seed; differences in this score therefore equal differences in the corresponding mean-field variational objective. The score is distinct from the world-derived joint-attention $F_d$. Because the fixed perceptual distributions and world-derived reference would be common across these methods, this experiment assesses how sign-mediated learning fits category models to fixed private representations, rather than a reduction of the original JA objective. The stochastic GMM updates using fixed latent samples are not assumed to be coordinate descent of the evaluated regularization score.\par\endgroup

\section{Discussion}

\label{sec:discussion}

The central contribution is a mechanistic connection between two perspectives on communication. Tomasello's account emphasizes joint attention and shared intentionality as foundations of cooperative communication \cite{tomasello2005understanding,tomasello2008origins};  Friston's free-energy principle provides a variational framework for explaining observations through a generative model \cite{friston2010free}. Our construction connects them through a specific question:  how can joint attention couple private inference and support the emergence of shared referential symbols when individuals have access only to their own observations?  Our derivation identifies an interaction induced by the common cause of private experiences and, in a categorical world, a mixture structure whose index provides a candidate shared variable for referential symbols.

\paragraph{Joint attention as a constraint on private inference.}
\begingroup\setlength{\emergencystretch}{4em}

The shared-object constraint formalizes a statistical relation between distinct private experiences. Under marginal calibration, marginalizing the common object yields a log Bayes factor that measures compatibility between private states through the world-derived joint prior. Bayesian causal-inference models examine whether sensory cues have common or separate sources \cite{kording2007causal}. Here, the common referent is assumed, allowing us to derive its consequences for private inference. Joint attention thereby supplies a relation between perspectives without requiring identical observations or representations.\par\endgroup

\paragraph{What the free-energy principle contributes.}
\begingroup\setlength{\emergencystretch}{4em}

The free-energy principle supplies the variational framework, while the shared-cause and categorical-world assumptions specify the model structure. Once the collective prior is defined, its mean-field free energy combines local prediction and regularization with the expected interaction and the prior normalization. The coupling between private inferences is thus expressed within a generative model. For more than two agents, the specified pairwise composition and a model retaining one common cause for all agents are distinct constructions (Supplementary Text~\ref{sec:global-joint-attention}).

\par\endgroup

\paragraph{From a common referent to a reusable shared variable.}
\begingroup\setlength{\emergencystretch}{4em}

For the two-agent triad, categorical world structure gives the induced joint prior an exact mixture representation. Its index $w$ is a candidate shared variable, with component marginals defining agent-specific category distributions reusable across objects. Communication and learning can make this variable usable as a shared referential sign. This separates the origin of the model structure from the established naming-game procedures that infer and coordinate signs.

The exact mixture can retain within-category dependence. Neglecting that dependence yields the conditionally independent CPC model and its product-form collective posterior \cite{taniguchi2024collective}. Shared categories thus retain a coarse aspect of the common referent while omitting finer correlations, quantified in Supplementary Text~\ref{subsec:category-residual}. Summing out $w$ recovers the joint-attention prior for the exact augmentation; it does not undo the CPC approximation. Supplementary Text~\ref{app:physics:sec:physics} illustrates this connection from joint attention to CPC with a physical example, using the Curie--Weiss model to show how a shared order parameter indexes the mixture components of the world distribution.\par\endgroup

\paragraph{What the numerical experiments establish.}
\begingroup\setlength{\emergencystretch}{4em}

The two-agent toy experiment examines sign coordination from independently initialized private category distributions. MHNG improves mean category recovery and world-referenced $F_d$ relative to the communication controls in the tested setting, while the population extension examines learning in larger agent populations. The seed-level results also distinguish word agreement from correct category recovery. The external $F_d$ evaluates compatibility with a world-derived reference; it is separate from the learned CPC model's inference objective. The recognition reference, method-specific readouts, and computational budgets are specified in Materials and Methods and constrain the interpretation of these comparisons. In particular, toy $F_d$ comparisons use the component-conditioned recognition channel to construct the world reference.

 The image experiment isolates category learning and sign coordination from changes in private visual representations. Each agent's VAE is trained for 1,000 updates and then held fixed, together with its recognition distributions and sampled latent states.  In the ten-seed comparison, MHNG improved retained-word ARI and raw word agreement and reduced $I+C$ relative to All-accept. Of the three measures, only raw word agreement differed significantly between MHNG and Central Gibbs, favoring Central Gibbs (Fig.~\ref{fig:exp3-results}(d--f)). The sum of individual and collective regularization evaluates the learned category distributions under the common fixed recognition family. It is distinct from absolute image CFE and the world-derived $F_d$. These are retained-training-sign results on distinct views of the same scans, rather than held-out predictions or generalization to new objects.\par\endgroup

\paragraph{Implications and open questions.}
The construction gives joint attention a computational role in connecting private inference to reusable shared symbols. Future tests can vary the reliability of the common referent and within-category dependence to examine when this connection supports useful communication. Developmental evidence distinguishes pointing from monitoring a partner's gaze \cite{matthews2012pointing}, motivating extensions that model how a common referent is established through action and social inference. Active gaze, pointing, and communicative intentions offer concrete directions for such extensions.

The present model treats the shared referent and finite category structure as given. Mutual awareness and cooperative motivation, emphasized in accounts of shared intentionality \cite{tomasello2005understanding,carpenter2007shared}, remain questions for an expanded model. Our experiments examine perceptual inference and learning under specified exchange schedules; action selection and feedback to the environment would connect the account to active inference and cultural niche construction \cite{parr2022active,constant2018niche,veissiere2020thinking}. The numerical evidence is specific to the tested settings and does not establish general convergence of adaptive naming games. The scope is the computational emergence of shared reference, rather than a historical explanation of language origins or the evolution of word forms and syntax.

\paragraph{Concluding perspective.}
\begingroup\setlength{\emergencystretch}{3em}
  This account connects joint attention to a variational framework by showing how a common referent can couple private inference. Under the stated assumptions, marginalizing the common cause induces an interaction between private states, while categorical world structure provides a reusable shared-variable candidate. Neglecting within-category dependence yields the effective CPC model, linking this construction to established naming-game procedures for decentralized inference. The resulting bridge makes explicit how joint attention can supply structure for shared referential symbols, connecting distinct private experiences through a common world.\par\endgroup

\section{Materials and Methods}
\label{sec:methods}

\subsection{World, private generative models, and recognition}
\label{methods:world}
\subsubsection{A Generative Assumption for an Unobservable External World}
\label{subsec:world-generation}
\begingroup

\endgroup
A common external object generates the private observations of the agents. The object exists independently of any agent's perception. For object $d$, the world draws $\xi_d\in\cX$ from $\Pideal$. Agent $i$ receives only its private observation $o_{i,d}\in\mathcal O$. Its observation model $P_i$ captures its sensory perspective, so agents can observe the same object differently.
\[
\xi_d\sim\Pideal,
\qquad
o_{i,d}\mid \xi_d \sim P_i(\cdot\mid \xi_d).
\]
We use uppercase $P$ for world distributions and lowercase $p$ for the internal models introduced below.

We assume that the private observations are conditionally independent given the shared object $\xi_d$. Marginalizing that unobserved object generally makes the observations dependent. This dependence is the source of the coupling between the agents' private inferences in the joint-attention model.

\subsubsection{Generative and Inference Models for Cognitively Closed Agents}
\label{subsec:local-variational}

Each agent $i$ receives only its private observation $o_{i,d}$ and infers its underlying cause through an internal state
\[
z_{i,d}\in\mathcal Z_i
\]
The internal-state space $\mathcal Z_i$ need not have the same dimension as the space $\cX$ of external objects; it may differ across agents, for example $\mathcal Z_i=\mathbb R^{m_i}$.

Each agent has the following local cognitive generative model:
\[
z_i\sim p_i(z_i),
\qquad
o_{i,d}\mid z_i\sim p_i(\cdot\mid z_i),
\qquad
p_i(o_{i,d},z_i)=p_i(z_i)\,p_i(o_{i,d}\mid z_i).
\]
That is, the agent generates an internal state $z_i$ from the prior $p_i(z_i)$ and then generates $o_{i,d}$ from that state. This cognitive generative model need not be a complete mapping of the world's generative process. It is an internal model through which the agent predicts private observations.

In equations, $z_i$ denotes an integration variable over $\mathcal Z_i$. For this model, introduce the local variational distribution over the internal state after observing $o_{i,d}$,
\[
q_i(z_i\mid o_{i,d}),
\qquad
\int_{\mathcal Z_i}q_i(z_i\mid o_{i,d})\,dz_i=1
\]
Treating the internal state as a distribution rather than a point permits uncertainty about observations and interactions with other agents to be described within one variational principle. Define the variational free energy of the local cognitive generative model as
\begin{equation}
F_{i,d}^{\mathrm{local}}[q_i(z_i\mid o_{i,d})]
=
\underbrace{
\E_{q_i(z_i\mid o_{i,d})}
\left[
-\log p_i(o_{i,d}\mid z_i)
\right]
}_{\text{prediction error}}
+
\underbrace{
\mathrm{KL}\!\left(
q_i(z_i\mid o_{i,d})
\middle\|
p_i(z_i)
\right)
}_{\text{regularization}}
\label{eq:local-fe}
\end{equation}
The first term is the prediction error for the private observation $o_{i,d}$ under the internal state $z_i$. The second regularizes the local variational distribution against large departures from the prior $p_i(z_i)$.

Throughout, we use \emph{sign} as a semiotic technical term for a discrete form that can refer to an object through an agent's private perceptual interpretation. Its computational realization is an integer-valued \emph{code symbol}; the corresponding sign system can thus be described as a code in information-theoretic terminology. We use \emph{symbol} when emphasizing a sign's shared referential use. A sign $w_{i,d}$ is distinguished from the private perceptual state $z_{i,d}$. Category distributions $p_i(z_i\mid w)$ describe private perceptual states conditional on a sign; their means, covariances, and other model-specific coefficients are category parameters, distinct from the sign itself.

\subsection{Toy-world design, methods, and uncertainty}
\label{methods:toy}

\label{subsec:exp-design}

The world is a $K_\xi=4$-component Gaussian mixture in $\mathbb R^{3}$. Component means form a regular tetrahedron of side length one, mixing proportions are unequal $(0.4,0.3,0.2,0.1)$, and within-component spread is $\sigma_\xi$. For each seed, we independently draw a two-dimensional orthogonal projection $A_i$ for each agent ($A_iA_i^\top=I_2$). The projection pair and observations are shared across methods within that seed. The world mixture is held constant across seeds. Figure~\ref{fig:exp1-world} shows one seed in the primary cell.

The primary experiment uses $\sigma_\xi=0.1$, $\sigma_{\mathrm{obs}}=0.10$, and $M=4$. The number of words $M$ matches the number of world components. With center distance fixed at one, the principal source of ambiguity is $\sigma_{\mathrm{obs}}$. This cell retains ambiguity in each private observation. Each learning condition uses $D=256$ episodes and 16 seeds. The separate fixed-vocabulary mechanism check crosses $\sigma_\xi\in\{0.1,0.2\}$ with $\sigma_{\mathrm{obs}}\in\{0.05,0.10,0.20\}$; it does not test  learning from independently initialized private category distributions across these six conditions.

Seeds are the unit of scientific uncertainty; episodes in one environment, redraws of points, and learning initializations are not counted as independent samples.  The two-agent toy experiment uses seed-level bootstrap intervals, whereas  the population extension displays individual seeds and their observed minimum--maximum range. In the two-agent toy experiment, we additionally use two-sided exact paired sign-flip tests for MHNG versus Central Gibbs, All-accept, and No-comm. on $F_d$, word ARI, and $F_{\mathrm{CPC}}$. We enumerate all $2^{16}$ sign assignments and apply Holm correction across these nine exploratory comparisons. The tests assume symmetry of paired seed differences under the null. All methods are evaluated by Eq.~\eqref{eq:Fd}.  For two agents, this is the negative ELBO for the specified JA evaluation model, with no prior-normalization offset. All methods within a seed are scored using the same reference density and quasi-Monte Carlo points.

 The two-agent toy experiment compares six conditions, ordered from the Exact JA reference through centralized and decentralized learning to the controls.  The population extension compares Central VI, Central Gibbs, RMHNG, and No-comm. under the same latent model and prior. Unlike  the two-agent toy experiment, all four receive an initial category correspondence, and the multi-agent communication schedule is balanced by listener updates.

\begin{enumerate}
\item \textbf{Exact joint attention (Exact JA)}:  A centralized observer computes the private-state posterior marginals using both observations under the component-conditioned JA reference and its supplied world parameters. The label Exact JA refers to inference within that reference model.
\item \textbf{Centralized variational inference (Central VI)}: Learns the latent CPC model below by structured coordinate-ascent variational inference, using both observations.
\item \textbf{Centralized Gibbs sampling (Central Gibbs)}: Samples a common word using both agents' latent states, then  updates each agent's category parameters and perceptual state under the same model and prior as Central VI.
\item \textbf{MH naming game (MHNG)}: Retains separate private words. A speaker proposes an integer from its local posterior; only the listener's stored word is subject to acceptance, followed by its local parameter and perceptual updates.
\item \textbf{Always-accept naming game (All-accept)}: Has the same initialization, directed exchange schedule, and local updates as MHNG, but the listener unconditionally adopts the proposed word.
\item \textbf{No communication (No-comm.)}: Each agent independently samples its private word, parameters, and perceptual state. It receives neither the other agent's observations nor its words.
\end{enumerate}

The primary metric is the raw $F_d$ in Eq.~\eqref{eq:Fd}; lower is better. In  the two-agent toy experiment, world ARI uses the terminal private words for MHNG, All-accept, and No-comm., the common sampled word for Central Gibbs, and posterior-mode categories for Central VI and Exact JA. Private-word ARIs are averaged over agents within each seed. Unpermuted Cohen's $\kappa$ measures agreement between agents' integer signs; it is not invariant to renaming, unlike ARI. True labels are used only for evaluation. Ground-truth world centers and projections are not supplied to learned inference; known sensor noise is supplied to every learned method. The evaluation functional itself uses the world-derived joint-attention prior and is not a training objective.

In  the two-agent toy experiment and its population extension, perceptual states are inferred explicitly under the latent Gaussian model rather than obtained by subtracting sensor variance from an observation-space fit. The latent model, normal--inverse-Wishart (NIW) prior, and two-agent updates are specified next;  the population extension extends them to multiple agents, with supplied initial correspondence and a balanced random-pair schedule as described in Sec.~\ref{subsec:exp2}.

An additional two-agent image experiment is described in Sec.~\ref{subsec:exp3-images}.  The two-agent toy experiment and its population extension isolate the physical and inferential mechanism in a tractable setting; the image experiment examines whether a corresponding naming procedure can operate on distinct rendered views of a common 3D object.

\subsubsection{Two-agent initialization and inference updates}

\label{subsec:exp1}

The experiment uses two agents. As shown in Proposition~\ref{prop:two-agent-exactness}, composition is the identity and no tempering occurs.  All five learned methods use the same local Gaussian families and priors from the latent CPC specification below. Centralized inference uses a common word per object, while the private-sign methods retain one word per agent and object.

\begin{align}
w_d&\sim\operatorname{Cat}(1/M), &
z_{i,d}\mid w_d,\theta_i&\sim\mathcal N(\mu_{i,w_d},\Sigma_{i,w_d}), &
o_{i,d}\mid z_{i,d}&\sim\mathcal N(z_{i,d},\sigma_{\mathrm{obs}}^2I_{d_z}),\\
\Sigma_{iw}&\sim\operatorname{IW}(\nu_0,\Psi_0), &
\mu_{iw}\mid\Sigma_{iw}&\sim\mathcal N(m_0,\Sigma_{iw}/\kappa_0).
\label{eq:exp1-latent-model}
\end{align}
$\operatorname{Cat}(1/M)$ denotes the uniform categorical distribution over $M$ words, and $\operatorname{IW}(\nu,\Psi)$ is the inverse-Wishart distribution with degrees of freedom $\nu$ and scale matrix $\Psi$. Here $d_z$ is the private-state dimension, $\sigma_{\mathrm{obs}}$ is the observation-noise standard deviation, and $(m_0,\kappa_0,\nu_0,\Psi_0)$ are the normal--inverse-Wishart hyperparameters. We set $d_z=2$, $\sigma_{\mathrm{obs}}=0.10$, $m_0=0\in\mathbb R^{d_z}$, $\kappa_0=0.01$, $\nu_0=4$, and $\Psi_0=0.1I_{d_z}$.

Write $\theta_i=\{\mu_{iw},\Sigma_{iw}\}_{w=1}^{M}$ for  agent $i$'s category parameters and $\theta=(\theta_1,\theta_2)$. Covariances are full $2\times2$ matrices, parameters are reused across episodes, and the word prior remains uniform rather than learning the unequal world frequencies. This inference model is an effective approximation, not the exact world model supplied to the references.

Each agent initializes from one k-means run on its own observations (at most 40 iterations), independently permuting its label names. Initial centers are randomly selected observations, without restarts. Local hard assignments and the observations, provisionally treated as latent states, initialize normal--inverse-Wishart factors. Initial means are their posterior means; initial covariances are their inverse-Wishart expectations. Initial perceptual states are then drawn from $p(z_i\mid o_i,w_i,\theta_i)$, combining the Gaussian sensor likelihood and the local category prior. There is no cross-agent label matching or prefitting to agent 0's partition. The four sampling methods start from the same local partitions, parameters, and perceptual states for each seed. Central VI starts from the corresponding local parameter factors. In its first round, Central Gibbs draws a common word from both latent likelihoods, rather than adopting agent 0's labels.

MHNG retains $w_{i,d}$ separately for each agent. For every object, speaker $s$ proposes $w'\sim p(w\mid z_{s,d},\theta_s)$ and listener $l$ accepts with probability $\min\!\left\{1,\frac{\mathcal N(z_{l,d};\mu_{l,w'},\Sigma_{l,w'})}{\mathcal N(z_{l,d};\mu_{l,w_{l,d}},\Sigma_{l,w_{l,d}})}\right\}$. Only the listener's word changes; rejection retains its current word. After the sweep over objects, the listener samples its parameters from their conjugate posterior given its private words and latent states, then samples its perceptual states given its observations, words, and new parameters. The speaker's stored state is unchanged. Directions alternate for 2,000 sweeps, giving each agent 1,000 parameter/perception updates. All-accept uses exactly this schedule and these local updates but always adopts the proposal. No-comm. instead performs 1,000 independent local Gibbs rounds: private word given local latent state and parameters, parameters given local words and latent states, then perceptual states given local observations, words, and parameters. Central Gibbs performs 1,000 rounds of common-word sampling from both latent likelihoods, followed by each agent's parameter and perceptual updates. Thus the sampling methods match local update counts, not communication or computational cost. No burn-in-based posterior averaging or convergence certification is claimed. Supplementary Algorithm~\ref{alg:toy-mhng} gives the common update. The accompanying text specifies the initialization, exchange schedule, and latent-state updates for each experiment.

Central VI uses the structured family $q(\theta)\prod_d[r_d(w_d)\prod_iq(z_{i,d}\mid w_d)]$, where $r_d(w_d)$ is the categorical variational word factor, with Gaussian conditional perceptual factors and conjugate parameter factors. It stops when successive ELBO values differ by at most $10^{-10}$ times $\max(1,|\mathrm{ELBO}|)$, with a cap of 300 iterations; all 16 seeds stop after 27--106 iterations. The automated stopping criterion does not use $F_d$ or world ARI. The 1,000-update sampling budget extends an earlier 30-update comparison after inspecting its learning behavior; it is exploratory rather than a preregistered confirmatory budget. All seeds are extended without seed-specific stopping, and the learning curves include that checkpoint. All 16 seeds are retained in the reported summaries, without excluding or restarting runs on the basis of their results.

Comparisons use common random numbers: for a given seed and agent, the sampling methods share parameter and perception random-number-generator streams. MHNG and All-accept also share proposal and acceptance-uniform streams, although All-accept ignores the gate. This couples methods for paired comparisons; they are not independent chains. Once word assignments coincide, some trajectories can synchronize. World realizations remain independent across seeds. Evaluation uses four scrambled Sobol replicates of 512 points, with common points across methods; inverse-normal inputs are clipped to $[10^{-12},1-10^{-12}]$. Intervals quantify seed variability under this paired design, not equilibrium Monte Carlo uncertainty.

\subsubsection{World-referenced readout and comparison limits}
\label{methods:toy-readout}

The JA reference used for the two-agent comparison in Fig.~\ref{fig:exp1-main} constructs the joint prior with the component-conditioned linear-Gaussian recognition channel described in Supplementary Text~\ref{subsec:exp2}. The generating component enters the definition of each channel in the reference mixture; the observed objects' true component labels are not supplied to the learned methods. This reference differs from the channel obtained by integrating the full soft local posterior $p_i(z_i\mid o_i)$.

For each sampling method, $F_d$ is evaluated using the exact Gaussian conditional $q_i(z)=p(z_i\mid o_i,w_i^{\mathrm{terminal}},\theta_i^{\mathrm{terminal}})$, with the common terminal word used by Central Gibbs. This integrates out the perceptual draw but conditions on one terminal word and parameter state. It is neither the free energy of a Dirac perceptual sample nor that of the full empirical sample posterior. Central VI instead uses its optimized soft mixture $q_i(z)=\sum_w r_d(w)q(z_i\mid w)$, retaining its variational parameter uncertainty. The learned methods therefore share a PGM and prior but not identical readout families or compute budgets. Their $F_d$ ordering is empirical: $F_d$ uses the world-derived prior, not the latent model's ELBO, and Exact JA is a reference rather than a claimed minimum of $F_d$.

For the CPC likelihood readout $F_{\mathrm{CPC}}$ in Fig.~\ref{fig:exp1-main}(b), we fix the fitted category parameters and integrate out both the shared word and the two private latent states in Eq.~\eqref{eq:exp1-latent-model}. For each training pair, the resulting score is
\[
-\log\left[\frac{1}{M}\sum_{w=1}^M\prod_{i=1}^2
\mathcal N\bigl(o_{i,d};\widehat\mu_{iw},\widehat\Sigma_{iw}+\sigma_{\mathrm{obs}}^2I_2\bigr)\right].%
\]
Here $\sigma_{\mathrm{obs}}=0.10$ is the observation-noise standard deviation. We use the terminal category parameters for each sampler and the variational means of the category means and covariances for Central VI. Scores are averaged over the 256 training pairs within each seed. This in-sample measure assesses the fit of the shared-word observation model at the fitted parameters. Whereas $F_d$ is unchanged by relabeling that preserves each perceptual readout, $F_{\mathrm{CPC}}$ combines the agents’ categories by their stored word indices and can depend on how those indices correspond, including in No-comm.

The centralized learned methods use both observations and estimate category parameters from the data. Exact JA uses the supplied world parameters to obtain posterior marginals under the evaluation model.

\subsection{Two-agent learning from multiview images}
\label{methods:images}
\subsubsection{Objects, rendering, and view split}
The image experiment follows the established combination of private deep generative representations and MH naming interactions \cite{hoang2024multimodal}, using the following dataset and update protocol.

The dataset contained ten scanned objects from the KIRI fruit-and-vegetable asset pack \cite{kiriFruitVegAssets}: banana (A143), pear (A112), apple (A34), dragon fruit (A107), broccoli (A87), eggplant (A7), carrot (A31), ginger (A9), pumpkin (A16), and mushroom (A79). There was one scan per object type. The assets are distributed under the Creative Commons CC0 license; we used the mid-resolution meshes and their color textures. Blender Cycles rendered $128\times128$ RGB images with fixed lighting and a plain background; the images were resized to $64\times64$ for the VAEs. Base azimuths were spaced by $10^\circ$ around a full rotation, with base elevations $10^\circ$, $25^\circ$, and $40^\circ$. Agent B's camera was offset from A's by $60^\circ$ in azimuth and $15^\circ$ in elevation, yielding 108 paired views per scan. The renderer uses orthographic cameras, and the image loader uses Lanczos resizing. Figure~\ref{fig:exp3-design}(a) shows the camera geometry and example stimuli. Supplementary Fig.~\ref{fig:exp3-stimulus-construction} shows the source scans and view split.

The view split was fixed before learning. For base-azimuth index $j\in\{0,\ldots,35\}$, views with $j\bmod6=0$ were assigned to validation, those with $j\bmod6=3$ to testing, and all others to training, at every elevation. Each scan supplied 72 training, 18 validation, and 18 test pairs, for totals of 720, 180, and 180. These are held-out views of the same scans, not held-out object identities.  Within each scan, the A and B training views were independently reshuffled at every minibatch round before the subset of communication targets was selected. Scan identity therefore supplied the common-referent pairing constraint; category labels were used for evaluation, not as VAE or GMM fitting targets.

\subsubsection{Private visual models and fixed representations}

Let $x_{i,d}$ denote agent $i$'s normalized RGB image for a paired observation of object $d$. The effective shared-word model was
\begin{equation}
p(w_d)\prod_{i\in\{A,B\}}p_{\theta_i}(z_{i,d}\mid w_d)\,p_{\psi_i}(x_{i,d}\mid z_{i,d}),
\label{eq:exp3-image-model}
\end{equation}
with a uniform prior over ten integer words,  private full-covariance Gaussian category distributions, and private diagonal-Gaussian encoders $q_{\phi_i}(z_i\mid x_i)$ in 12 dimensions. Here $\theta_i$, $\psi_i$, and $\phi_i$  denote the category, decoder, and encoder parameters, respectively; $\theta=(\theta_A,\theta_B)$  collects both agents' category parameters for the joint prior. Each agent used a residual convolutional RGB VAE with a sigmoid decoder, He initialization of the applicable convolutional layers, and zero initialization of the final convolution in each residual branch. Reconstruction used the original summed pixelwise binary cross-entropy (BCE) on normalized RGB values. We retain this fractional-target BCE convention rather than reinterpret it as a normalized continuous-RGB likelihood.

For each agent and word, the  Normal--Wishart prior over category parameters was
\[
\Lambda_{iw}\sim\mathrm{Wishart}(\nu_0,W_0),
\qquad
\mu_{iw}\mid\Lambda_{iw}\sim\mathcal N(m_0,(\kappa_0\Lambda_{iw})^{-1}),
\]
where $\Lambda_{iw}$ is the component precision matrix, $d_z$ the latent dimension, $\nu_0$ the degrees of freedom, $W_0$ the Wishart scale matrix, $m_0$ the prior mean, and $\kappa_0$ the mean-precision multiplier. We set $d_z=12$, $\nu_0=48$, $W_0=I_{d_z}/35$, $m_0=0$, and $\kappa_0=1$. With this scale-matrix parameterization, $\mathbb E[\Lambda_{iw}^{-1}]=W_0^{-1}/(\nu_0-d_z-1)=I_{d_z}$. The  independent initial category parameters were drawn using the retained initialization rule $\Lambda_{iw}^{(0)}\sim\mathrm{Wishart}(12,0.05I_{12})$, $\mu_{iw}^{(0)}\mid\Lambda_{iw}^{(0)}\sim\mathcal N(0,(\Lambda_{iw}^{(0)})^{-1})$; these initialization draws were not replaced by draws from the updated prior. Initial private words were independently uniform. No cross-agent label matching was supplied.

For each seed $s\in\{0,\ldots,9\}$, the two private VAEs were trained independently for 1,000 full-batch Adam updates on their own 720 training images, with learning rate $3\times10^{-4}$. The objective was summed pixelwise BCE per image plus the analytic KL from the diagonal-Gaussian encoder to a standard-normal latent prior, with KL coefficient one and no warmup. This yielded 20 trained VAEs in ten independent seed pairs. After training, the encoder and decoder parameters, all encoder means and variances, and one sampled latent vector $z_{i,d}$ for each training image were held fixed. The same pair of VAEs, recognition distributions, and sampled latent vectors was shared across all four methods within each seed. No further VAE training, reset, feedback-prior fitting, or resampling of these latent states occurred. This fixed-representation design isolates sign coordination and GMM learning from the mutual representation learning studied in earlier image-based naming games \cite{taniguchi2023emergent,hoang2024multimodal}.

The visual model and training hyperparameters were selected empirically during development and held fixed across methods and seeds in the reported comparison.

\subsubsection{Minibatch sign exchange and category updates}
All methods shared the same seed-specific initial GMM parameters, independently uniform private signs, and target-pairing schedule. Each round independently permuted A's and B's 72 views within each scan, then selected 128 of the 720 paired targets without replacement; the last round selected 64. Selection was renewed across rounds. There were 563 rounds, corresponding to 72,000 target visits and 563 GMM updates per agent. A GMM update sampled category parameters from their Normal--Wishart posterior computed from all 720 fixed private latent states and current retained signs, using the fixed hyperprior specified above. Unselected targets retained their signs.

In MHNG, A first proposed signs from its local Gaussian-category posterior for the selected targets without overwriting A's retained signs. B accepted or rejected each proposal using the ratio of its latent-state likelihoods under the proposed and current signs, then updated its GMM from all its memories. B next proposed using this updated GMM, followed by A's acceptance decisions and full-memory GMM update. All-accept used exactly this private-sign procedure and update timing but accepted every proposal. Central Gibbs sampled the selected shared signs once from the product of both current latent likelihoods under the uniform prior, wrote them to both agents' selected slots, and then updated both GMMs. No-comm.\ instead sampled selected private signs from each agent's own posterior and updated its own GMM. All methods matched target visits and GMM-update counts; proposal, communication, and computational costs were not identical. Central Gibbs was a centralized comparator, not a guaranteed optimum of the adaptive learning problem.

\begin{table}[!tb]
\centering
\caption{Two-agent image experiment with fixed visual representations. All four methods share the same seed-specific VAE pair and latent samples.}
\label{tab:exp3-design}
\small
\begin{tabular}{p{0.27\linewidth}p{0.65\linewidth}}
\toprule
Design element & Specification \\
\midrule
Objects and data & Ten scans, one per type; 720/180/180 train/validation/test pairs. Only training targets enter the reported readouts. \\
Private representations & RGB $64\times64$, no crop; residual VAE, 12 latent dimensions; BCE plus KL to a standard normal. \\
VAE learning & 1,000 full-batch Adam updates per individual, learning rate $3\times10^{-4}$; then encoder, decoder, recognition distributions, and latent samples fixed. \\
Category learning & Ten Gaussian categories; 563 minibatch rounds; 72,000 target visits and 563 full-memory GMM updates per agent. \\
Comparison & MHNG, All-accept, Central Gibbs, No-comm.; ten independent, paired seeds. \\
Readouts & Retained-word ARI; raw within-scan cross-view word agreement; individual and collective regularization and their sum. \\
\bottomrule
\end{tabular}
\end{table}

\subsubsection[Retained-word readouts and statistical comparisons]{Retained-word readouts and statistical comparisons}
ARI compares the current retained integer sign of each of the 720 training images with its object-type label, separately for A and B; the primary seed-level ARI averages the two agents. Raw agreement is the fraction of equal retained integers over all $72\times72$ A/B training-view combinations within each of ten scans, or 51,840 equally weighted pairs. No post-hoc relabeling is performed. This distinguishes category recovery, which is invariant to a permutation of sign names, from agreement on the names. No-comm.'s agreement can therefore be low even for similar private partitions. The reported readouts are not frozen predictions on the reserved validation or test images.

All ten seeds and the terminal 72,000-visit state were retained, with no outcome-based exclusion or selection of intermediate checkpoints. Curves were evaluated initially and approximately once per 720 target visits. The independent unit was a seed-specific pair of private VAEs together with its GMM-learning realization, not an image or an individual agent. Summaries use the mean and sample SD across ten seeds. 

For the final values in Fig.~\ref{fig:exp3-results}(d--f), the two-sided exact paired sign-flip tests enumerate all $2^{10}$ sign configurations, use the absolute mean difference, and assume sign symmetry under the null. Holm correction covers all 18 method-pair and endpoint combinations. These analyses are exploratory.

\subsubsection{Individual and collective regularization with fixed recognition}
\label{methods:image-elbo}
\label{methods:image-regularization}
For an A/B observation pair, define the analytic Gaussian KL cost for shared sign $w$ as
\begin{equation}
a_w(x_A,x_B)=\sum_{i\in\{A,B\}}\mathrm{KL}\!\left(q_{\phi_i}(z_i\mid x_i)\middle\|p_{\theta_i}(z_i\mid w)\right).
\label{eq:exp3-category-cost}
\end{equation}
We optimize an evaluation-only mean-field shared-sign factor, distinct from either agent's retained sign, using
\begin{equation}
\rho^*(w\mid x_A,x_B)=\frac{p(w)\exp(-a_w)}{\sum_v p(v)\exp(-a_v)},\qquad p(w)=1/10.
\label{eq:exp3-evaluation-rho}
\end{equation}
The two reported terms and their sum are
\begin{equation}
\begin{aligned}
I&=\sum_w\rho^*(w)a_w,\qquad C=\mathrm{KL}(\rho^*\|p(w)),\\
I+C&=-\log\sum_w p(w)\exp(-a_w).
\end{aligned}
\label{eq:exp3-regularization}
\end{equation}
The individual term $I$ measures compatibility of the fixed private recognition distributions with the learned sign-conditioned categories; $C$ is collective regularization of the evaluation-only shared-sign factor. All 51,840 within-scan cross-view training pairs are equally weighted. Encoder variances and full category precision matrices enter the Gaussian KL analytically; no decoder sampling or numerical latent integration is needed. Calculations use float64 and log-sum-exp. Each learned GMM's terminal sampled parameters are plug-in values; parameter uncertainty is not integrated out.

We call $I+C$ a regularization score, not an absolute image CFE. For a common normalized observation model, fixed encoder and decoder parameters make the prediction-error contribution $R_s=\sum_i\mathbb E_{q_{s,i}}[-\log p_{\psi_{s,i}}(x_i\mid z_i)]$ common across methods $m$ in seed $s$. Consequently, $F^{\mathrm{CPC}}_{s,m}=R_s+(I+C)_{s,m}$ implies that method differences in $I+C$ equal differences in the corresponding mean-field variational objective. This cancellation does not turn fractional-target RGB BCE into a normalized continuous likelihood.

The score evaluates the fitted CPC category models under a common recognition family. It is distinct from the world-derived JA functional $F_d$, for which this experiment does not directly construct the induced joint prior. Fixing the recognition distributions and world reference would also fix that JA functional across methods. Moreover, the learning procedure conditions GMM updates on a single fixed latent sample per image, whereas evaluation integrates each full Gaussian recognition distribution analytically. Its stochastic updates are therefore not assumed to monotonically decrease the evaluated score. Final configurations, frozen-state hashes, all 40 saved GMM states, readouts, and deterministic figure-generation code were retained for reproducibility.

\begingroup
We used OpenAI GPT-6.1 Sol, GPT-5.6 Sol, and GPT-5.6 Terra through Codex to write and revise code for the numerical experiments, evaluation, and plotting. We used Anthropic Claude Sonnet 5.5 through Claude Code for code review and computational cross-checks.\par\endgroup

\bibliographystyle{ScienceAdvancesCPC}
\bibliography{common/references,references-context}

\section*{Acknowledgments}

\begingroup
\noindent\textbf{Use of generative AI:} Generative AI assisted with brainstorming during development of the theoretical framework, manuscript drafting, Japanese-to-English translation, structural revision, and refinement of visualizations of research results. The tools included OpenAI GPT-6.1 Sol, GPT-5.6 Sol, and GPT-5.6 Terra through Codex and Anthropic Claude Opus 5.0, Claude Opus 5.5, and Claude Sonnet 5.5 through Claude Code. Materials and Methods describes AI assistance with research code and computational checks. The authors are responsible for the theoretical claims, interpretation of the results, and final manuscript.\par\endgroup

\noindent\textbf{Funding:} This work was supported by JSPS KAKENHI Grant Number JP23H04835 (Grant-in-Aid for Transformative Research Areas (A), ``Symbol Emergence Systems Theory for Qualia Structure'').

\noindent\textbf{Author contributions:} T.T. developed the theoretical framework, with contributions from K.H. T.T. implemented and performed the numerical experiments, analyzed the results, and drafted the manuscript. K.H. and R.I. contributed to manuscript review and revision.

\noindent\textbf{Competing interests:} The authors declare that they have no competing interests.

\noindent\textbf{Data, code, and materials availability:} The research code, experimental configurations, numerical data, and rendered image stimuli supporting this study are planned for deposit in a permanent public repository. Sources and licenses for the third-party 3D assets and the stimulus-generation procedures are described in Materials and Methods.

\clearpage
\appendix
\renewcommand{\thesection}{S\arabic{section}}
\setcounter{equation}{0}
\renewcommand{\theequation}{S\arabic{equation}}
\renewcommand{\theHequation}{S\arabic{equation}}
\setcounter{algorithm}{0}
\renewcommand{\thealgorithm}{S\arabic{algorithm}}
\providecommand{\theHalgorithm}{\arabic{algorithm}}
\renewcommand{\theHalgorithm}{S\arabic{algorithm}}
\setcounter{table}{0}
\renewcommand{\thetable}{S\arabic{table}}
\setcounter{figure}{0}
\renewcommand{\thefigure}{S\arabic{figure}}
\renewcommand{\theHfigure}{S\arabic{figure}}
\renewcommand{\theHtable}{S\arabic{table}}
\begin{center}\Large\textbf{Supplementary Materials}\end{center}

\section{Common-cause interactions and free energy}
\label{supp:common-cause}
\subsection{A Common-Cause Model of Joint Attention}
\label{subsec:free-energy-functional}

Our aim is to show how symbol emergence can be grounded in the common-referent aspect of joint attention. We assume a shared-object constraint on two cognitively closed agents; mutual awareness and its acquisition are not modeled. Similar observations alone do not establish joint attention. Here the hypothesis specifies that distinct private observations arise from one cause, the same object $\xi_d$. Let $H_{\mathrm{IND}}$ denote the hypothesis that the two observations originate from independent sources, and $H_{\mathrm{JA}}$ the shared-cause hypothesis. We introduce $H_{\mathrm{JA}}$ as a constraint that binds the distinct private observations $o_{i,d}$ and $o_{j,d}$ to a common cause, and derive an effective interaction between internal states from it.

In developmental psychology and accounts of language origins, joint attention involves two agents $i,j$, an object $\xi_d$, and mutual awareness of their shared attention \cite{tomasello2005understanding,carpenter2007shared,tomasello2008origins}. The present model retains the common-referent constraint without modeling that mutual awareness. For a population, the pairwise interaction sum developed below composes such shared-cause constraints; this is a modeling choice, not a psychological definition of collective joint attention.

Under $H_{\mathrm{JA}}$, the world draws one object for both agents. Each agent receives an observation through its own channel $P_\ell$.
\[
\xi_d\sim\Pideal,
\qquad
o_{\ell,d}\mid\xi_d\sim P_\ell(\cdot\mid\xi_d),
\qquad \ell\in\{i,j\}.
\]
We assume conditional independence given the object,
\[
o_{i,d}\perp o_{j,d}\mid \xi_d.
\]
Marginalizing $\xi_d$ generally makes the observations dependent. Private recognition draws are also assumed independent across agents conditional on their respective observations. Together, these assumptions give the product $r_i(z_i\mid\xi)r_j(z_j\mid\xi)$.

The common-cause assumption for $\xi_d$ concerns the world's generative process and does not by itself determine a concrete interaction between internal states. We therefore make explicit the probabilistic path from the world, through observations, to internal states. We write $P$ for world distributions, including $\Pideal$ and $P_i(o\mid\xi)$, and $p$ for model distributions over observations and internal states. Let the posterior induced by the local cognitive generative model be
\[
p_i(z_i\mid o)
=
\frac{p_i(z_i)p_i(o\mid z_i)}
{\int p_i(z_i')p_i(o\mid z_i')\,dz_i'}
\]
and define the distribution of agent $i$'s internal state induced by a given external object $\xi$ as
\[
r_i(z_i\mid \xi)
:=
\int_{\mathcal O}
P_i(o\mid \xi)
p_i(z_i\mid o)
\,do
\]
Both $P_i(o\mid \xi)$ and $p_i(z_i\mid o)$ are normalized, so $r_i(z_i\mid \xi)$ is also a normalized conditional distribution. The distinct notation $r_i$ identifies the composition of an external observation channel $P_i$ with a private model posterior $p_i$. Although it could mathematically be denoted by $p_i(z_i\mid\xi)$, doing so would obscure its dependence on both processes. It is also distinct from the variational approximation $q_i(z_i\mid o)$: $r_i$ averages recognition over possible observations of a given external object, rather than being a variational factor optimized for one observed input.

Under $H_{\mathrm{JA}}$, which states that agents $i$ and $j$ jointly attend to the same object, their internal states share the same external cause. Write their joint distribution under this hypothesis as
\[
p_{ij}^{\mathrm{JA}}(z_i,z_j)
:=
p(z_i,z_j\mid H_{\mathrm{JA}})
=
\int_{\cX}
\Pideal(\xi)
r_i(z_i\mid \xi)
r_j(z_j\mid \xi)
\,d\xi
\]
In contrast, under the independent-cause hypothesis $H_{\mathrm{IND}}$, the external causes of the two internal states are independent.

Under the independent-cause hypothesis, $\xi_i,\xi_j\overset{\mathrm{i.i.d.}}{\sim}\Pideal$ are the respective external causes for agents $i$ and $j$. The two internal states are therefore not bound by a common cause, and their joint distribution is the product of marginals, $m_i(z_i)m_j(z_j)$. This hypothesis provides a comparison for the joint-attention hypothesis; it does not require the two causes to take different values.

\[
\begin{aligned}
p_{ij}^{\mathrm{IND}}(z_i,z_j)
&:=
p(z_i,z_j\mid H_{\mathrm{IND}})
=
m_i(z_i)m_j(z_j),
\\
m_i(z_i)
&:=
\int_{\cX}
\Pideal(\xi)r_i(z_i\mid \xi)
\,d\xi
\end{aligned}
\]
We assume the marginal calibration condition that the marginal observation distribution supplied by the world to agent $i$ agrees with the prior predictive distribution of that agent's cognitive generative model,
\begin{equation}
\int_{\cX}
\Pideal(\xi)P_i(o\mid \xi)
\,d\xi
=
\int_{\mathcal Z_i}
p_i(z_i)p_i(o\mid z_i)
\,dz_i
=:
p_i(o)
\label{eq:marginal-calibration}
\end{equation}
By the law of total probability, this condition gives $m_i(z_i)=p_i(z_i)$. Hence, in the two-agent system, the generative model that uses $p_{ij}^{\mathrm{JA}}(z_i,z_j)$ as the joint-attention prior over internal states is
\begin{equation}
\begin{aligned}
p^{\mathrm{JA}}(z_i,z_j,o_{i,d},o_{j,d})
&:=
p_{ij}^{\mathrm{JA}}(z_i,z_j)
\,p_i(o_{i,d}\mid z_i)p_j(o_{j,d}\mid z_j)
\end{aligned}
\label{eq:pairwise-JA-generative}
\end{equation}
Equation~\eqref{eq:pairwise-JA-generative} defines an effective cognitive model by attaching the local likelihoods to the induced prior. Equation~\eqref{eq:marginal-calibration} matches each agent's marginal observations. Equality of the joint observation distribution with that of the external world would require an additional assumption.
Under $H_{\mathrm{JA}}$, define the usual negative ELBO that approximates the two-agent posterior by an arbitrary variational distribution $q_{ij,d}(z_i,z_j\mid o_{i,d},o_{j,d})$ as
\begin{equation}
\begin{aligned}
F_{ij,d}^{\mathrm{JA}}[q_{ij,d}]
&=
\underbrace{
\E_{q_{ij,d}(z_i,z_j\mid o_{i,d},o_{j,d})}
\bigl[
-\log p_i(o_{i,d}\mid z_i)
-\log p_j(o_{j,d}\mid z_j)
\bigr]
}_{\text{prediction error}}
\\
&\quad+
\underbrace{
\mathrm{KL}\!\left(
q_{ij,d}(z_i,z_j\mid o_{i,d},o_{j,d})
\middle\|
p_{ij}^{\mathrm{JA}}(z_i,z_j)
\right)
}_{\text{regularization}}
\label{eq:pairwise-JA-elbo}
\end{aligned}
\end{equation}
The first term is the sum of the two agents' prediction errors for their private observations. The second regularizes the joint variational distribution against large departures from the joint-attention prior $p_{ij}^{\mathrm{JA}}(z_i,z_j)$. This is the same decomposition as in Eq.~\eqref{eq:local-fe}, now written for the two-agent generative model~\eqref{eq:pairwise-JA-generative}.

To represent the fact that cognitively closed agents cannot directly access one another's internal states, we restrict the variational distribution to the mean-field family
\begin{equation}
q_{ij,d}^{\mathrm{MF}}(z_i,z_j\mid o_{i,d},o_{j,d})
:=
q_i(z_i\mid o_{i,d})q_j(z_j\mid o_{j,d})
\label{eq:pairwise-mf}
\end{equation}
Expanding the joint-attention prior as a log density ratio against the independent-cause prior gives
\begin{equation}
\begin{aligned}
\log p_{ij}^{\mathrm{JA}}(z_i,z_j)
={}&
\log p_i(z_i)+\log p_j(z_j)
+ V_{ij}(z_i,z_j),
\\
V_{ij}(z_i,z_j)
:={}&
\log\frac{p_{ij}^{\mathrm{JA}}(z_i,z_j)}
{p_{ij}^{\mathrm{IND}}(z_i,z_j)}
=
\log\frac{p(z_i,z_j\mid H_{\mathrm{JA}})}
{p(z_i,z_j\mid H_{\mathrm{IND}})}
\label{eq:Vij-general}
\end{aligned}
\end{equation}
Thus $V_{ij}$ is not an externally introduced coupling. It arises as the residual that separates the joint-attention prior from the product of local priors. Substituting the mean-field family~\eqref{eq:pairwise-mf} into~\eqref{eq:pairwise-JA-elbo} splits both terms. The prediction error factors because each likelihood depends on only one internal state,
\begin{equation}
\begin{aligned}
&\E_{q_{ij,d}^{\mathrm{MF}}}
\bigl[
-\log p_i(o_{i,d}\mid z_i)
-\log p_j(o_{j,d}\mid z_j)
\bigr]
\\
&\qquad=
\E_{q_i(z_i\mid o_{i,d})}
\bigl[-\log p_i(o_{i,d}\mid z_i)\bigr]
+
\E_{q_j(z_j\mid o_{j,d})}
\bigl[-\log p_j(o_{j,d}\mid z_j)\bigr].
\end{aligned}
\label{eq:pairwise-pe-split}
\end{equation}
For the regularization term, Eq.~\eqref{eq:Vij-general} and $p_{ij}^{\mathrm{IND}}(z_i,z_j)=p_i(z_i)p_j(z_j)$ give
\begin{equation}
\begin{aligned}
&\mathrm{KL}\!\left(
q_{ij,d}^{\mathrm{MF}}
\middle\|
p_{ij}^{\mathrm{JA}}
\right)
=
\E_{q_{ij,d}^{\mathrm{MF}}}
\bigl[
\log q_i(z_i\mid o_{i,d})
+\log q_j(z_j\mid o_{j,d})
-\log p_{ij}^{\mathrm{JA}}(z_i,z_j)
\bigr]
\\
&\qquad=
\mathrm{KL}\!\bigl(q_i(z_i\mid o_{i,d})\,\|\,p_i(z_i)\bigr)
+
\mathrm{KL}\!\bigl(q_j(z_j\mid o_{j,d})\,\|\,p_j(z_j)\bigr)
-
\E_{q_{ij,d}^{\mathrm{MF}}}
\!\left[V_{ij}(z_i,z_j)\right].
\end{aligned}
\label{eq:pairwise-kl-split}
\end{equation}
Adding~\eqref{eq:pairwise-pe-split} and~\eqref{eq:pairwise-kl-split} recovers the two local free energies~\eqref{eq:local-fe} and leaves the coupling as a single expectation,
\begin{equation}
F_{ij,d}^{\mathrm{JA}}[q_{ij,d}^{\mathrm{MF}}]
=
F_{i,d}^{\mathrm{local}}[q_i(z_i\mid o_{i,d})]
+
F_{j,d}^{\mathrm{local}}[q_j(z_j\mid o_{j,d})]
-
\E_{q_{ij,d}^{\mathrm{MF}}}
\!\left[V_{ij}(z_i,z_j)\right].
\label{eq:pairwise-elbo-decomposition}
\end{equation}
The resulting $V_{ij}$ is the log Bayes factor expressing how much more likely two internal states are to arise from the same object than from independent causes. Under marginal calibration, it equals the pointwise mutual information under the pairwise distribution induced by $H_{\mathrm{JA}}$.

Specifically, under the independent-cause hypothesis, $p_{ij}^{\mathrm{IND}}(z_i,z_j)=p_i(z_i)p_j(z_j)$, so
\[
V_{ij}(z_i,z_j)
=
\log\frac{p_{ij}^{\mathrm{JA}}(z_i,z_j)}
{p_i(z_i)p_j(z_j)}
=
\log p_{ij}^{\mathrm{JA}}(z_i,z_j)
-\log p_i(z_i)
-\log p_j(z_j).
\]
This is the standard pointwise-mutual-information form: at each pair $(z_i,z_j)$ it measures how much more probable the joint occurrence is than independent occurrence under $p_{ij}^{\mathrm{JA}}$. 
  The interaction is thus the log density ratio obtained by comparing the joint-attention prior with the product of local priors.

Reversing the order gives $V_{ij}(z_i,z_j)=V_{ji}(z_j,z_i)$. Only distinct-agent pairs contribute to the interaction sums below.

\subsection{Multi-Agent Joint Attention and the Free-Energy Functional}
\label{subsec:multi-agent-ja-free-energy}

We treat a triad comprising two agents, such as a caregiver and infant, and an object as the basic unit of joint attention. When several agents attend to the same object, we represent the overall interaction by superposing these agent--object relations. For $N\geq2$, our primary convention is
\begin{equation}
\frac{\beta_N}{2}\sum_{i\neq j}V_{ij}(z_i,z_j),
\qquad
\beta_N:=\frac{1}{N-1}.
\label{eq:beta-kac}
\end{equation}

We choose this weight when constructing the composite prior for $N>2$. The $N-1$ pairwise terms involving a given agent then have weights that sum to one.

For $N>2$, the composite prior using Eq.~\eqref{eq:beta-kac} is a tempered surrogate with weakened couplings, not a simple sum of pairwise log Bayes factors. For two agents, $\beta_2=1$, so it coincides exactly with the original log Bayes factor. This two-agent level anchors  the two-agent toy experiment and is stated as Proposition~\ref{prop:two-agent-exactness}. Pairwise composition is the main route for multi-agent interactions. Global joint attention, discussed in Sec.~\ref{sec:global-joint-attention}, is a reference system in which the construction of a shared index $w$ remains closed in the same form for $N>2$. We retain $\beta_N$ explicitly below to make the location of the scaling clear.

The pairwise joint-attention density ratio derived in Eq.~\eqref{eq:Vij-general} is
\[
\frac{p_{ij}^{\mathrm{JA}}(z_i,z_j)}
{p_i(z_i)p_j(z_j)}
=
\exp\!\left[V_{ij}(z_i,z_j)\right]
\]
This density ratio is precisely the effective interaction that appeared in the pairwise ELBO decomposition, Eq.~\eqref{eq:pairwise-elbo-decomposition}. We apply it to every pair as an overlap of triadic relations. Define the unnormalized $N$-agent prior arising from this compositional assumption by
\begin{equation}
\widetilde p_{1:N}^{\mathrm{JA}}(z_{1:N})
:=
\prod_{i=1}^{N}p_i(z_i)
\prod_{1\leq i<j\leq N}
\left[
\frac{p_{ij}^{\mathrm{JA}}(z_i,z_j)}
{p_i(z_i)p_j(z_j)}
\right]^{\beta_N}
=
\prod_{i=1}^{N}p_i(z_i)\,
\exp\!\left[
\frac{\beta_N}{2}
\sum_{\substack{i,j=1\\i\neq j}}^{N}
V_{ij}(z_i,z_j)
\right]
\label{eq:pairwise-JA-unnormalized}
\end{equation}
Normalizing Eq.~\eqref{eq:pairwise-JA-unnormalized} gives
\begin{equation}
p_{1:N}^{\mathrm{JA}}(z_{1:N})
:=
\frac{\widetilde p_{1:N}^{\mathrm{JA}}(z_{1:N})}
{\mathcal Z^{\mathrm{JA}}},
\qquad
\mathcal Z^{\mathrm{JA}}
:=
\int \widetilde p_{1:N}^{\mathrm{JA}}(z_{1:N})\,dz_{1:N}
<\infty
\label{eq:pairwise-JA}
\end{equation}
This is a composite prior formed by superposing pairwise joint-attention density ratios, not a prior directly induced by every agent sharing one external cause. Its pairwise marginals generally differ from the respective $p_{ij}^{\mathrm{JA}}$, so it must also be distinguished from $H_{\mathrm{GJA}}$ in Sec.~\ref{sec:global-joint-attention}. The choice $\beta_N=1/(N-1)$ normalizes the total coupling received by each agent independently of population size, and we assume integrability. Only for $N=2$ does Eq.~\eqref{eq:pairwise-JA} reduce to the original $p_{12}^{\mathrm{JA}}$ under the calibration condition (Proposition~\ref{prop:two-agent-exactness}).

For the generative model consisting of this composite prior and the local likelihoods,
\[
p_{1:N}^{\mathrm{JA}}(z_{1:N})
\prod_{i=1}^{N}p_i(o_{i,d}\mid z_i)
\]
we expand the negative ELBO using a product-form variational distribution. This extends the two-agent expressions in Eqs.~\eqref{eq:pairwise-JA-elbo}--\eqref{eq:pairwise-elbo-decomposition} to the composite prior in Eq.~\eqref{eq:pairwise-JA}. Up to the normalization constant $\log\mathcal Z^{\mathrm{JA}}$, the free-energy functional for object $d$ is
\begin{equation}
\begin{aligned}
F_d[\{q_i(\cdot\mid o_{i,d})\};o_{1:N,d}]
:={}&
\sum_{i=1}^{N}
F_{i,d}^{\mathrm{local}}[q_i(z_i\mid o_{i,d})]
\\
&-
\frac{\beta_N}{2}
\sum_{\substack{i,j=1\\i\neq j}}^{N}
\iint
q_i(z_i\mid o_{i,d})q_j(z_j\mid o_{j,d})
V_{ij}(z_i,z_j)
\,dz_i\,dz_j
\end{aligned}
\label{eq:Fd}
\end{equation}
The ordinary negative ELBO equals $F_d+\log\mathcal Z^{\mathrm{JA}}$, so both yield the same minimization problem over variational distributions. The second term averages the coupling by which each pair supports the joint-attention hypothesis. By Eq.~\eqref{eq:Vij-general}, $V_{ij}$ is pointwise mutual information. Given the composite prior in Eq.~\eqref{eq:pairwise-JA}, Eq.~\eqref{eq:Fd} is an ordinary mean-field variational free energy  for the chosen composite model, up to the prior normalization. For two agents, it additionally agrees term by term with the pairwise ELBO decomposition using the original log Bayes factor (Proposition~\ref{prop:two-agent-exactness}). Minimizing $F_d$ means that agents explain their private observations $o_{i,d}$ while making their internal states consistent across agents for the jointly attended object.

The target approximated by Eq.~\eqref{eq:Fd} is the observation-conditional distribution of the generative model consisting of the composite joint-attention prior and local likelihoods,
\[
p_{1:N}^{\mathrm{JA}}(z_{1:N})
\prod_i p_i(o_{i,d}\mid z_i)
\]
Indeed, substituting Eq.~\eqref{eq:pairwise-JA} into Bayes' rule gives the posterior

\begin{equation}
p_{1:N}^{\mathrm{JA}}(z_{1:N}\mid o_{1:N,d})
:=
\frac{
\widetilde p_{1:N}^{\mathrm{JA}}(z_{1:N})
\prod_{i=1}^{N}p_i(o_{i,d}\mid z_i)
}{
\mathcal Z_d^{\mathrm{JA}}(o_{1:N,d})
}
\label{eq:JA-posterior}
\end{equation}
Given the composite joint-attention prior in Eq.~\eqref{eq:pairwise-JA}, Eq.~\eqref{eq:JA-posterior} is an ordinary posterior derived by Bayes' rule for every $N$. Thus Eq.~\eqref{eq:Fd} is the variational free energy for approximating this $N$-agent posterior  with a product-form distribution, up to the prior normalization. For $N=2$ under calibration, it returns to the observation-conditional distribution of the generative model $p_{12}^{\mathrm{JA}}(z_1,z_2)\prod_i p_i(o_{i,d}\mid z_i)$. Its normalization constant is
\[
\mathcal Z_d^{\mathrm{JA}}(o_{1:N,d})
:=
\int
\widetilde p_{1:N}^{\mathrm{JA}}(z_{1:N})
\prod_{i=1}^{N}
p_i(o_{i,d}\mid z_i)
\,dz_{1:N}
\]
where $dz_{1:N}:=\prod_i dz_i$, and we assume $\mathcal Z_d^{\mathrm{JA}}(o_{1:N,d})<\infty$. This integral uses the unnormalized prior $\widetilde p_{1:N}^{\mathrm{JA}}$. The ordinary model evidence of the normalized generative model is
\[
p^{\mathrm{JA}}(o_{1:N,d})
:=
\int
p_{1:N}^{\mathrm{JA}}(z_{1:N})
\prod_{i=1}^{N}
p_i(o_{i,d}\mid z_i)
\,dz_{1:N}.
\]
Substituting $p_{1:N}^{\mathrm{JA}}=\widetilde p_{1:N}^{\mathrm{JA}}/\mathcal Z^{\mathrm{JA}}$ from Eq.~\eqref{eq:pairwise-JA} therefore factors out the observation-independent constant $\mathcal Z^{\mathrm{JA}}$:
\[
\mathcal Z_d^{\mathrm{JA}}(o_{1:N,d})
=
\mathcal Z^{\mathrm{JA}}\,
p^{\mathrm{JA}}(o_{1:N,d}).
\]
The offset $\log\mathcal Z^{\mathrm{JA}}$ does not depend on observations or variational distributions, so it does not affect minimization of $F_d$. Let $q_{1:N,d}(z_{1:N}):=\prod_{i=1}^{N}q_i(z_i\mid o_{i,d})$ denote the product-form variational distribution.

This calculation evaluates the ordinary ELBO for the composite joint-attention generative model under $q_{1:N,d}$. Expanding the KL divergence between this product-form variational distribution and the $N$-agent posterior in Eq.~\eqref{eq:JA-posterior} recovers Eq.~\eqref{eq:Fd} as the negative ELBO, namely the variational free energy:
\[
\begin{aligned}
&\mathrm{KL}\!\left(
q_{1:N,d}
\middle\|
p_{1:N}^{\mathrm{JA}}(\cdot\mid o_{1:N,d})
\right)
-
\log\mathcal Z_d^{\mathrm{JA}}(o_{1:N,d})
\\
&=
\sum_{i=1}^{N}
\underbrace{
\E_{q_i(z_i\mid o_{i,d})}
\left[
-\log p_i(o_{i,d}\mid z_i)
\right]
}_{\text{prediction error}}
\\
&\quad+
\underbrace{
\sum_{i=1}^{N}
\mathrm{KL}\!\left(
q_i(z_i\mid o_{i,d})
\middle\|
p_i(z_i)
\right)
-
\frac{\beta_N}{2}
\sum_{\substack{i,j=1\\i\neq j}}^{N}
\E_{q_i(z_i\mid o_{i,d})q_j(z_j\mid o_{j,d})}
\left[
V_{ij}(z_i,z_j)
\right]
}_{\text{regularization}}
\\
&=
\sum_{i=1}^{N}
F_{i,d}^{\mathrm{local}}[q_i(z_i\mid o_{i,d})]
-
\frac{\beta_N}{2}
\sum_{\substack{i,j=1\\i\neq j}}^{N}
\E_{q_i(z_i\mid o_{i,d})q_j(z_j\mid o_{j,d})}
\left[
V_{ij}(z_i,z_j)
\right]
\\
&=
F_d[\{q_i(\cdot\mid o_{i,d})\};o_{1:N,d}],
\end{aligned}
\]
The second grouping is the regularization toward the composite joint-attention prior $p_{1:N}^{\mathrm{JA}}$, not toward the product of local priors. The local KL terms alone would penalize departures from $\prod_i p_i(z_i)$; subtracting the expected couplings $\E[V_{ij}]$ completes that KL against $p_{1:N}^{\mathrm{JA}}$, up to the observation-independent constant $\log\mathcal Z^{\mathrm{JA}}$. The next equality uses the definition of local free energy in Eq.~\eqref{eq:local-fe}, and the final equality the definition of the interaction-bearing free-energy functional in Eq.~\eqref{eq:Fd}. We have therefore not introduced a new free energy, but rewritten the previously defined $F_d$ in KL form for the approximate posterior:
\begin{equation}
F_d[\{q_i(\cdot\mid o_{i,d})\};o_{1:N,d}]
=
\mathrm{KL}\!\left(
q_{1:N,d}(z_{1:N})
\middle\|
p_{1:N}^{\mathrm{JA}}(z_{1:N}\mid o_{1:N,d})
\right)
-
\log
\mathcal Z_d^{\mathrm{JA}}(o_{1:N,d})
\label{eq:Fd-as-KL}
\end{equation}
\begingroup\setlength{\emergencystretch}{3em}
Equation~\eqref{eq:Fd-as-KL} uses the unnormalized evidence; the normalized generative model has negative ELBO $F_d+\log\mathcal Z^{\mathrm{JA}}$. Since the second term is independent of $\{q_i\}$, minimizing $F_d$ is equivalent to mean-field variational inference that approximates the posterior $p_{1:N}^{\mathrm{JA}}(z_{1:N}\mid o_{1:N,d})$ of the joint-attention generative model by $q_{1:N,d}=\prod_iq_i(z_i\mid o_{i,d})$.\par\endgroup

The interaction theory is stated for general $N$ under the normalization in Eq.~\eqref{eq:beta-kac}. The construction of the shared categorical variable and  the two-agent toy experiment are given at the two-agent level, the triadic basic unit of joint attention. At that level the composite prior is not an approximation.

\begin{proposition}[Exactness for two agents]
\label{prop:two-agent-exactness}
\label{sec:two-agent-exactness}
Let $N=2$. Then $\beta_2=1$ by Eq.~\eqref{eq:beta-kac}, and the unnormalized composite prior in Eq.~\eqref{eq:pairwise-JA-unnormalized} coincides with the original pairwise joint-attention prior,
\[
\widetilde p_{1:2}^{\mathrm{JA}}(z_1,z_2)
=
p_{12}^{\mathrm{JA}}(z_1,z_2).
\]
Since $p_{12}^{\mathrm{JA}}$ is normalized, $\mathcal Z^{\mathrm{JA}}=1$ and $p_{1:2}^{\mathrm{JA}}=p_{12}^{\mathrm{JA}}$. Marginal calibration in Eq.~\eqref{eq:marginal-calibration} identifies the two marginals of $p_{12}^{\mathrm{JA}}$ with the local priors in the interaction denominator. Moreover, $F_d$ agrees term by term with the two-agent ELBO decomposition in Eq.~\eqref{eq:pairwise-elbo-decomposition}, and equals the ordinary mean-field variational free energy in Eq.~\eqref{eq:pairwise-JA-elbo} for the generative model in Eq.~\eqref{eq:pairwise-JA-generative}.
\end{proposition}

\begin{proof}
For $N=2$, Eq.~\eqref{eq:beta-kac} gives $\beta_2=1$. Therefore
\[
\widetilde p_{1:2}^{\mathrm{JA}}(z_1,z_2)
=
p_1(z_1)p_2(z_2)
\frac{p_{12}^{\mathrm{JA}}(z_1,z_2)}{p_1(z_1)p_2(z_2)}
=
p_{12}^{\mathrm{JA}}(z_1,z_2),
\]
and normalization of $p_{12}^{\mathrm{JA}}$ gives $\mathcal Z^{\mathrm{JA}}=1$. The coupling term in Eq.~\eqref{eq:Fd} reduces to
\[
\frac{\beta_2}{2}\sum_{i\neq j}
\E\!\left[V_{ij}\right]
=
\E_{q_1q_2}\!\left[V_{12}(z_1,z_2)\right].
\]
By Eq.~\eqref{eq:Vij-general}, this is the expectation of the log Bayes factor for $H_{\mathrm{JA}}$ against the independent-cause hypothesis. Hence $F_d$ matches the pairwise decomposition, and no tempering occurs.
\end{proof}

In this sense, the two-agent case is the anchor of the framework.  For $N>2$, the interaction in the collective free energy follows from the chosen pairwise composite prior in Eq.~\eqref{eq:pairwise-JA}. For two agents, moreover, the prior is obtained directly from the original triad. The construction of the shared categorical variable in Sec.~\ref{sec:shared-variable} is likewise exact at this level. This is why  the two-agent toy experiment uses two agents.

\subsection{Global Joint Attention}
\label{sec:global-joint-attention}

Call the stronger hypothesis that every agent sees the same object $\xi_d$ in one episode $d$ global joint attention, $H_{\mathrm{GJA}}$. The joint distribution of internal states induced by this hypothesis is
\begin{equation}
p_{1:N}^{\mathrm{GJA}}(z_{1:N})
:=
\int_{\cX}
\Pideal(\xi_d)
\prod_{i=1}^{N}
r_i(z_i\mid \xi_d)
\,d\xi_d
\label{eq:pGJA-N-app}
\end{equation}
Because $\Pideal$ and every $r_i(\cdot\mid \xi_d)$ are normalized, Eq.~\eqref{eq:pGJA-N-app} is normalized. This is not an assumption that agents share the data set $\xi=\{\xi_d\}$; it states that all agents see the same $\xi_d$ in one episode. The associated generative model uses this already-normalized prior and the local likelihoods,
\begin{equation}
p^{\mathrm{GJA}}(z_{1:N},o_{1:N,d})
:=
p_{1:N}^{\mathrm{GJA}}(z_{1:N})
\prod_{i=1}^{N}
p_i(o_{i,d}\mid z_i).
\label{eq:GJA-generative}
\end{equation}
Under $H_{\mathrm{GJA}}$, the ordinary negative ELBO for the same product-form family $q_{1:N,d}=\prod_i q_i$ is therefore
\begin{equation}
\begin{aligned}
F_d^{\mathrm{GJA}}[\{q_i(\cdot\mid o_{i,d})\};o_{1:N,d}]
:={}&
\sum_{i=1}^{N}
\underbrace{
\E_{q_i(z_i\mid o_{i,d})}
\bigl[
-\log p_i(o_{i,d}\mid z_i)
\bigr]
}_{\text{prediction error}}
\\
&+
\underbrace{
\mathrm{KL}\!\left(
q_{1:N,d}
\middle\|
p_{1:N}^{\mathrm{GJA}}
\right)
}_{\text{regularization}}.
\end{aligned}
\label{eq:Fd-GJA}
\end{equation}
The second term regularizes toward the global-joint-attention prior $p_{1:N}^{\mathrm{GJA}}$, not toward the product of local priors. Expanding the log density ratio
\begin{equation}
V^{\mathrm{GJA}}(z_{1:N})
:=
\log p_{1:N}^{\mathrm{GJA}}(z_{1:N})
-
\sum_{i=1}^{N}
\log p_i(z_i)
\label{eq:V-GJA}
\end{equation}
gives
\[
F_d^{\mathrm{GJA}}
=
\sum_{i=1}^{N}
F_{i,d}^{\mathrm{local}}[q_i(z_i\mid o_{i,d})]
-
\E_{q_{1:N,d}}\!\left[V^{\mathrm{GJA}}(z_{1:N})\right].
\]
The ordinary model evidence of Eq.~\eqref{eq:GJA-generative} is
\[
\mathcal Z_d^{\mathrm{GJA}}(o_{1:N,d})
:=
\int
p_{1:N}^{\mathrm{GJA}}(z_{1:N})
\prod_{i=1}^{N}
p_i(o_{i,d}\mid z_i)
\,dz_{1:N}
=
p^{\mathrm{GJA}}(o_{1:N,d}),
\]
and the same expansion as in Eq.~\eqref{eq:Fd-as-KL} yields
\begin{equation}
F_d^{\mathrm{GJA}}[\{q_i(\cdot\mid o_{i,d})\};o_{1:N,d}]
=
\mathrm{KL}\!\left(
q_{1:N,d}
\middle\|
p_{1:N}^{\mathrm{GJA}}(\cdot\mid o_{1:N,d})
\right)
-
\log
\mathcal Z_d^{\mathrm{GJA}}(o_{1:N,d}).
\label{eq:Fd-GJA-as-KL}
\end{equation}

For $N=2$, $H_{\mathrm{JA}}=H_{\mathrm{GJA}}$, $V^{\mathrm{GJA}}=V_{12}$, and $F_d^{\mathrm{GJA}}$ coincides with $F_d$ in Eq.~\eqref{eq:Fd}. For $N>2$, however, overlapping triads do not automatically imply that all agents see the same $\xi_d$; pairwise $H_{\mathrm{JA}}$ also permits configurations in which the object differs across pairs. Thus $H_{\mathrm{GJA}}$ is a stronger common-cause hypothesis than pairwise composition: every pair shares one object because all $N$ agents share one object. Its prior in Eq.~\eqref{eq:pGJA-N-app} is not the composite prior in Eq.~\eqref{eq:pairwise-JA}, and $V^{\mathrm{GJA}}$ is not the $\beta_N$-weighted sum of pairwise $V_{ij}$. Accordingly $F_d^{\mathrm{GJA}}$ and the pairwise $F_d$ are distinct functionals.

When the world is categorical (Sec.~\ref{sec:shared-variable}), $p_{1:N}^{\mathrm{GJA}}$ remains a $K$-component mixture, so a single shared index $w_d$ can be read by the same construction as in the two-agent case. The composite prior in Eq.~\eqref{eq:pairwise-JA} does not preserve mixture structure under fractional powers, so reduction to one shared index remains open. The main route for $V_{ij}$ and $F_d$ remains pairwise composition. The $F_d$ evaluated in  the population extension belongs to that pairwise composition; $F_d^{\mathrm{GJA}}$ is the variational free energy of the GJA generative model itself. GJA is a reference system in which construction of $w$ closes in the same form.

We have obtained the joint-attention prior $p_{ij}^{\mathrm{JA}}$, the interaction $V_{ij}$ that separates it from the product of local priors, and its mean-field free energy $F_d$. When the world is a finite mixture, the joint-attention prior inherits that mixture structure. The shared variable $w$ introduced next indexes its components, while the interaction remains the log ratio of the joint prior to the product of local priors.


\section{Shared-variable construction and effective CPC}
\label{supp:shared-variable}
\subsection{A Categorical World and the Mixture Structure of the Derived Prior}
\label{subsec:categorical-world}

We impose the following structure on the world's generative process.

\begin{assumption}[Categorical world]
\label{assump:categorical-world}
The distribution of external objects is a finite mixture,
\[
\Pideal(\xi)=\sum_{k=1}^{K}\pi_k\,P_k(\xi),
\qquad
\pi_k>0,\quad\sum_k\pi_k=1
\]
where each component $P_k$ represents a collection of objects of the same kind.
\end{assumption}

This assumption is not an ontological claim that the world intrinsically consists of finitely many categories. It models a shareable coarse-graining by a finite mixture. As motivation for this representation, Bacharoglou establishes finite Gaussian-mixture approximation in $L^1$ and $L^\infty$ for continuous compactly supported densities on $\mathbb R$, and notes analogous multivariate extensions \cite{bacharoglou2010approximation}. Infinite Gaussian mixtures, or Dirichlet-process mixtures, provide a Bayesian extension that does not fix $K$ in advance \cite{rasmussen2000infinite}. Posterior consistency results for density estimation require the stated conditions on the true density and the prior; stronger consistency results impose additional restrictions \cite{ghosal1999posterior}. Neither density approximation nor asymptotic posterior consistency, by itself, bounds the KL or free-energy error of the finite-component model used here. These results also do not establish a true finite number of categories or socially shared categories. The finite mixture remains an explicit assumption used to derive a reusable discrete shared variable $w$.

Let $r_i(z_i\mid\xi)$ denote agent $i$'s recognition channel, the map from an object in the world to its internal state after marginalizing observations. The pairwise joint-attention prior is obtained by marginalizing the common cause:
\[
p^{\mathrm{JA}}_{ij}(z_i,z_j)
=\int_{\cX}\Pideal(\xi)\,r_i(z_i\mid\xi)\,r_j(z_j\mid\xi)\,d\xi
\]
For two agents, this marginalization is the content of $H_{\mathrm{JA}}$ itself; the same form holds for multi-agent global joint attention (Sec.~\ref{sec:global-joint-attention}). Substituting Assumption~\ref{assump:categorical-world} separates the integral by component:
\begin{equation}
p^{\mathrm{JA}}_{ij}(z_i,z_j)
=\sum_{k=1}^{K}\pi_k\,p^{(k)}_{ij}(z_i,z_j),
\qquad
p^{(k)}_{ij}(z_i,z_j)
:=\int_{\cX}P_k(\xi)\,r_i(z_i\mid\xi)\,r_j(z_j\mid\xi)\,d\xi
\label{eq:mixture-pair}
\end{equation}
Letting $p^{(k)}_i(z_i):=\int P_k(\xi)r_i(z_i\mid\xi)d\xi$ denote the component marginal, the local prior is $p_i=\sum_k\pi_kp^{(k)}_i$, consistent with the marginal calibration condition in Eq.~\eqref{eq:marginal-calibration}.

Thus, the derived joint-attention prior is exactly a $K$-component mixture that inherits the world's categorical structure. No approximation has entered so far.  The mixture represents both the choice of a generating component and the dependence that may remain within that component. The shared variable introduced below records the component index.

\subsection{The Shared Categorical Variable as a Mixture Index}
\label{subsec:w-construction}

Define the shared categorical variable as a reading of the mixture index in Eq.~\eqref{eq:mixture-pair}:

\begin{equation}
p(w=k):=\pi_k,
\qquad
p_i(z_i\mid w=k):=p^{(k)}_i(z_i)
\qquad(k=1,\ldots,K)
\label{eq:w-construction}
\end{equation}

This is neither estimation nor learning, but a construction from the prior derived in Sec.~\ref{sec:derivation}. Directly from the definition, for every $i$,
\[
\sum_{w}p(w)\,p_i(z_i\mid w)=p_i(z_i)
\]
holds exactly.  Introducing $w$ through the exact augmentation preserves the joint prior and hence its marginals. The subsequent CPC factorization can change the coupling. Returning to the definition in Eq.~\eqref{eq:Vij-general},
\begin{equation}
V_{ij}(z_i,z_j)
=
\log\frac{p_{ij}^{\mathrm{JA}}(z_i,z_j)}{p_{ij}^{\mathrm{IND}}(z_i,z_j)}
=
\log\frac{p_{ij}^{\mathrm{JA}}(z_i,z_j)}{p_i(z_i)p_j(z_j)}
=
\log\frac{\sum_{k}\pi_k\,p^{(k)}_{ij}(z_i,z_j)}{p_i(z_i)p_j(z_j)}.
\label{eq:Vij-mixture}
\end{equation}
The first equality is the definition. The second uses $p_{ij}^{\mathrm{IND}}(z_i,z_j)=p_i(z_i)p_j(z_j)$ under the marginal calibration condition in Eq.~\eqref{eq:marginal-calibration}. The third substitutes the mixture form of the derived prior, Eq.~\eqref{eq:mixture-pair}.  The world-referenced $F_d$ in Sec.~\ref{sec:numerical-experiment} uses a fixed $V_{ij}$. Exchange and learning can change the recognition distributions used to evaluate the local terms and the expected interaction.

The relation between $w$ and $\xi_d$ requires care.  The variable $w$ records the generating component of the specified world mixture. It does not determine the individual object, and when the components overlap, an object can have nonzero posterior probability under more than one component. The decisive difference is accessibility. Cognitive closure prevents anyone from conditioning directly on $\xi_d$, whereas $w$ is finite and discrete and can be externalized as a public sign.  This discrete index can serve as a shared sign once agents learn corresponding uses.

\subsection{Within-Category Residual Correlations and the Effective CPC Generative Model}
\label{subsec:category-residual}

\begingroup\setlength{\emergencystretch}{4em}

 The mixture index gives the exact augmented prior shown below. Introducing this variable preserves the joint distribution over private states; conditional independence within a category requires a further approximation.

\[
p(w=k,z_i,z_j)
=
\pi_k\,p^{(k)}_{ij}(z_i,z_j)
=
p(w=k)\,p(z_i,z_j\mid w=k).
\]

Although private states are independent conditional on both $w$ and the common object $\xi_d$, conditioning on $w$ alone can leave dependence through the within-category position of that object. We quantify this dependence by conditional mutual information:

\begin{equation}
\epsilon^{\mathrm{cat}}_{ij}
:=\sum_{k}\pi_k\,
\mathrm{KL}\!\left(p^{(k)}_{ij}\,\middle\|\,p^{(k)}_{i}\,p^{(k)}_{j}\right)
=I(z_i;z_j\mid w)
\label{eq:eps-cat}
\end{equation}
For two agents, the KL divergence from the exact augmented prior to the CPC prior is exactly $\epsilon^{\mathrm{cat}}_{ij}$, by conditioning on their common $p(w)$. Marginalizing $w$ can only reduce this divergence. This bound concerns the priors and does not bound the free-energy error for a specified observation and variational distribution.

 For two-agent readouts, we suppress the pair subscript and write $\epsilon^{\mathrm{cat}}$. It measures the discrepancy between the exact category-conditional joint distribution and the product of its marginals, averaged over categories. A sign can retain category identity while omitting this finer dependence.

 The effective CPC model replaces each category-conditional joint distribution by the product of its marginals:

\[
p(z_i,z_j\mid w=k)
\approx
p^{(k)}_i(z_i)\,p^{(k)}_j(z_j)
=
p_i(z_i\mid w=k)\,p_j(z_j\mid w=k).
\]

Attaching the local observation models already used in Eq.~\eqref{eq:pairwise-JA-generative} then yields
\[
p(w,z_i,z_j,o_i,o_j)
=
p(w)\,p_i(z_i\mid w)\,p_j(z_j\mid w)\,p_i(o_i\mid z_i)\,p_j(o_j\mid z_j).
\]
The same replacement for every agent---equivalently, the global-joint-attention mixture after dropping within-component dependence---is the standard CPC generative model
\begin{equation}
p(w_d,z_{1:N,d},o_{1:N,d})
=
p(w_d)
\prod_{i=1}^{N}
p_i(z_{i,d}\mid w_d)\,
p_i(o_{i,d}\mid z_{i,d})
\label{eq:cpc-generative}
\end{equation}
The graph is $w\to z_i\to o_i$; given $w$, the agents factor, and given $z_i$ the observation does not depend on $w$.  This factorization does not require a Gaussian distributional family.

\paragraph{Special case: a component-conditioned linear-Gaussian channel.}

 A closed form is available for the special reference channel in Sec.~\ref{subsec:exp2}, where recognition is conditioned on the generating component, the channels are linear Gaussian, and world components are isotropic with common covariance. Here $A_i,A_j\in\mathbb R^{2\times3}$ are the agents' observation projection matrices defined in Materials and Methods, Sec.~\ref{methods:toy}, and illustrated in Fig.~\ref{fig:exp1-world}; they map the three-dimensional object state $\xi$ to the two-dimensional observations and satisfy $A_iA_i^\top=A_jA_j^\top=I_2$. Writing $C_{ij}=A_iA_j^\top$ and $s=\sigma_\xi^2/(\sigma_\xi^2+\sigma_{\mathrm{obs}}^2)$, its normalized within-component cross-covariance is $s^2C_{ij}$. Under these assumptions,

\begin{equation}
\epsilon^{\mathrm{cat}}_{ij}
=-\tfrac12\log\det\!\left(I-s^{4}C_{ij}C_{ij}^{\top}\right)
\label{eq:within-category-mi}
\end{equation}

 The value is the same for every component under this covariance structure. This example evaluates the residual for the specified component-conditioned Gaussian channel; it is not a general expression for the full soft recognition channel or for learned image representations. The mixture construction and CPC approximation above do not require this Gaussian specialization. Choosing Gaussian category distributions $p_i(z\mid w)$ gives the Inter-GMM--Gaussian form.\par\endgroup

\subsection[Inference and communication through the shared variable]{ Inference and communication through the shared variable}
\label{subsec:generative}

\begingroup\setlength{\emergencystretch}{4em}
 With the world distribution and recognition channels held fixed across episodes, the construction in Eq.~\eqref{eq:w-construction} supplies the same sign prior $p(w)$ and agent-specific category distributions $p_i(z\mid w)$ for each episode. These distributions can therefore be reused when inferring signs for different objects within this setting.\par\endgroup

Inference within each episode is ordinary Bayesian computation. Define $p_i(o\mid w):=\int p_i(z\mid w)p_i(o\mid z)dz$.

The exact observation-conditioned sign posterior is $p(w\mid o_{1:N,d})\propto p(w)\prod_i p_i(o_{i,d}\mid w)$. The corresponding private-state marginal is $p(z_i\mid o_{1:N,d})=\sum_w p(w\mid o_{1:N,d})p_i(z_i\mid w,o_{i,d})$. We reserve $q_d$ for the variational sign factor in Sec.~\ref{subsec:cfe}.
 Conditional independence is exact within the effective CPC model. Given fixed private states, Bayes' rule gives a separate product-form sign posterior
\begin{equation}
p(w_d\mid z_{1:N,d})
\propto
p(w_d)
\prod_{i=1}^{N}
p_i(z_{i,d}\mid w_d)
\label{eq:product-posterior}
\end{equation}
The naming channel of each agent is likewise derived by the same Bayes rule, $p_i(w\mid z_i)\propto p(w)p_i(z_i\mid w)$; it is not learned independently.

 \begingroup\setlength{\emergencystretch}{2em}
 In the naming-game protocol, agents exchange only the sign $w$. 
 The naming games in this paper condition word exchange on the agents' private perceptual states. Observations inform those states through local perceptual inference. During a two-agent word update, fix $z_{s,d},z_{l,d}$ and the category parameters $\theta_s,\theta_l$. The speaker proposes $w'$ from $g_s(w)\propto p(w)p_s(z_{s,d}\mid w,\theta_s)$, and the listener accepts with probability $\alpha=\min\{1,p_l(z_{l,d}\mid w',\theta_l)/p_l(z_{l,d}\mid w_{l,d},\theta_l)\}$. The listener compares the proposal with its currently retained word $w_{l,d}$. For fixed states and parameters, this MH word-update kernel leaves $p(w\mid z_{s,d},z_{l,d},\theta_s,\theta_l)\propto p(w)p_s(z_{s,d}\mid w,\theta_s)p_l(z_{l,d}\mid w,\theta_l)$ invariant, since the prior and speaker likelihood cancel in the acceptance ratio \cite{taniguchi2023emergent}. This is the two-agent instance of Eq.~\eqref{eq:product-posterior}. This fixed-evidence result applies to a listener word update. It does not establish agreement between the two retained words or invariance of the centralized CPC posterior under the full adaptive learning schedule.\par\endgroup

\subsection{The Collective Variational Distribution and Collective Free Energy}
\label{subsec:cfe}

For a finite vocabulary $\mathcal W$, the conditional distribution $p(w_d\mid z_{1:N,d})$ of the effective CPC model can be evaluated by normalizing $|\mathcal W|$ terms in Eq.~\eqref{eq:product-posterior}. Inference from observations additionally requires integration over the private latent states. As one approximation to this observation-conditioned posterior, introduce a collective variational distribution
\[
q_d(w_d\mid o_{1:N,d}),
\qquad
\sum_{w_d\in\mathcal W}q_d(w_d\mid o_{1:N,d})=1
\]
The corresponding mean-field inference model is
\begin{equation}
q(w_d,z_{1:N,d}\mid o_{1:N,d})
=
q_d(w_d\mid o_{1:N,d})
\prod_{i=1}^{N}
q_i(z_{i,d}\mid o_{i,d})
\label{eq:inference-model}
\end{equation}
Here $q_d$ is independently varied to approximate the posterior distribution of the category label $w_d$.

For the effective CPC generative and inference models constructed above, define collective free energy as the negative ELBO:
\begin{equation}
\begin{aligned}
F_d^{\mathrm{CPC}}
:={}&
\E_q
\Bigl[
\log q_d(w_d\mid o_{1:N,d})
+
\sum_{i=1}^{N}
\log q_i(z_i\mid o_i)
\\
&\qquad
-
\log p(w_d)
-
\sum_{i=1}^{N}
\log p_i(o_i\mid z_i)
-
\sum_{i=1}^{N}
\log p_i(z_i\mid w_d)
\Bigr]
\end{aligned}
\label{eq:cfe}
\end{equation}
This satisfies
\[
F_d^{\mathrm{CPC}}
=
\mathrm{KL}\!\left(
q_d(w_d\mid o_{1:N,d})\prod_iq_i(z_i\mid o_i)
\middle\|
p(w_d,z_{1:N}\mid o_{1:N})
\right)
-
\log p(o_{1:N})
\]
and therefore $F_d^{\mathrm{CPC}}\geq-\log p(o_{1:N})$.
Using the factorization gives
\begin{equation}
\begin{aligned}
F_d^{\mathrm{CPC}}
={}&
\underbrace{
\mathrm{KL}\!\left(
q_d(w_d\mid o_{1:N,d})\middle\|p(w_d)
\right)
}_{\text{collective regularization}}
+
\sum_{i=1}^{N}
\underbrace{
\E_{q_i(z_i\mid o_i)}
\!\left[
-\log p_i(o_i\mid z_i)
\right]
}_{\text{prediction error}}
\\
&+
\sum_{i=1}^{N}
\underbrace{
\E_{q_d(w_d\mid o_{1:N,d})}
\!\left[
\mathrm{KL}\!\left(
q_i(z_i\mid o_i)
\middle\|
p_i(z_i\mid w_d)
\right)
\right]
}_{\text{individual regularization}}
\end{aligned}
\label{eq:cfe-decomposition}
\end{equation}
In customary CPC terminology, the first term is collective regularization. It measures the departure of the posterior $q_d(w_d\mid o_{1:N,d})$ of the shared variable from its prior $p(w_d)$. This is a complexity term for a variable inferred collectively.

The second term is prediction error; it measures how well internal state $z_i$ predicts or reconstructs each agent's private observation $o_i$.

The third term is individual regularization; it aligns each agent's local posterior $q_i(z_i\mid o_i)$ with the generative distribution $p_i(z_i\mid w_d)$ conditioned on the shared variable. It is a category-consistency term linking local internal representations to the collective shared variable.

The stationary condition for $q_d$ is
\begin{equation}
q_d^*(w\mid o_{1:N,d})
\propto
p(w)
\exp\!\left[
\sum_{i=1}^{N}
\E_{q_i(z_i\mid o_i)}
\log p_i(z_i\mid w)
\right]
\label{eq:qd-star}
\end{equation}
 If each local distribution concentrates at $z_{i,d}$ and its expected log category likelihood converges to the log likelihood at that point, the limiting variational factor satisfies
\[
q_d^*(w\mid o_{1:N,d})
\propto
p(w)
\prod_{i=1}^{N}
p_i(z_{i,d}\mid w)
\]
\begingroup\setlength{\emergencystretch}{3em}
 which agrees with Eq.~\eqref{eq:product-posterior} in that limit. For nondegenerate recognition distributions, Eq.~\eqref{eq:qd-star} defines the variational factor used for the image regularization readout. The implemented MHNG word update uses fixed private latent values at the time of each proposal and acceptance decision.\par\endgroup


\subsection{From a Categorical World to Symbol Emergence}
\label{subsec:mapping}

The preceding derivation links the structure of the world to a computational role for shared symbols. Table~\ref{tab:main-path} organizes this argument into five stages: a categorical world, joint attention, a shared variable and its effective CPC model, decentralized inference through MHNG, and the emergence of shared referential symbols. The sequence expresses explanatory dependencies. A finite mixture of unobservable kinds makes discrete reference reusable across episodes, while joint attention supplies the common-cause relation through which a later shared sign can coordinate private inferences.

\begin{table}[tbp]
\centering
\caption{From a categorical world to shared referential symbols. The first three stages specify the world, derive its joint-attention prior, and construct an effective model. The fourth supplies a decentralized inference mechanism. The fifth states the proposed role of this mechanism in symbol emergence. The mixture construction is exact for two agents and extends to global joint attention; CPC neglects within-category correlations.}
\label{tab:main-path}
\small
\renewcommand{\arraystretch}{1.18}
\begin{tabular}{@{}p{0.20\linewidth}p{0.43\linewidth}p{0.31\linewidth}@{}}
\toprule
\raggedright Stage & \raggedright Mathematical or computational content & \raggedright Role in the explanation \tabularnewline
\midrule
\raggedright 1. Categorical world &
\raggedright $\Pideal(\xi)=\sum_k\pi_k P_k(\xi)$ (Assumption~\ref{assump:categorical-world}). &
\raggedright Unobservable kinds recur across episodes, so a discrete sign can later be reused. \tabularnewline
\addlinespace
\raggedright 2. Joint attention &
\raggedright Marginalizing the same object gives $p_{ij}^{\mathrm{JA}}=\sum_k\pi_k p_{ij}^{(k)}$ and the interaction $V_{ij}$ (Eqs.~\eqref{eq:mixture-pair} and \eqref{eq:Vij-general}). &
\raggedright Private inferences become coupled through a common cause. \tabularnewline
\addlinespace
\raggedright 3. Shared $w$ and CPC &
\raggedright Read $w$ from the mixture index. Neglect within-category correlations to obtain the CPC model and its collective posterior (Eqs.~\eqref{eq:cpc-generative} and \eqref{eq:product-posterior}). &
\raggedright An externalizable category variable carries the discrete part of this coupling. \tabularnewline
\addlinespace
\raggedright 4. Decentralized inference: MHNG &
\raggedright A speaker proposes $w'$ from its private posterior; a listener accepts or rejects it using a local likelihood ratio (Sec.~\ref{subsec:generative}). &
\raggedright  For two agents, the fixed-evidence word kernel preserves the collective sign posterior. \tabularnewline
\addlinespace
\raggedright 5. Symbol emergence &
\raggedright  The experiments test whether repeated exchange and local learning support shared category use (Sec.~\ref{subsec:exp1}). &
\raggedright A reusable public sign acquires a shared referential function, providing a possible precursor of language. \tabularnewline
\bottomrule
\end{tabular}
\end{table}

The first transition establishes why private predictions are coupled. Joint attention identifies a common object behind distinct observations. Marginalizing that object yields the effective interaction $V_{ij}$ and the associated free energy $F_d$. Under the categorical-world assumption, the induced prior retains an exact mixture structure. Its index defines $w$, and its component marginals define $p_i(z_i\mid w)$. The effective CPC model follows upon neglecting the within-category dependence measured by $\epsilon^{\mathrm{cat}}_{ij}=I(z_i;z_j\mid w)$. This conditional mutual information is the category-averaged KL divergence from the exact within-category joint distribution to the product of its marginals. It measures the distributional error of the conditional-independence approximation, rather than a fixed offset in the evaluated free energy. Bayes' rule then specifies how evidence from different agents should be combined to infer the shared category.

The next transition concerns implementation. Cognitive closure prevents an agent from directly evaluating another agent's private evidence.  For two agents with fixed private evidence and model parameters, the MHNG word update preserves the collective sign posterior: the prior and speaker likelihood cancel in the Metropolis--Hastings ratio, leaving a decision based only on the listener's likelihood. Thus an exchange of public signs can perform collective Bayesian inference while each participant uses only locally available information. In an evolutionary or cultural interpretation, acquiring such a proposal-and-acceptance practice would constitute the invention of a decentralized inference procedure. MHNG provides a constructive realization of this possibility; the present model specifies the procedure and tests its consequences, but does not model the process by which a population acquires it.

The final transition connects this inferential role to reference. A discrete variable available for exchange becomes useful as a shared sign when participants associate it with corresponding categories in their private models and reuse those associations across episodes. The learning experiments examine this step by allowing  independently initialized private category distributions without cross-agent label matching to change during exchange. They assess both agreement in sign use and correspondence to world categories. These are distinct requirements: shared use alone does not establish reference, and a decrease in $F_d$ alone does not establish semantic agreement (Results, Sec.~\ref{results:toy}). The proposed connection to language emergence is therefore the formation of reusable shared referential symbols through decentralized inference. The emergence of word forms, syntax, and the evolutionary origin of the exchange protocol remain outside the model.

The scope of each transition is explicit. The shared-index construction is exact for a two-agent joint-attention prior and also applies to the global-joint-attention reference system. Reduction of the tempered pairwise prior for $N>2$ to a single shared index remains open. Likewise, the fixed-parameter two-agent stationarity result does not imply that pairwise RMHNG samples the $N$-agent collective posterior or monotonically decreases $F_d$.  The population extension examines the behavior of that multi-agent extension. These distinctions make the proposed route a sequence of a derived representation, an effective model, a decentralized implementation, and an experimentally examined account of shared reference.


\section{Two-agent toy-learning results}
\label{supp:toy-learning-summary}

\subsection[Learning from independent initialization]{Learning from independent initialization}
MHNG achieved lower mean $F_d$ and higher world ARI than No-comm.\ and All-accept across the 16 seeds. The paired $F_d$ interval for MHNG versus Central Gibbs includes zero, leaving that difference uncertain. Table~\ref{tab:toy-learning-summary} gives the scores and paired contrasts. Exact JA provides supplied-model reference scores alongside the five learned methods.

\begin{table}[b]
\centering
\caption{Two-agent toy-learning scores and paired differences across 16 seeds. Top: mean world-referenced $F_d$, world ARI, and private-sign agreement measured by Cohen's $\kappa$. Dashes indicate quantities not reproduced in this numerical summary. Bottom: paired differences, defined as MHNG minus the comparison method; negative values favor MHNG. Intervals are 95\% percentile seed-bootstrap intervals using 20,000 resamples. Evaluation distributions and update budgets follow Fig.~\ref{fig:exp1-main} and Materials and Methods, Sec.~\ref{methods:toy-readout}.}
\label{tab:toy-learning-summary}
\begingroup\small
\begin{tabular}{lrrr}
\toprule
Method & Mean $F_d$ [nat] & Mean world ARI & Mean $\kappa$ \\
\midrule
Exact JA & $-0.915$ & $0.992$ & --- \\
Central VI & $-0.695$ & $0.907$ & --- \\
Central Gibbs & $-0.593$ & $0.908$ & --- \\
MHNG & $-0.603$ & $0.916$ & $0.973$ \\
All-accept & $+0.123$ & $0.466$ & $0.503$ \\
No-comm. & $-0.068$ & $0.653$ & $0.033$ \\
\bottomrule
\end{tabular}
\par\vspace{8pt}
\begin{tabular}{lrr}
\toprule
Paired comparison & Mean $\Delta F_d$ [nat] & 95\% interval \\
\midrule
MHNG $-$ No-comm. & $-0.535$ & $[-0.649,-0.422]$ \\
MHNG $-$ All-accept & $-0.726$ & $[-0.893,-0.570]$ \\
MHNG $-$ Central Gibbs & $-0.010$ & $[-0.089,+0.070]$ \\
\bottomrule
\end{tabular}
\endgroup
\end{table}

High agreement on signs can coexist with poor category recovery. Figure~\ref{fig:exp1-main} shows individual runs, and all 16 seeds contribute to the summaries.

\clearpage
\begingroup\setlength{\emergencystretch}{3em}
Algorithm~\ref{alg:toy-mhng} gives the update shared by the two-agent toy experiment, the population extension, and the image experiment. It follows the speaker--listener exchange of the original MHNG~\cite{taniguchi2023emergent}. A directed sweep specifies a speaker, a listener, and a set of paired targets. We write $\ell_i(d,v)=\log p_i(z_{i,d}\mid v,\theta_i)$ for the local category log likelihood.\par\endgroup

\begin{algorithm}[H]
\caption[MHNG with a protocol-specific exchange schedule]{MHNG with a protocol-specific exchange schedule}
\label{alg:toy-mhng}
\small
\algrenewcommand\algorithmicrequire{\textbf{Input:}}
\algrenewcommand\algorithmicensure{\textbf{Output:}}
\begin{algorithmic}[1]
\Require Private observations $o_i$, uniform sign prior, and initial local states $(w_i,z_i,\theta_i)$
\Require Schedule $(s_t,l_t,\mathcal B_t)_{t=1}^{T}$ of directed sweeps and paired targets
\Ensure Updated private signs, latent states, and category parameters for each agent
\For{$t=1,\ldots,T$}
    \State $(s,l,\mathcal B)\gets(s_t,l_t,\mathcal B_t)$
    \For{each $(d_s,d_l)\in\mathcal B$}
        \State Speaker $s$ draws $v$ with probability proportional to $\exp\ell_s(d_s,v)$
        \State Speaker $s$ sends $v$ and keeps its stored sign $w_{s,d_s}$
        \State $a\gets\exp\{\min[0,\ell_l(d_l,v)-\ell_l(d_l,w_{l,d_l})]\}$
        \State Draw $u\sim\operatorname{Uniform}[0,1)$
        \If{$u<a$}
            \State $w_{l,d_l}\gets v$
        \EndIf
    \EndFor
    \State Draw $\theta_l$ from its conjugate posterior given all local $z_l,w_l$
    \If{the protocol resamples latent states}
        \State Draw $z_{l,d}\sim p_l(z_{l,d}\mid o_{l,d},w_{l,d},\theta_l)$ for all local targets $d$
    \EndIf
\EndFor
\end{algorithmic}
\end{algorithm}

\begingroup\setlength{\emergencystretch}{3em}
\textit{Two-agent toy experiment.} Each agent initializes from its own observation clusters with independently permuted sign names. The agents alternate directions and exchange signs for all objects on each sweep. The listener samples its category parameters and then its latent states (Sec.~\ref{subsec:exp1}). The All-accept control adopts every proposed sign.

\textit{Population extension (RMHNG condition).} All agents start from the category correspondence supplied by agent 0. At each block, randomly partition the population into disjoint pairs. For each pair, choose its first speaker at random and alternate directions for five round trips, using all objects in each direction. Only the current listener updates its category parameters and latent states. Each exchange uses evidence from the current pair (Supplementary Text~\ref{subsec:exp2}).

\textit{Image experiment.} Initialize the two private visual models and signs as in Sec.~\ref{methods:images}. For each round, independently reshuffle the two agents' views within scan and select a minibatch. Use the same pairs for A-to-B and B-to-A exchanges. The listener samples its GMM parameters from all 720 local latent states and retained signs after each direction, so B uses its updated GMM when proposing back to A. Skip latent-state resampling throughout naming because both VAEs and latent samples are fixed.\par\endgroup

\FloatBarrier

\section{Population extension: protocol and results}
\label{supp:population-results}
\subsection{Initialization, update schedule, and evaluation}
\label{subsec:exp2}

We examine multi-agent inference and learning after supplying an initial category correspondence. Central VI, Central Gibbs, RMHNG, and No-comm. use the latent Gaussian PGM and NIW prior of  the two-agent toy experiment, with a uniform word prior. All four methods receive agent 0's initial observation-clustering labels for the same $D$ objects. Each agent initializes its own NIW factor using its own observations under those supplied labels. The sampling methods start from identical parameter and perceptual states. This initialization supplies correspondence; it does not use true world categories or establish correspondence through communication. After initialization, No-comm. accesses only the agent's own observations and private variables.

The population condition here uses random-pair MH exchanges. It is distinct from the recursive Metropolis--Hastings naming game of Inukai et al.\ \cite{inukai2023recursive}, which recursively combines evidence across agents; that method's inference guarantees are not transferred to the present update schedule.

RMHNG retains a private word $w_{i,d}$, perceptual state $z_{i,d}$,  and category parameters for every agent. A speaker proposes a word from its own latent-state likelihood, while the listener applies its own likelihood ratio to accept or reject the proposal. The proposal does not overwrite the speaker's stored word. Only the listener then  samples its category parameters conditional on its current private words and perceptual states, followed by its perceptual states conditional on its observations, words, and new parameters. There is no public word sequence and no all-agent refitting after an utterance.

The communication schedule balances local updates across agents. At each block, a random permutation partitions the even-sized population into disjoint pairs. Each pair performs five round trips, with its first speaker chosen by a fair coin. After 200 blocks, every agent has received exactly 1000 listener updates and has acted as speaker exactly 1000 times; directed communication therefore totals $1000N$ utterances, each covering all $D$ objects. Partners vary across blocks, and individual pair-visit counts need not be equal. This schedule fixes individual update counts rather than communication per pair.

\begingroup\setlength{\emergencystretch}{2em}
Central Gibbs draws a common word from the product of all $N$ agents' latent-state likelihoods, then  updates every agent's category parameters and perceptual states. No-comm. samples each agent's  private word, category parameters, and perceptual states using only its local model. Both run for 1000 local parameter/perception updates per agent, as does RMHNG. Central VI uses the same structured variational family generalized to $N$ agents, with a 300-iteration cap and relative ELBO stopping tolerance $10^{-10}$; all 64 population-size/seed combinations satisfy this criterion in 23--100 iterations. VI iterations are not equated with sampling updates, and computational or communication costs are not matched. Parameter and perception random streams are shared across sampling methods within each seed, as in  the two-agent toy experiment.\par\endgroup

The primary cell remains $(\sigma_\xi,\sigma_{\mathrm{obs}})=(0.1,0.10)$, $M=4$, $D=256$, with all 16 seeds $0,\ldots,15$ at each $N=2,4,8,16$. We score every method using the same world-derived pairwise $F_d$, normalized by $N$. For the samplers, the readout is the terminal Gaussian conditional on observations, private or common words, and sampled parameters; it is not the free energy of the empirical MCMC distribution. VI uses its soft Gaussian mixture. ARI is averaged over agents within a seed and then over seeds. The seed, not an object or an agent, is the independent replicate. The shared initialization means these $N=2$ results are not interchangeable with  the two-agent toy experiment's independent-initialization results.

The evaluation reference uses a component-conditioned linear recognition channel. Writing $m_k$ for the mean of world component $k$ and $s=\sigma_\xi^2/(\sigma_\xi^2+\sigma_{\mathrm{obs}}^2)$, this channel is $r_i(z\mid\xi,k)=\mathcal N(z;sA_i\xi+(1-s)A_im_k,s(1+s)\sigma_{\mathrm{obs}}^2I_2)$. Marginalizing $\xi\mid k$ gives diagonal covariance blocks $\sigma_\xi^2I_2$ and cross-block $s^2\sigma_\xi^2A_iA_j^\top$ for the reference pair prior. This construction conditions the recognition component on the generating category; it is not the exact channel obtained by integrating the full soft mixture posterior $p_i(z\mid o)$. The resulting reference functional is common to all methods. Its world parameters and component labels are used only to construct the evaluation density, not supplied to learned inference.

\subsection{Population-extension results}
\begin{figure}[tb!p]
\centering
\makebox[0.49\linewidth]{\small (a) Collective free energy per agent}\hfill
\makebox[0.49\linewidth]{\small (b) World-category agreement}\par\smallskip
\includegraphics[width=\linewidth]{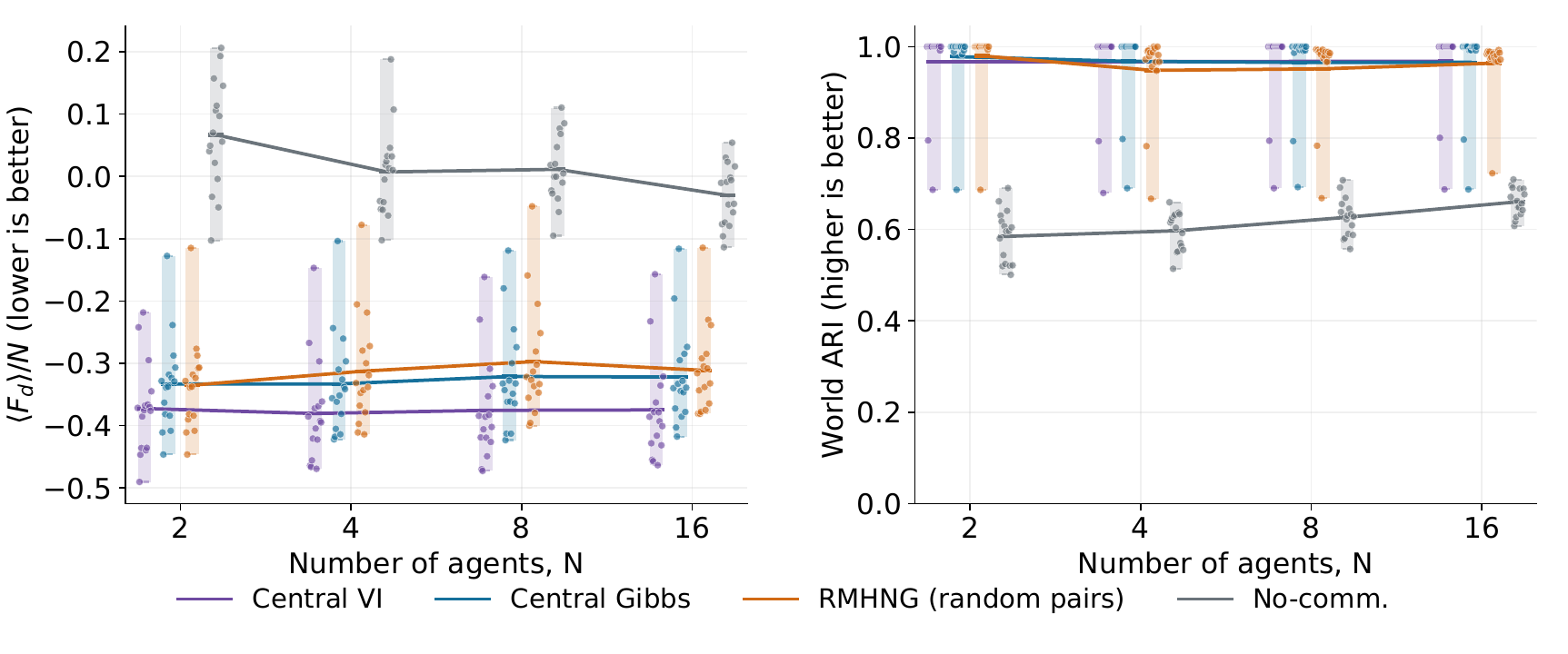}
\caption{ The population extension: matched latent-PGM methods after shared-label initialization. Cell $(\sigma_\xi,\sigma_{\mathrm{obs}})=(0.1,0.10)$, $M=4$, $D=256$; all 16 seeds at each population size. The initial correspondence is supplied to every method, unlike independent initialization in  the two-agent toy experiment. Samplers perform 1000 parameter/perception updates per agent; Central VI uses ELBO stopping. (a) Mean pairwise $F_d/N$ (lower is better), using terminal conditional Gaussians for samplers and the soft mixture for VI. (b) World ARI, averaged over agents within a seed. Dots are individual seeds, connected lines and short horizontal bars are seed means, and pale ranges span the seed minimum and maximum, not confidence intervals. Individual seeds are not connected. Horizontal positions are offset for readability; the axis is logarithmic with ticks $2,4,8,16$. The legend is outside the plotting boxes. }
\label{fig:exp2-rmhng}
\end{figure}

RMHNG achieved lower mean $F_d/N$ and higher mean world ARI than No-comm.\ at every tested population size (Fig.~\ref{fig:exp2-rmhng}). In paired comparisons, its $F_d/N$ was lower in every run, and its ARI was higher in all but one. RMHNG and the centralized methods had similar mean ARI, while Central VI gave the lowest mean $F_d/N$ at every size.

RMHNG also achieved higher private-sign agreement than No-comm.\ at every tested size. The centralized methods use a common sign per object and therefore agree by construction. Neither category recovery nor sign agreement improved monotonically with population size. The summaries include all seeds, including the lower-ARI RMHNG run at $N=16$ (seed 6).

For $N>2$, pairwise communication is not asserted to sample the full collective posterior, and neither the adaptive sampler nor VI directly minimizes the external pairwise $F_d$. Reduction of the tempered pairwise prior to one shared index remains open.  The population extension did not test robustness across observation geometries generated by redrawing the projections.

\clearpage 
\section{Construction of the multiview image stimuli}
\label{supp:image-stimuli}
\begin{figure}[H] 
\centering
{\small\textbf{(a) Ten textured 3D scans: source-asset previews}\par\smallskip}
\includegraphics[width=\linewidth]{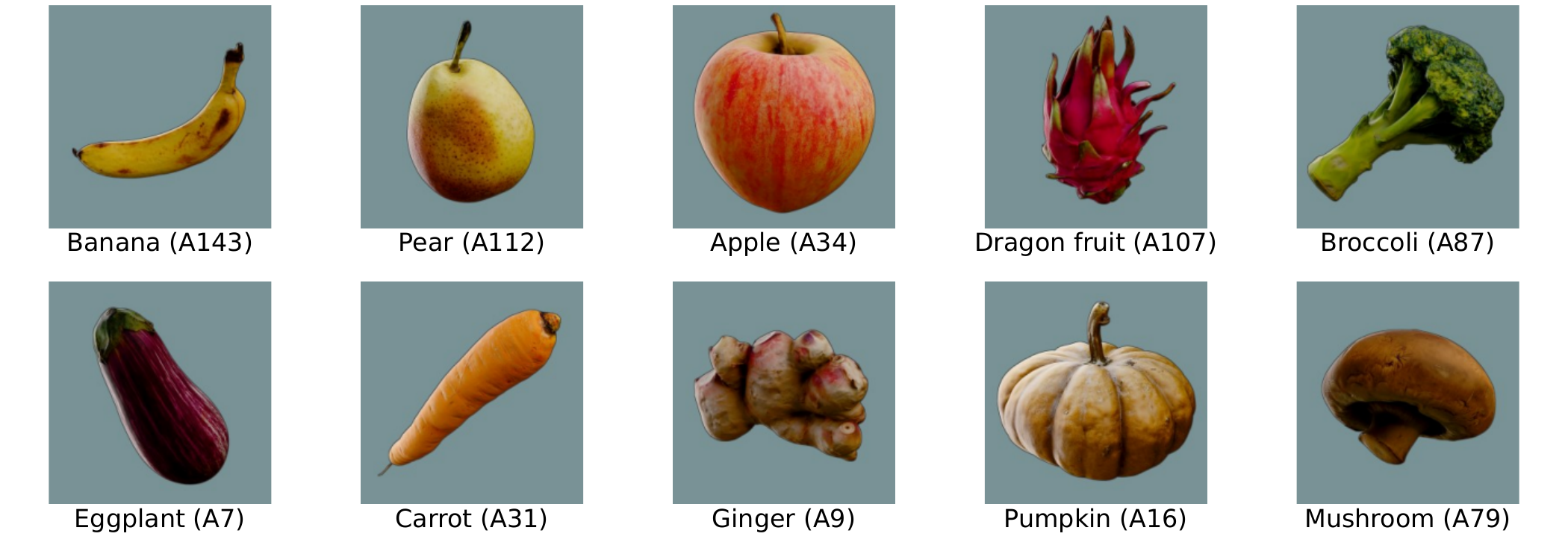}\par\smallskip
{\small\textbf{ (b) Fixed azimuth split, repeated at each base elevation}\par\smallskip}
\includegraphics[width=\linewidth]{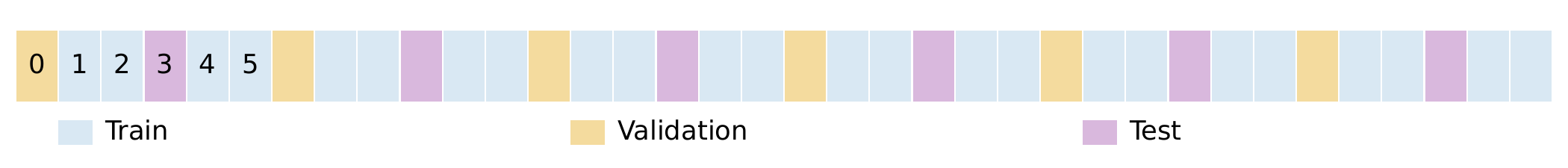}
\caption[Source scans and view split for the image experiment]{ Source scans and view split for the image experiment. (a) Source-pack previews of the ten textured 3D scans, one per object type, from the KIRI Engine asset pack (CC0) \cite{kiriFruitVegAssets}. (b) The 36 base azimuths, with the first six indices labeled. The train/validation/test pattern repeats every six positions at each of three base elevations. Each scan supplies 108 camera pairs; the complete dataset contains 720/180/180 training/validation/test pairs. The split holds out views of the same scans, not object identities.}
\label{fig:exp3-stimulus-construction}
\end{figure}

Figure~\ref{fig:exp3-design}(a) shows an example camera pair and its private RGB images. Supplementary Fig.~\ref{fig:exp3-stimulus-construction} shows the full set of source scans and the view split. The same textured scan generates both private views. Camera geometry, rendering, resizing, and the view split are fixed before learning; the agents receive the resulting images, not the 3D assets or camera poses. After centering and uniformly scaling the mesh, Blender Cycles renders orthographic RGB views with fixed lighting and background at 16 samples per pixel. Agent A uses base azimuth $\alpha\in\{0^\circ,10^\circ,\ldots,350^\circ\}$ and elevation $\epsilon\in\{10^\circ,25^\circ,40^\circ\}$; B uses $(\alpha+60^\circ,\epsilon+15^\circ)$. For $j=\alpha/10^\circ$, $j\bmod6=0$ is validation, $j\bmod6=3$ is test, and all other indices are training. This gives 72/18/18 training/validation/test view pairs per scan, or 720/180/180 overall. Held-out views depict the same scans, rather than new object identities.  During category learning, each agent's training views are independently reshuffled within each scan at every minibatch round, then a subset of targets is selected for communication; the illustrated camera pair does not remain fixed throughout learning.

\begin{figure}[!tb]
\centering
\includegraphics[width=\linewidth]{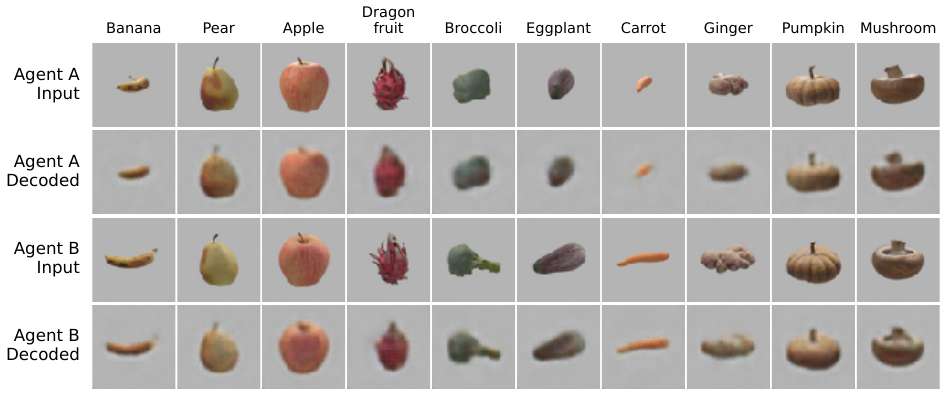}
\caption[Private observations and reconstructions]{Private observations and reconstructions from the fixed seed-0 VAEs after 1,000 updates. Columns follow the ten scans in dataset order. Rows show Agent A's input and posterior-mean reconstruction, followed by Agent B's input and reconstruction. Each column uses the first stored training view of that scan. All panels come from saved images and VAE outputs; the corresponding checkpoint hashes match those used in the ten-seed experiment. These are training views, and the displayed reconstructions are common to all four methods.}
\label{fig:exp3-images}
\end{figure}

\section{Image readouts with fixed visual representations}
\label{supp:image-frozen-results}
All four methods are reported across ten independent seed pairs of private VAEs. The VAEs and sampled latent states remain fixed during GMM and sign learning. The following tables use the terminal 72,000-target state for every seed; all readouts concern retained training-image signs and fixed encoder distributions, not held-out predictions.
Table~\ref{tab:exp3-results} summarizes category recovery and cross-agent sign agreement. MHNG and Central Gibbs show similar mean ARI, while Central Gibbs has higher mean sign agreement; both outperform All-accept and No-comm. on these descriptive readouts.

\begin{table}[!tb]
\centering
\caption{Retained-sign readouts across all ten seeds (mean $\pm$ sample SD). ARI is reported for A and B separately and for their within-seed average. Agreement uses raw integers over all 51,840 same-scan training-view combinations without post-hoc alignment.}
\label{tab:exp3-results}
\small
\begin{tabular}{lrrrr}
\toprule
Method & ARI A & ARI B & Mean ARI & Agreement \\
\midrule
MHNG & $0.834\pm0.050$ & $0.834\pm0.051$ & $0.834\pm0.050$ & $0.931\pm0.030$ \\
All-accept & $0.538\pm0.067$ & $0.543\pm0.069$ & $0.540\pm0.067$ & $0.611\pm0.048$ \\
Central Gibbs & $0.842\pm0.045$ & $0.843\pm0.043$ & $0.843\pm0.044$ & $0.987\pm0.006$ \\
No-comm. & $0.569\pm0.067$ & $0.529\pm0.074$ & $0.549\pm0.046$ & $0.107\pm0.059$ \\
\bottomrule
\end{tabular}
\end{table}

Table~\ref{tab:exp3-regularization} decomposes the fixed-representation score into individual regularization $I$ and collective regularization $C$. MHNG has a lower mean total $I+C$ than All-accept and No-comm., and a mean close to Central Gibbs. Because the visual representations are fixed within each seed, this comparison concerns the regularization contribution to CFE rather than an absolute image CFE.

\begin{table}[!tb]
\centering
\caption{Individual and collective regularization and their sum across all ten seeds (mean $\pm$ sample SD), in nats per observation pair. The sum is a fixed-representation regularization score and is not reported as an absolute image CFE.}
\label{tab:exp3-regularization}
\small
\begin{tabular}{lrrr}
\toprule
Method & Individual $I$ & Collective $C$ & Sum $I+C$ \\
\midrule
MHNG & $21.536\pm0.382$ & $2.273\pm0.012$ & $23.809\pm0.373$ \\
All-accept & $22.844\pm0.501$ & $1.948\pm0.069$ & $24.792\pm0.445$ \\
Central Gibbs & $21.394\pm0.360$ & $2.292\pm0.003$ & $23.687\pm0.360$ \\
No-comm. & $28.287\pm0.953$ & $1.955\pm0.044$ & $30.242\pm0.938$ \\
\bottomrule
\end{tabular}
\end{table}

\begin{figure}[!tb]
\centering
\includegraphics[width=\linewidth]{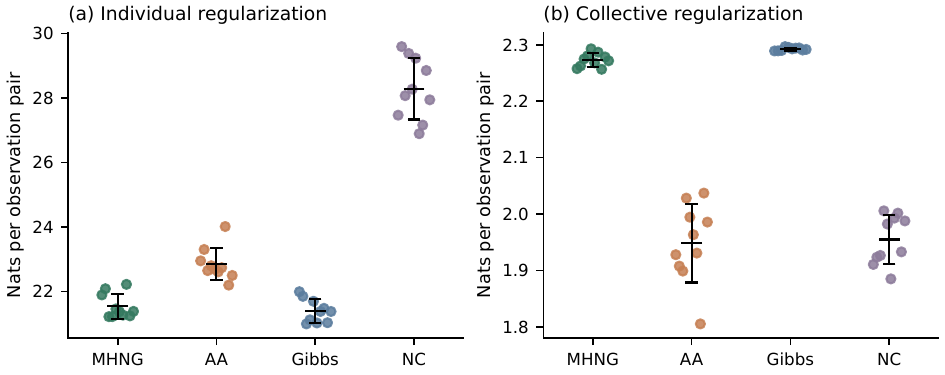}
\caption{Components of the fixed-representation regularization score. Points show all ten seeds and black marks the means with sample SD. AA, Gibbs, and NC denote All-accept, Central Gibbs, and No-comm. (a) Individual regularization, the shared-sign-weighted sum of Gaussian recognition-to-category KL divergences. (b) Collective regularization, the optimized shared-sign factor's KL divergence from the uniform prior. MHNG has a lower individual term but a higher collective term than All-accept; the sum, not either component alone, is the reported model-compatibility score.}
\label{fig:exp3-regularization-terms}
\end{figure}

Figure~\ref{fig:exp3-regularization-terms} shows the individual seeds for the two regularization terms. Relative to All-accept, MHNG reduces the individual term despite increasing the collective term, yielding the lower total reported in Table~\ref{tab:exp3-regularization}. The components must therefore be interpreted together.

Table~\ref{tab:exp3-all-contrasts} reports all pairwise tests for retained-word ARI, raw word agreement, and $I+C$, with Holm correction across all 18 comparisons. MHNG differs from All-accept and No-comm. in the favorable direction on all three endpoints. The comparison with Central Gibbs detects a difference only in word agreement, which is lower for MHNG; the other two comparisons do not establish equivalence.

\begin{table}[!tb]
\centering
\caption{All 18 pairwise comparisons of the final values in Fig.~\ref{fig:exp3-results}(d--f). Mean differences are calculated as method A minus method B across the ten paired seeds. Differences in $I+C$ are in nats per observation pair. The two-sided exact paired sign-flip tests use Holm correction across all 18 comparisons. \textsuperscript{*} indicates Holm-adjusted $p<0.05$. The smallest attainable raw $p$ value in the two-sided exact sign-flip test with ten paired seeds is $2/2^{10}\approx0.00195$.}
\label{tab:exp3-all-contrasts}
\small
\begin{tabular}{lllrrr}
\toprule
Endpoint & Method A & Method B & Mean diff. & Raw $p$ & Holm $p$ \\
\midrule
Retained-word ARI & MHNG & All-accept & $+0.294$ & 0.00195 & 0.0352\rlap{\textsuperscript{*}} \\
 & MHNG & Central Gibbs & $-0.009$ & 0.650 & 1.00 \\
 & MHNG & No-comm. & $+0.285$ & 0.00195 & 0.0352\rlap{\textsuperscript{*}} \\
 & All-accept & Central Gibbs & $-0.302$ & 0.00195 & 0.0352\rlap{\textsuperscript{*}} \\
 & All-accept & No-comm. & $-0.008$ & 0.717 & 1.00 \\
 & Central Gibbs & No-comm. & $+0.294$ & 0.00195 & 0.0352\rlap{\textsuperscript{*}} \\
\midrule
Raw word agreement & MHNG & All-accept & $+0.320$ & 0.00195 & 0.0352\rlap{\textsuperscript{*}} \\
 & MHNG & Central Gibbs & $-0.056$ & 0.00391 & 0.0352\rlap{\textsuperscript{*}} \\
 & MHNG & No-comm. & $+0.824$ & 0.00195 & 0.0352\rlap{\textsuperscript{*}} \\
 & All-accept & Central Gibbs & $-0.376$ & 0.00195 & 0.0352\rlap{\textsuperscript{*}} \\
 & All-accept & No-comm. & $+0.504$ & 0.00195 & 0.0352\rlap{\textsuperscript{*}} \\
 & Central Gibbs & No-comm. & $+0.880$ & 0.00195 & 0.0352\rlap{\textsuperscript{*}} \\
\midrule
$I+C$ & MHNG & All-accept & $-0.982$ & 0.00195 & 0.0352\rlap{\textsuperscript{*}} \\
 & MHNG & Central Gibbs & $+0.122$ & 0.189 & 0.568 \\
 & MHNG & No-comm. & $-6.433$ & 0.00195 & 0.0352\rlap{\textsuperscript{*}} \\
 & All-accept & Central Gibbs & $+1.105$ & 0.00195 & 0.0352\rlap{\textsuperscript{*}} \\
 & All-accept & No-comm. & $-5.451$ & 0.00195 & 0.0352\rlap{\textsuperscript{*}} \\
 & Central Gibbs & No-comm. & $-6.555$ & 0.00195 & 0.0352\rlap{\textsuperscript{*}} \\
\bottomrule
\end{tabular}
\end{table}

\FloatBarrier
\begingroup
\makeatletter
\Hy@hypertexnamesfalse
\makeatother

\section{Physics scheme of joint attention and CPC}
\label{app:physics:sec:physics}

The central idea of the main text is that joint attention couples private inference and supports the emergence of shared symbols that mediate this coupling.
The dependence between private observations of the same object induces a joint prior that couples local states.
The categorical structure of the world then allows this dependence to be represented through a shared index $w$
(main-text Eqs.~\eqref{eq:results-common-cause}--\eqref{eq:results-cpc}).
This structure---a common world, distinct local states, and a collective description through a shared variable---has a natural counterpart in the physical description of systems through order parameters.

We first identify the Gibbs distribution determined by a Hamiltonian with the world distribution $P_{\mathrm{ideal}}$ in the main text (Sec.~\ref{app:physics:sec:gibbs}).
Using the Curie--Weiss model, we show explicitly how interactions between spins can be represented through a shared order parameter, and how the world distribution separates into a mixture with weights $\pi_w$ and conditional configuration distributions $P_w(S)$.
Building on this example, Sec.~\ref{app:physics:sec:setup} sets up a physical system in which a population of local distributions interacts.

 The population average gives an exact rewriting of the interaction under a common-coordinate kernel. An additional effective model treats the shared distribution as an independent variable and exhibits a CPC-like structure.
Section~\ref{app:physics:sec:classicalfield} then shows how this physical system can be mapped explicitly to a classical scalar field theory.

To do so, we introduce the square roots of the probability densities as amplitude fields and  express the effective free energy in terms of couplings between local predictive fields and a shared order field.
A Fisher-information term that measures distributional smoothness, together with a local interaction kernel, yields the correspondence with classical scalar field theory.
This progression, from organizing the world distribution through an order parameter to describing the population through a shared variable and then to a field-theoretic representation, clarifies the physical basis of the main text's route from joint attention to shared symbols and of the broad applicability of CPC.

\begin{samepage}
The table below summarizes the correspondence between the main text and Supplementary Text and the physical description developed here. It can be consulted when relating the individual sections of this appendix to the main text.
\begingroup
\small
\begin{center}
\begin{tabular}{>{\raggedright\arraybackslash}p{0.27\linewidth}>{\raggedright\arraybackslash}p{0.60\linewidth}}
\toprule
Reference in the main text or Supplementary Text & Physical interpretation in this appendix\\
\midrule
Main-text Eq.~\eqref{eq:results-common-cause}; Eq.~\eqref{eq:marginal-calibration} & Joint prior obtained by locally reading out a common Gibbs configuration\\
Main-text Eq.~\eqref{eq:results-Fd}; Eqs.~\eqref{eq:pairwise-JA-unnormalized}--\eqref{eq:Fd-as-KL} & Mean-field free energy coupling private distributions\\
Main-text Eq.~\eqref{eq:results-mixture}; Eqs.~\eqref{eq:mixture-pair}--\eqref{eq:Vij-mixture} & Categories and weights $\pi_w$ defined by regions of an order parameter\\
Main-text Eq.~\eqref{eq:results-cpc}; Eqs.~\eqref{eq:eps-cat}--\eqref{eq:cpc-generative} & Shared order coordinate and the CPC approximation that summarizes within-phase structure\\
Main-text Eq.~\eqref{eq:results-cfe-decomposition} & Variational principle aligning private prediction and a shared description\\
Eqs.~\eqref{eq:pGJA-N-app}--\eqref{eq:GJA-generative} & Collective system referring to a single common microscopic configuration\\
\bottomrule
\end{tabular}
\end{center}
\endgroup
\end{samepage}

\subsection{The world distribution in physics: Gibbs distributions and order-parameter mixtures}
\label{app:physics:sec:gibbs}
\label{app:physics:sec:cw}
\label{app:physics:sec:category}
In physics, the probability distribution corresponding to the world distribution $P_{\mathrm{ideal}}$ in the main text is a standard object for a system in thermal equilibrium with a specified Hamiltonian and temperature.
For a microscopic configuration $S$, it is the Gibbs distribution\footnote{We denote the physical inverse temperature by $\beta_{\mathrm{phys}}=(k_{\mathrm B}T)^{-1}$ and the population normalization in the main text by $\beta_N=1/(N-1)$. Equations in this appendix are prefixed with P.}
\begin{equation}
 P_{\mathrm{ideal}}(S)=\frac{e^{-\beta_{\mathrm{phys}} H(S)}}{Z_{\mathrm{phys}}},
 \qquad Z_{\mathrm{phys}}=\sum_S e^{-\beta_{\mathrm{phys}} H(S)}.
 \tag{P1}\label{app:physics:eq:gibbs}
\end{equation}
Specifying the interactions in the physical system therefore also specifies the world distribution that generates its configurations.
The distribution $P_{\mathrm{ideal}}$ in the main text has a natural counterpart in this established physical probability distribution.

In many physical systems, an order parameter describes the collective structure of this distribution.
Specifying the order-parameter state gives a distribution over microscopic configurations, and the full distribution is a weighted mixture of these conditional distributions.
The categorical mixture $P_{\mathrm{ideal}}(S)=\sum_w\pi_w P_w(S)$ in the main text corresponds to this physical description.
Here $\pi_w$ is the weight of a state identified by the order parameter, and $P_w(S)$ is the configuration distribution in that state.
We verify this structure by a direct algebraic calculation in the simplest zero-field Curie--Weiss model.

For $n$ spins $s_i\in\{-1,+1\}$, consider
\begin{equation}
 H_{\mathrm{CW}}(S)=-\frac{J}{2n}\left(\sum_i s_i\right)^2,
 \qquad J>0,\qquad m(S)=\frac{1}{n}\sum_i s_i.
 \tag{P2}\label{app:physics:eq:cw}
\end{equation}
We introduce a continuous order parameter $\varphi\in\mathbb{R}$ common to all spins.
The variable $\varphi$ corresponds to the shared variable $w$ in the main text in its role of connecting local states through a common reference.
We use $\varphi$ for this continuous description and $w$ for the categories subsequently obtained by coarse-graining.
Completing the square in a Gaussian integral, or equivalently applying the Hubbard--Stratonovich transformation, gives
\begin{align}
 e^{\frac{\beta_{\mathrm{phys}} J}{2n}(\sum_i s_i)^2}
 &=C_n\int_{\mathbb{R}}d\varphi\,
 e^{-\frac{n\beta_{\mathrm{phys}} J}{2}\varphi^2+\beta_{\mathrm{phys}} J\varphi\sum_i s_i}
 \tag{P3}\label{app:physics:eq:hs}\\
 &=C_n e^{\beta_{\mathrm{phys}} Jn/2}\int_{\mathbb{R}}d\varphi\,
 e^{-\frac{\beta_{\mathrm{phys}} J}{2}\sum_i(\varphi-s_i)^2},
 \qquad C_n=\sqrt{\frac{n\beta_{\mathrm{phys}} J}{2\pi}}.
 \nonumber
\end{align}
The interaction between spins is thus reproduced by integrating over couplings defined by squared differences between a common variable $\varphi$ and the individual spins.
Substituting this identity into the Gibbs distribution and normalizing the conditional distribution of each spin gives
\begin{align}
 P(\varphi)&=\frac{C_n}{Z_{\mathrm{CW}}}
 e^{-n\beta_{\mathrm{phys}} J\varphi^2/2}[2\cosh(\beta_{\mathrm{phys}} J\varphi)]^n,
 \tag{P4}\label{app:physics:eq:fieldprior}\\
 P(S\mid\varphi)&=\prod_i\frac{e^{\beta_{\mathrm{phys}} J\varphi s_i}}{2\cosh(\beta_{\mathrm{phys}} J\varphi)},
 \qquad
 P_{\mathrm{ideal}}(S)=\int_{\mathbb{R}}P(\varphi)P(S\mid\varphi)d\varphi.
 \tag{P5}\label{app:physics:eq:cwfactor}
\end{align}
Here $Z_{\mathrm{CW}}=\sum_S e^{-\beta_{\mathrm{phys}} H_{\mathrm{CW}}(S)}$.
The same completion of the square gives $\mathbb{E}[\varphi\mid S]=m(S)$, so $\varphi$ can be read as a shared order coordinate representing magnetization.
The world distribution therefore has an exact mixture representation in terms of the order-parameter distribution and the configuration distribution conditional on that order parameter.

The connection to the discrete categories $w$ in the main text follows by grouping this integral over regions $B_w$ that partition the order-parameter space.
For example, at low temperature, $\beta_{\mathrm{phys}} J>1$, the two peaks corresponding to positive and negative magnetization can be read as $w=+,-$.
For each region, define
\begin{equation}
 \pi_w=\int_{B_w}P(\varphi)d\varphi,\qquad
 P_w(S)=\frac{1}{\pi_w}\int_{B_w}P(\varphi)P(S\mid\varphi)d\varphi.
 \tag{P6}\label{app:physics:eq:phasecomponents}
\end{equation}
Then $\sum_w\pi_w=1$, $\sum_S P_w(S)=1$, and
\begin{equation}
 P_{\mathrm{ideal}}(S)=\sum_w\pi_w P_w(S)
 \tag{P7}\label{app:physics:eq:worldmixture}
\end{equation}
holds exactly. At zero field, a symmetric choice of the positive and negative regions gives $\pi_+=\pi_-=1/2$.

Conditional independence at fixed $\varphi$ need not survive averaging over a region $B_w$. A product of spin marginals at fixed $w$ would therefore be a further approximation.
This representation first selects an order-parameter state $w$ with weight $\pi_w$ and then generates a configuration from $P_w(S)$.
It provides a physical realization of the categorical-world assumption underlying main-text Eq.~\eqref{eq:results-mixture}, namely the world mixture used in Supplementary Text~\ref{subsec:categorical-world} and Eq.~\eqref{eq:mixture-pair}.
The mixture structure in the main text thus agrees with the statistical description of a physical system organized by an order parameter; the calculation above makes this correspondence explicit for the Curie--Weiss model.

Hashimoto and Taniguchi's preprint \emph{Atom as an Agent: Emergent Order by Collective Predictive Coding}~\cite{hashimoto2026atomagent} has already discussed the relation between CPC and the emergence of order parameters from microscopic configurations.
It presents an example in which CPC learns collective order coordinates from local observations of configurations, providing prior work that supports the use of shared variables to describe physical systems.

\subsection{A physical setup of interacting local distributions gives rise to CPC}
\label{app:physics:sec:setup}
This section shows that a physical formulation of the CPC scheme naturally admits a shared variable $w$ that summarizes interactions between local degrees of freedom.
The Curie--Weiss model in Sec.~\ref{app:physics:sec:gibbs} provides a concrete example through the relation between the world distribution and an order parameter.
That section concerned $P_{\mathrm{ideal}}(S)$, the distribution over configurations $S$ of the common world. We now turn to interactions between local inference distributions $q_i(z_i)$ that capture this world from local perspectives.
Starting from local distributions whose interactions can be expressed in a common coordinate system, we redescribe the population through a shared distribution and obtain a local--global variational structure corresponding to CPC.
The distribution $q_i(z_i)$ denotes a local variational distribution with observations held fixed, and the shared coordinate is continuous, $w\in\mathbb{R}$. The discrete categories $w$ in the main text correspond to the coarse-graining of the order parameter described in Sec.~\ref{app:physics:sec:gibbs}.
This physical structure, in which relations between local states are summarized in a shared description, supports the main text's account of the route from joint attention to shared symbols and provides a natural physical interpretation of the broad applicability of CPC.

\subsubsection{Interacting physical systems with probabilistic variables and CPC}

The claim here is that a free-energy description of an interacting system with probabilistic variables gives a physical realization of the variational structure of CPC. The central structure is that local probabilistic variables interact and the system as a whole is redescribed through a collective variable. We represent the state of each degree of freedom $i$ by $z_i\in\mathbb{R}$ and examine how this structure provides a mathematical basis for CPC.

\paragraph{Local probability distributions.}

Represent the state of each degree of freedom $i$ by $z_i\in\mathbb{R}$ and use its probability distribution
\[
q_i(z_i),\quad z_i \in \mathbb{R},\quad \int_{\mathbb{R}} q_i(z_i)\,dz_i = 1
\]
as the local probabilistic variable. Treating each degree of freedom as a distribution rather than a point naturally introduces a physical counterpart of the variational distributions in CPC.

\paragraph{Free-energy functional.}

For the collection of local distributions, consider a free-energy functional consisting of entropy and interaction terms,

\begin{equation}
F[\{q_i\}] =
\underbrace{\sum_{i=1}^{N}\int_{\mathbb{R}} q_i(z_i)\ln q_i(z_i)\,dz_i}_{\text{ Negative individual entropy}}
-
\underbrace{\frac{1}{2N}\sum_{i,j=1}^{N}\int_{\mathbb{R}}\int_{\mathbb{R}} q_i(z_i)q_j(z_j)V(z_i,z_j)\,dz_i\,dz_j}_{\text{Interaction term}}.
\tag{P8}\label{app:physics:eq:local-free-energy}
\end{equation}

Here $V(z_i,z_j)$ is a general interaction kernel. In the reductions below, we assume the symmetry $V(z_i,z_j)=V(z_j,z_i)$ and normalize the coupling strength to one. In the main text, attending to a common object induces an interaction between private states and couples local inference (Eqs.~\eqref{eq:results-common-cause} and \eqref{eq:results-Fd}). Here we describe such coupling between local states as a physical model with a general interaction kernel.  The first term is the negative differential entropy, and the second describes coupling between degrees of freedom. Equation (P8) defines a separate physical model with self terms and its own coupling normalization. In particular,
\[
\int_{\mathbb{R}} q_i(z_i)\ln q_i(z_i)\,dz_i
\]
is a basic term that determines the shape of each local distribution. Physically, it acts as a self-term for the local distribution in the free energy. Whereas the interaction term couples degrees of freedom, this term assigns an intrinsic variational cost to each $q_i$ and controls how broad or concentrated the distribution becomes. The functional thus expresses the CPC balance between a term that shapes each local distribution and a term that connects different degrees of freedom.

\subsubsection{From a collective variable to the CPC structure}

We next introduce a collective variable for the population of local distributions and rewrite the physical system as a two-level structure of global and local variables. This makes explicit how an interacting physical system with probabilistic variables can exhibit the two-level variational structure shared with CPC.

\paragraph{Why introduce a collective variable?}

The original interaction term is
\[
\frac{1}{2N}\sum_{i,j=1}^{N}\iint q_i(z_i)q_j(z_j)V(z_i,z_j)\,dz_i\,dz_j.
\]
Evaluating it requires $O(N^2)$ operations and access to each pair $q_i,q_j$. We therefore introduce a macroscopic probabilistic variable $q(w)$ that summarizes the contributions of the other degrees of freedom and represents the overall context by a single distribution.

The local coordinate system for each $z_i$ and the reference coordinate system for the shared variable $w$ need not coincide. In particular, the shared coordinate system need not be identical to the local coordinates of each agent. For each degree of freedom, we therefore introduce an invertible coordinate transformation from the shared coordinate $w$ to the local coordinate $z_i$,
\[
z_i=h_i(w).
\]
We assume that $h_i$ is sufficiently smooth, with Jacobian $|h_i'(w)|$.

\paragraph{Self-consistent definition of the collective distribution.}

First, introduce the density obtained by expressing each local distribution in the shared coordinate,
\[
\tilde q_j(w)
:=
q_j(h_j(w))|h_j'(w)|
=
\int_{\mathbb{R}} q_j(z_j)\,\delta(h_j(w)-z_j)\,|h_j'(w)|\,dz_j.
\]
This transfers a local probability density expressed in the $z_j$ coordinate to the common $w$ coordinate system. We then define the collective distribution by
\[
q(w)=\frac{1}{N}\sum_{j=1}^{N}\tilde q_j(w)
=
\frac{1}{N}\sum_{j=1}^{N}\int_{\mathbb{R}} q_j(z_j)\,\delta(h_j(w)-z_j)\,|h_j'(w)|\,dz_j.
\]
This is a global probabilistic variable that directly represents the population-averaged distribution in the shared coordinate system, giving equal weight to each $q_j$ after its coordinate transformation. In this sense, $q(w)$ can be read as a candidate shared-context variable in CPC.\footnote{The following transformation rewrites the population average using the definition of the collective distribution.}

\paragraph{Reorganization into a two-level free energy.}

We rewrite the original free energy
\[
F_0[\{q_i\}] = \sum_{i=1}^{N}\int q_i(z_i)\ln q_i(z_i)\,dz_i - \frac{1}{2N}\sum_{i,j}\iint q_i(z_i)q_j(z_j)V_{ij}(z_i,z_j)\,dz_i\,dz_j
\]
as a coupled system of local and collective variables. Assume that the interaction kernels are induced by a symmetric kernel on the shared coordinates,
\[
\mathcal V(u,v)=\mathcal V(v,u)
\]
through
\[
V_{ij}(z_i,z_j):=\mathcal V\!\bigl(h_i^{-1}(z_i),h_j^{-1}(z_j)\bigr).
\]

\begin{enumerate}
\item Using the symmetry of the index pair, rewrite the interaction term as
\[
\begin{aligned}
\frac{1}{2N}\sum_{i,j}\iint q_i(z_i)q_j(z_j)V_{ij}(z_i,z_j)\,dz_i\,dz_j
&= \frac{1}{2}\sum_{i=1}^{N}\int_{\mathbb{R}} q_i(z_i)\,dz_i \\
&\qquad \times
\left(
\frac{1}{N}\sum_{j=1}^{N}\int_{\mathbb{R}} q_j(z_j)V_{ij}(z_i,z_j)\,dz_j
\right).
\end{aligned}
\]

\item Applying the change of variables $z_j=h_j(w)$ and the definition of the collective distribution gives
\[
\begin{aligned}
\frac{1}{N}\sum_{j=1}^{N}\int_{\mathbb{R}} q_j(z_j)V_{ij}(z_i,z_j)\,dz_j
&=
\frac{1}{N}\sum_{j=1}^{N}\int_{\mathbb{R}} q_j(h_j(w))|h_j'(w)|\,\mathcal V\!\bigl(h_i^{-1}(z_i),w\bigr)\,dw \\
&=
\int_{\mathbb{R}} q(w)\,\mathcal V\!\bigl(h_i^{-1}(z_i),w\bigr)\,dw.
\end{aligned}
\]
For brevity, write
\[
V_i(z_i,w):=\mathcal V\!\bigl(h_i^{-1}(z_i),w\bigr).
\]
The interaction term is therefore
\[
\frac{1}{2}\sum_{i=1}^{N}\int_{\mathbb{R}}\int_{\mathbb{R}} q_i(z_i)q(w)V_i(z_i,w)\,dz_i\,dw.
\]

\item Combining this expression with the individual entropy terms gives the exact dependent-variable representation

\[
\widetilde F_{\mathrm{exact}}[q,\{q_i\}] =
\underbrace{\sum_{i=1}^{N}\int_{\mathbb{R}} q_i(z_i)\ln q_i(z_i)\,dz_i}_{\text{ Negative local entropy}}
-
\frac{1}{2}\underbrace{\sum_{i=1}^{N}\int_{\mathbb{R}}\int_{\mathbb{R}} q(w)q_i(z_i)V_i(z_i,w)\,dz_i\,dw}_{\text{Local--global coupling}}.
\]

\end{enumerate}

At this stage, $q(w)$ is a dependent variable defined by
\[
q(w)=\frac{1}{N}\sum_{j=1}^{N}\int_{\mathbb{R}} q_j(z_j)\,\delta(h_j(w)-z_j)\,|h_j'(w)|\,dz_j
\]
and $\widetilde F_{\mathrm{exact}}[q,\{q_i\}]$ remains an exact mean-field redescription of the original model. The entropy terms remain those of the original collection of local distributions.

\paragraph{CPC-type effective extension and KL consistency term.}

So far, we have rewritten the average of the local distributions as a shared distribution. We now extend this description by treating a shared distribution that approximately represents this population average as an independently updated variational degree of freedom. These two stages lead to a two-level effective theory that describes both the population state and its shared representation. In doing so, we introduce a term that maintains consistency between the shared distribution and the original population distribution,
\[
\bar q(w):=\frac{1}{N}\sum_{j=1}^{N}\int_{\mathbb{R}} q_j(z_j)\,\delta(h_j(w)-z_j)\,|h_j'(w)|\,dz_j.
\]
Here $\bar q(w)$ is the average obtained by expressing each local distribution in the common $w$ coordinate system through its transformation $h_j$. We therefore consider the CPC-type two-level effective free energy

\[
\begin{aligned}
F_{\mathrm{eff}}[q,\{q_i\}]
&=
\underbrace{\int_{\mathbb{R}} q(w)\ln q(w)\,dw}_{\text{ Negative global entropy}}
+
\underbrace{\sum_{i=1}^{N}\int_{\mathbb{R}} q_i(z_i)\ln q_i(z_i)\,dz_i}_{\text{ Negative local entropy}} \\
&\quad
- \underbrace{\sum_{i=1}^{N}\int_{\mathbb{R}}\int_{\mathbb{R}} q(w)q_i(z_i)V_i(z_i,w)\,dz_i\,dw}_{\text{Local--global coupling}} \\
&\quad
+ \underbrace{\gamma\,\mathrm{KL}\!\left(q(w)\middle\|\bar q(w)\right)}_{\text{Consistency constraint}}.
\end{aligned}
\]
The overall normalization of the local--global coupling is absorbed into the definition of the effective coupling. Here
\[
\mathrm{KL}\!\left(q(w)\middle\|\bar q(w)\right)
=
\int_{\mathbb{R}} q(w)\ln\frac{q(w)}{\bar q(w)}\,dw
\]
and $\gamma>0$ controls the strength of the consistency constraint.

This model is a two-level effective theory with an independent shared field $q(w)$. Physically, a finite cost is assigned to deviations of the shared field $q(w)$ from the actual population distribution $\bar q(w)$. In the limit $\gamma\to\infty$, the constraint $q(w)=\bar q(w)$ makes the shared field coincide with the average of the local distributions. Even under the hard constraint, the global negative entropy remains, so the effective functional differs from the exact rewriting. For finite $\gamma$, the shared field is a macroscopic variable subject to a soft constraint relative to the collection of local distributions.

In this sense, the KL term implements an effective closure approximation with an independent global variable. Varying the local and shared distributions together aligns the population state with its shared description.

\paragraph{Mapping to CPC.}

The local and global entropy terms and the local--global coupling in the effective free energy correspond to the following structure in the CPC model. Observations are held fixed, and we use the mean-field distribution $q(w)\prod_i q_i(z_i)$.

\[
F[q(w),\{q(z_k)\}] =
\underbrace{\mathbb{E}_{q(w)}[\ln q(w)]}_{\text{ Negative global entropy}}
+
\underbrace{\sum_{k}\mathbb{E}_{q(z_k)}[\ln q(z_k\mid o_k)]}_{\text{ Negative local entropy}}
- \underbrace{\sum_{k}\mathbb{E}_{q(w,z_k)}[\ln p(z_k\mid w)]}_{\text{Local--global consistency}}.
\]
Here $q(w)$ corresponds to the variational distribution over the shared variable, and $q_i(z_i)\leftrightarrow q(z_k)$.
Choosing the local--global coupling as $V_i(z_i,w)=\ln p_i(z_i\mid w)$ and adding the shared-prior term $-\mathbb{E}_{q(w)}\ln p(w)$ and local observation-model terms $-\sum_i\mathbb{E}_{q_i}\ln p_i(o_i\mid z_i)$ gives a correspondence with the variational free energy in main-text Eq.~\eqref{eq:results-cfe-decomposition}.
A physical system with probabilistic variables can thus be read as an effective implementation of the mathematical structure underlying CPC: local distributions, a collective distribution, and their alignment through free energy.

The correspondence to Eq.~\eqref{eq:results-cfe-decomposition} concerns the terms shared by the two functionals. It uses normalized conditional densities and sets aside the additional closure penalty.

The physical model's $\gamma\,\mathrm{KL}(q\|\bar q)$ is an effective closure term that aligns the shared distribution with the average distribution of the local population. The collective regularization $\mathrm{KL}(q\|p(w))$ in the main text describes the relation between the approximate posterior over the shared variable and its prior.

This construction shows how a collective description of interactions between local distributions naturally introduces a shared distribution $q(w)$, and how consistency between local and shared distributions yields a CPC-type effective theory.
The Curie--Weiss model in Sec.~\ref{app:physics:sec:gibbs} is a concrete realization of this local--shared structure as a mixture of world distributions organized by an order parameter.
In the main text, joint attention to a common object induces the interaction (Eqs.~\eqref{eq:results-common-cause} and \eqref{eq:results-Fd}), while categorical structure gives a CPC model mediated by a shared variable (Eqs.~\eqref{eq:results-mixture} and \eqref{eq:results-cpc}).
Organizing interacting local degrees of freedom through a shared variable and aligning the two through free energy thus has a natural basis in physical systems whose collective state admits a common-coordinate description.
This correspondence provides a physical interpretation of how CPC-type descriptions naturally arise in a broad class of interacting systems, supporting the main text's route from joint attention to shared symbols.

\subsection{An explicit representation of CPC in classical scalar field theory}
\label{app:physics:sec:classicalfield}

We now express the CPC-type effective free energy from the preceding section in terms of classical scalar field theory.
The construction proceeds through a change of variables to amplitude fields, the introduction of a Fisher term for distributional smoothness, and an approximation by a local interaction kernel.
First, introduce amplitude fields defined as the square roots of the probability densities,
\[
q_i(z_i)=\psi_i(z_i)^2,\qquad q(w)=\Psi(w)^2.
\]
Here $w$ specifies a shared state, and $\Psi(w)^2=q(w)$ expresses the degree of support for that state. The normalization conditions become
\[
\int_{\mathbb{R}} \psi_i(z_i)^2\,dz_i = 1,\qquad \int_{\mathbb{R}} \Psi(w)^2\,dw = 1.
\]
The fields $\psi_i,\Psi$ can thus be viewed as the amplitudes of local predictive fields and a shared order field, respectively. As in Sec.~\ref{app:physics:sec:setup}, local and shared coordinates need not coincide. We therefore also use the amplitude field pulled back to the common coordinate,
\[
\tilde\psi_i(w):=\psi_i(h_i(w))\,|h_i'(w)|^{1/2}.
\]
Its square $\tilde\psi_i(w)^2$ is the density obtained by expressing the local distribution in the shared coordinate.

Under this change of variables, the entropy terms become
\[
q(w)\log q(w)\to \Psi(w)^2\log \Psi(w)^2,\qquad
q_i(z_i)\log q_i(z_i)\to \psi_i(z_i)^2\log \psi_i(z_i)^2.
\]
The free-energy terms originally introduced in information-theoretic terms therefore correspond to logarithmic self-potentials for the amplitude fields. Although these differ from polynomial potentials, they act as self-interaction potentials that locally constrain the magnitude of each field.

The CPC-type effective free energy in amplitude variables is
\[
\begin{aligned}
F_{\mathrm{amp}}[\Psi,\{\psi_i\}]
&=
\int_{\mathbb{R}} \Psi(w)^2\log \Psi(w)^2\,dw
+
\sum_{i=1}^{N}\int_{\mathbb{R}} \psi_i(z_i)^2\log \psi_i(z_i)^2\,dz_i \\
&\quad
- \sum_{i=1}^{N}\int_{\mathbb{R}}\int_{\mathbb{R}} \psi_i(z_i)^2\Psi(w)^2V_i(z_i,w)\,dz_i\,dw \\
&\quad
+ \gamma\,\mathrm{KL}\!\left(\Psi(w)^2\middle\|\bar q_{\psi}(w)\right),
\end{aligned}
\]
where
\[
\bar q_{\psi}(w):=\frac{1}{N}\sum_{j=1}^{N}\tilde\psi_j(w)^2
=
\frac{1}{N}\sum_{j=1}^{N}\int_{\mathbb{R}} \psi_j(z_j)^2\,\delta(h_j(w)-z_j)\,|h_j'(w)|\,dz_j.
\]
Here $\Psi$ corresponds to a collective predictive field or order parameter, $\psi_i$ to a local predictive field, and the coupling $\tilde\psi_i^2\Psi^2$ in the common coordinate represents a local--global predictive interaction.

\subsubsection{Fisher information and the kinetic term}

Next, add Fisher-information regularization to measure distributional smoothness,
\[
\frac{\kappa_{\Psi}}{4}\int_{\mathbb{R}} \frac{(\nabla q(w))^2}{q(w)}\,dw
+
\sum_{i=1}^{N}\frac{\kappa_i}{4}\int_{\mathbb{R}} \frac{(\nabla q_i(z_i))^2}{q_i(z_i)}\,dz_i.
\]
In one dimension, $\nabla$ can be read as $\partial$. In amplitude variables,
\[
\frac{(\nabla q)^2}{q}
=
\frac{\bigl(\nabla \Psi^2\bigr)^2}{\Psi^2}
=4(\nabla \Psi)^2,
\qquad
\frac{(\nabla q_i)^2}{q_i}
=4(\nabla \psi_i)^2,
\]
so this regularization becomes
\[
\kappa_{\Psi}\int_{\mathbb{R}} (\nabla \Psi)^2\,dw
+
\sum_{i=1}^{N}\kappa_i\int_{\mathbb{R}} (\nabla \psi_i)^2\,dz_i.
\]
The Fisher term can therefore be interpreted as regularization corresponding to the kinetic term in a scalar field theory.

\subsubsection{Ginzburg--Landau effective action}

Suppose further that the interaction kernel is local or short-ranged and can be approximated by
\[
\mathcal V(u,w)\simeq g\,\delta(u-w),
\]
or equivalently,
\[
V_i(z_i,w)\simeq g\,\delta(h_i^{-1}(z_i)-w).
\]
The coupling term then becomes
\[
- \sum_{i=1}^{N}\int_{\mathbb{R}}\int_{\mathbb{R}} \psi_i(z_i)^2\Psi(w)^2V_i(z_i,w)\,dz_i\,dw
\;\simeq\;
-g\sum_{i=1}^{N}\int_{\mathbb{R}} \tilde\psi_i(w)^2\Psi(w)^2\,dw.
\]
Combining this with the Fisher term gives

\[
\begin{aligned}
\Gamma_{\mathrm{GL}}[\Psi,\{\psi_i\}]
&=
\int_{\mathbb{R}}
\Bigl[
\kappa_{\Psi}(\nabla \Psi)^2 + U_{\Psi}(\Psi)
\Bigr]\,dw \\
&\quad
+ \sum_{i=1}^{N}\int_{\mathbb{R}}
\Bigl[
\kappa_i(\nabla \psi_i)^2 + U_i(\psi_i)
\Bigr]\,dz_i \\
&\quad
- g\sum_{i=1}^{N}\int_{\mathbb{R}} \tilde\psi_i(w)^2\Psi(w)^2\,dw\\
&\quad + \gamma\,\mathrm{KL}\!\left(\Psi^2\middle\|\bar q_{\psi}\right),
\end{aligned}
\]
which has the form of an effective action for a Ginzburg--Landau or interacting scalar field theory.

For example, the effective potentials can be written as

\[
U_{\Psi}(\Psi)
=
\Psi^2\log \Psi^2 + \Lambda \Psi^2 ,
\qquad
U_i(\psi_i)
=
\psi_i^2\log \psi_i^2 + \lambda_i \psi_i^2.
\]

Here $\Lambda,\lambda_i$ are multipliers for the normalization constraints.  The explicit KL term couples the shared and local fields and penalizes deviations between $\Psi^2$ and $\bar q_{\psi}$.  For a positive reference density and small relative deviations from self-consistency,
\[
\mathrm{KL}\!\left(\Psi^2\middle\|\bar q_{\psi}\right)
\approx
\frac12\int_{\mathbb{R}}
\frac{\bigl(\Psi(w)^2-\bar q_{\psi}(w)\bigr)^2}{\bar q_{\psi}(w)}\,dw,
\]
so this term can be viewed as a quartic-type effective potential that brings the shared order field and the mean distribution of the local population into alignment.

Under these changes of variables and physical conditions, CPC has an explicit representation as a mean-field classical field theory in which local predictive fields $\psi_i$ form a shared order field $\Psi$. Specifically,
\[
\Psi:\ \text{collective predictive field / order parameter},\qquad
\psi_i:\ \text{local predictive field},
\]
and
\[
\tilde\psi_i^2\Psi^2:\ \text{local-global predictive interaction},\qquad
\Psi^2\log \Psi^2:\ \text{logarithmic self-potential}.
\]

 With the added Fisher term and local-kernel approximation, the CPC-type physical model takes the form of an interacting scalar-field functional.

The structure in which local distributions form a shared field, and that field connects local states, corresponds to the role of the shared variable in the main text.
This physical description provides a natural interpretation of the main text's account of how joint attention gives rise to a referential structure for shared symbols.

\endgroup

\end{document}